\documentclass[11pt]{article}
\usepackage{geometry}                
\usepackage{amsmath}
\usepackage{vmargin}
\usepackage{graphicx,color}
\usepackage{subcaption}
\usepackage{overpic}
\usepackage{algorithm}
\usepackage{multirow}
\usepackage{natbib}

\newcommand{\inv}{^{-1} }
\newcommand{\itp}{^{\scriptsize -T} }
\newcommand{\trp}{^{\scriptsize T} }
\newcommand{\sbr}[1]{\ensuremath{_{\mathrm{#1}}}}
\newcommand{\spr}[1]{\ensuremath{^{\mathrm{#1}}}}

\newcommand{\cov}{\operatorname{cov}}

\def\trp{^{\mathrm{T}}}
\def\inv{^{-1}}
\def\itp{^{-\mathrm{T}}}
\title{Hybrid iterative ensemble smoother for history matching of hierarchical models}

\author{ Dean S. Oliver\footnote{ORCID: 0000-0001-6710-7791}\\
Energy and Technology Department \\
NORCE Norwegian Research Centre, Bergen, Norway
}

\begin{document}


\maketitle

\section{Abstract}

The choice of the prior model can have a large impact on the ability to assimilate data. In standard applications of ensemble-based data assimilation, all realizations in the initial ensemble  are generated from the same covariance matrix with the implicit assumption that this covariance is appropriate for the problem. In a hierarchical approach, the parameters of the covariance function, for example the variance, the orientation of the anisotropy and the ranges in two principal directions, may all be uncertain. Thus, the hierarchical approach is much more robust against model misspecification. 
In this paper, three approaches to sampling from the posterior for hierarchical parameterizations are discussed: an optimization-based sampling approach (RML), an iterative ensemble smoother (IES), and a novel hybrid of the previous two approaches (hybrid IES). 
The three approximate sampling methods are applied  to a linear-Gaussian inverse problem for which it is possible to compare results with an exact ``marginal-then-conditional'' approach.  Additionally, the IES and the hybrid IES methods are tested on a two-dimensional flow problem with uncertain anisotropy in the prior covariance. The standard IES method is shown to perform poorly in the flow examples because of the poor representation of the local sensitivity matrix by the ensemble-based method. The hybrid method, however, samples well even with a relatively small ensemble size.


\section{Introduction}

In Bayesian methods of data assimilation it is necessary to specify the prior joint probability density for the model parameters (the ``prior'' for short). The prior for parameters of a subsurface reservoir model (properties such as permeability and porosity) may be based partly on data that have already been assimilated such as log or core measurements and seismic surveys. 
The choice of a prior joint probability is also often influenced by the joint distributions of properties observed in modern geological analogues of ancient depositional environments.    

The choice of the prior is a challenge in Bayesian inference \citep{scales:01}.
In almost all applications of ensemble Kalman-based data assimilation methods, the prior is chosen to be multivariate normal with fully specified prior mean and prior covariance. In these cases, the prior mean  largely determines the average properties in regions where the data are not sensitive to parameters of the model while the choice of the covariance determines the smoothness of the spatial distributions, the variability of magnitudes, and the orientation and the range of correlation of the spatially distributed parameter fields. In most subsurface applications, it is difficult to select these parameters \citep{malinverno:04}.

In many applications of the Bayesian data assimilation methods to field cases, the prior is found to have been too narrowly specified -- it does not allow for the possibility of events that could (and perhaps will) occur in the future, and is often inconsistent with historical measurements of flow and transport. 
If a prior model that is inconsistent with data is used for data assimilation, the result will be biased estimates and forecasts, implausible parameter values  and 
unjustified reduction in uncertainty \citep{moore:05,oliver:18}. At the end of an expensive model building and calibration exercise, it will be apparent that the model is inadequate and will need to be rebuilt.

Although allowing for uncertainty in the prior mean is not common in ensemble-based history matching, its usefulness has been demonstrated on both synthetic and field cases.  Li et al.~\cite{li:10a} allowed for a uniform adjustment to the mean, while Zhang and Oliver \cite{zhang:11a} allowed the prior mean to be characterized by a trend surface whose coefficients were uncertain.  In some cases (such as the Brugge benchmark case), the use of a hierarchical model with uncertainty in the mean permeability was not necessary for matching production data, but the hierarchical model provided a more reasonable explanation of the bias in the initial forecasts \citep{chen:10a}.

Uncertainty in the prior covariance is widely recognized as being a key aspect of overall uncertainty, but careful treatment of covariance uncertainty in history matching is rare. A practical approach to parameterizing the uncertainty in the covariance is to assume a family of covariance functions with hyperparameters that control the smoothness, the covariance range, the variance, and the anisotropy.
If these hyperparameters are fixed at inappropriate values, which are then used in data assimilation, the ability to assimilate flow data will often be limited.  One approach to treating uncertainty in the hyperparameters  is to use the concept of scenarios to represent the  possibilities of a discrete number of alternative values for the hyperparameters. Park et al.~\cite{park:13} demonstrated the usefulness of this approach for a case in which the orientation of the anisotropy was assumed to be  one of two possible angles. The approach used by Emerick~\cite{emerick:16} was somewhat similar, except that weights for the scenarios were computed using an iterative ensemble smoother. Although Malinverno and Briggs \cite{malinverno:04} did not consider uncertainty in anisotropy, they allowed uncertainty in five hyperparameters of the the covariance for a problem of inferring compressional wave slowness in a one-dimensional earth model.

If the covariance is allowed to be uncertain in an ensemble Kalman-based approach, the parameterization of the model and the choice of the covariance model are both important to successful data assimilation.
Chada et al.~\cite{chada:18} investigated the effect of both centered and non-centered parameterizations (described in Sec.~\ref{sec:parameterizations}) of a hierarchical spatial model for use with ensemble Kalman iteration (EKI). 
In their experiments, which included nonstationary hyperparameters,  they concluded that using a hierarchical approach with a non-centered parameterization was significantly  better than a hierarchical approach with a centered parameterization and bothh hierarchical approaches were better than a non-hierarchical approach. 
Subsequently, Dunlop et al.~\cite{dunlop:20} showed that for MAP estimation of the hyperparameters in a linear inverse problem, the centered parameterization is to be preferred when the goal is MAP estimation as opposed to uncertainty quantification.

\section{The hierarchical model}

Gaussian priors are often used in data assimilation because they are relatively tolerant to errors in misspecification, but the choice of the parameters of the covariance still have an influence on the quality of the final data match. In Sec.~\ref{sec:2D_flow}, the feasibility of the hybrid IES to assimilate data in a model with an uncertain prior covariance is investigated. In that example, the `true' model that generated the data has an anisotropic covariance with a longer range in one direction. If one were fortunate enough to know the covariance model from which the truth was sampled, it would not be necessary to use a hierarchical model as a good average total squared data mismatch (e.g.~ $S_d^o = 365$ for the correct covariance in the two-dimensional flow problem) could be achieved without consideration of uncertainty in hyperparameters.  
On the other hand, if the prior covariance is mistakenly fixed at an incorrect orientation (off by $\pi/2$), the misfit to data would be substantially  worse ($S_d^o = 2142$) and the ability to accurately forecast future behavior would also suffer. 
Uncertainty in the  covariance can be easily accounted for by introducing  a hierarchical model for which the anisotropy and ranges are uncertain.  
Applying the hierarchical model in this problem and using the hybrid iterative ensemble smoother for data assimilation, it is possible to obtain a mean squared data mismatch  ($S_d^o = 1151$) that is intermediate between the value obtained using the correct covariance and a covariance with incorrect orientation. When hierarchical modeling is an appropriate approach to describing uncertainty, it provides more robust priors for history matching and reduces the need for rebuilding the model and re-history matching. Unfortunately, hierarchical models are more nonlinear than non-hierarchical models and data assimilation or history matching is more difficult. A straightforward application of an iterative ensemble smoother with localization to the hierarchical problem with uncertain anisotropy results in an extremely poor match to data ($S_d^o = 13000$). It appears that in order to use a hierarchical model with an ensemble-based data assimilation will generally require modification of the data assimilation methodology, as presented in Sec.~\ref{sec:hybrid-IES}.

The probability distribution of Gaussian random variables (denoted $m$) is completely determined by the mean and the covariance of the variable. In this study, we will assume that the mean $m_{pr}$ is known but that the covariance $C_m$ is uncertain. There are a number of parameters that could describe the uncertainty in $C_m$; for two-dimensional fields, we will focus on correlation range $\rho$, orientation of the anisotropy $\phi$ and the ratio of range in two principal directions. For a one-dimensional field, we will focus on problems in which the variance and the correlation range are uncertain.

\subsection{Hierarchical model parameterization -- centered and non-centered } \label{sec:parameterizations}

There are two possible parameterizations for Gaussian hierarchical problems: the \emph{centered} (or natural) parameterization and the \emph{non-centered} parameterization   \citep{papaspiliopoulos:03}. The centered parameterization utilizes the conditional independence of the data $d$ and the hyperparameters of the distribution $\theta$  given the observable parameters $m$. In the non-centered parameterization $\theta$ and $z$ are a priori independent. Thus we can write the entire set of parameters as $x = (m,\theta)$  in the centered parameterization  or as $x = (z, \theta)$ for the non-centered parameterization. In the non-centered relationship the relationship between the observable Gaussian variable $m$ and the parameters can be written
\begin{equation}
m = m\sbr{pr} + L(\theta) z
\label{eq:m_Lz}
\end{equation}
where $L$ is a ``square root'' of the model covariance matrix, i.e. $L L\trp = C_m$. Note that the pdf for $m$ is not Gaussian unless the hyperparameters $\theta$ are fixed.

Although Eq.~\ref{eq:m_Lz} is used in this paper to generate realizations of $m$,
Eq.~\ref{eq:m_Lz} will only be feasible for large grids if the square root of the covariance, $L$, is compact (as it is for the spherical covariance), or if the correlation range is sufficiently short that most elements of $L$ are effectively zero. Other methods could be more efficient for large grids. In particular, it may often be advantageous to use either the inverse of the covariance matrix (the precision matrix)  or a factorization of the precision matrix  \citep{stojkovic:17} or a Mat{\'e}rn-Whittle covariance model
\citep{gneiting:10} for which stochastic partial differential equations can be used efficiently to generate a random field \citep{roininen:19,zhou:18}.
Another relatively common approach to describing the uncertainty in  Gaussian hierarchical models is to assume that  the prior distribution for the covariance model matrix is inverse Wishart \citep{myrseth:10, tsyrulnikov:17}. While this approach has some computational advantages, it seems less likely to describe the prior uncertainty in the covariance.

\section{Data assimilation for Gaussian hierarchical models} \label{sec:data_assim_methods}

For large geoscience inverse problems, iterative ensemble smoothers are often an effective approach. These are all based on the original development of the ensemble Kalman filter \citep{evensen:94}, which has several advantages over Kalman filters or extended Kalman filters: a low-rank approximation of the covariance matrix is used instead of the full covariance, and the linearization of the relationship between predicted data and model parameters is approximated without requiring adjoints. For geoscience inverse problems, it has been found that it is more efficient to use a `smoother' to update the parameters of the inverse problem using all the data simultaneously. On the other hand, the problem of parameter estimation becomes more nonlinear when all data are assimilated simultaneously so iteration is almost always required when a smoother is used for history matching.
Although there are many variants of iterative ensemble smoothers, they can generally be classified into one of two approaches.  In multiple data assimilation (MDA), the same data are assimilated multiple times with an inflated observation error \citep{reich:11,emerick:13a}. This approaches reduces the nonlinearity in the update, but also requires updating the approximation of the covariance matrix at each iteration. That seems to be more difficult for hierarchical models than for models in which the prior covariance is assumed to be known.

The second class of iterative ensemble smoothers is based on the randomized maximum likelihood (RML) approach to approximate sampling from the posterior \citep{kitanidis:95,oliver:96e,oliver:08}. In the RML approach, samples from the prior are updated to become approximate samples from the posterior by minimizing a stochastic objective function. Like MDA, this method is exact for linear Gaussian data assimilation problems, but this method does not require updating of the covariance at each iteration. In the iterative ensemble smoother form of RML \citep{chen:12a}, an average sensitivity, computed from the ensemble of samples, is used to approximate the downhill direction. Because an ensemble average sensitivity will not provide an accurate sensitivity when the problem is highly nonlinear,  a hybrid RML-IES method is introduced in which some derivatives are computed analytically, while other derivatives are estimated from the ensemble. Finally, for the linear observation case, results will be compared with the marginal-then-conditional (MTC) approach introduced by Fox and Norton~\cite{fox:16}.

The unnormalized posterior pdf for the model parameters in the non-centered parameterization can be written as
\begin{multline}
p(z , \theta \mid d)
\propto \exp \left( -\frac{1}{2} (d^o-g(m(\theta , z)))\trp C_d\inv (d^o-g(m(\theta, z))) \right) \\
\times \exp \left(-\frac{1}{2} (\theta-\bar{\theta})\trp C_\theta\inv (\theta -\bar{\theta}) -\frac{1}{2} z\trp z \right)  
\end{multline}
or more simply
\begin{multline}
p(x \mid d) \propto \exp \left( -\frac{1}{2} (d^o-g(m(x)))\trp C_d\inv (d^o-g(m(x))) \right) 
\\ \exp \left(-\frac{1}{2} (x-\bar x)\trp C_x\inv (x -\bar x) \right).
\label{eq:p_noncentered}
\end{multline}
The non-centered parameterization appears to be well-suited to data assimilation using an iterative ensemble smoother when the prior pdf  for both $z$ and $\theta$ have been assumed Gaussian (perhaps after transformation), as the pdf for $x$ (Eq.~\ref{eq:p_noncentered}) is identical in form to the pdf for $m$ in traditional Bayesian history matching.  Nonlinearity is a result either of the relationship $d=g(m)$ being nonlinear, as is the case if the data are water rates and the model parameters are porosity and log-permeability, or a result of nonlinearity in the relationship between $m$ and $\theta$. Although Eq.~\ref{eq:p_noncentered} is written for the case in which the prior for the hyperparameters is Gaussian,  when the uncertainty in the orientation of the anisotropy is moderately large, a Gaussian approximation is not appropriate. We discuss that case in the Appendix \ref{sec:circular_hyperparameters}.

\subsubsection{Data assimilation -- RML}

The randomized maximum likelihood approach begins, like the perturbed observation form of the ensemble Kalman filter \citep{burgers:98,houtekamer:98}, by drawing samples $(x_i',\epsilon_i')$, $i=1,\ldots,N_s$, from the Gaussian
distribution
\begin{equation}
q_{X'\epsilon'} (x',\epsilon') \propto  \exp \left(-\frac{1}{2} (x-\bar x)\trp C_x\inv (x -\bar x) \right) \exp \left( - \frac{1}{2}
 \left(\epsilon' \right)\trp C_\epsilon\inv \epsilon'  \right)
 \label{eq:xdprop}
\end{equation}
for given $\bar x$.  The $i$th approximate  sample from the posterior is obtained by computing the minimizer of the cost functional
\begin{multline}
J_i(x) = \frac{1}{2} \left(x- x_i' \right)\trp C_x\inv \left(x- x_i' \right) \\
 + \frac{1}{2} \left( g(m(x)) + \epsilon'_i - d^o  \right)\trp C_d\inv \left( g(m(x)) + \epsilon'_i - d^o  \right).
\label{eq:Jix}
\end{multline}
This is usually done by solving
\begin{equation} \label{eq:critical}
 \nabla_x J_i(x) = 0
\end{equation}
for $x$. If Levenberg-Marquardt with a Gauss-Newton approximation of the Hessian  is used for the minimization, the $\ell$th update is of the form
\begin{multline}
\delta x_{\ell} = \frac{x'_i-x_{\ell}}{1+\lambda_{\ell}}-  C_x G_{\ell}\trp \biggl[  (1+\lambda_{\ell}) C_d + G_{\ell} C_x G_{\ell}\trp \biggr]\inv  \\
\times \biggl[(g(x_{\ell}) + \epsilon'_i - d^{o}) - \frac{  G_{\ell} (x_{\ell} - x'_i) }{1+\lambda_{\ell}}  \biggr].
\label{eq:RML}
\end{multline}
where
$G\trp = \nabla_{x} \big(g\trp \big)$ and $\lambda_{\ell}$ is the Levenberg-Marquardt regularization parameter. This method of sampling from the posterior distribution is only exact if the relationship between the data and the model parameters is linear. It will sample accurately from multimodal distributions in some situations, but exact sampling using RML requires computation of additional critical points and weighting of solutions \citep{ba:22}. For geoscience applications, the standard unweighted RML is almost always used.

\subsubsection{Data assimilation -- IES}

One disadvantage of RML is the need for computation of the gradient of the objective function. For many forward models of the data such as reservoir production simulation, the derivatives are not readily available. In those cases the iterative ensemble smoothers offer an alternative that avoids the need to compute $G$ directly. The basic idea is that terms that appear in Eq.~\ref{eq:RML} can often be computed efficiently using ensemble approximations. In an ensemble-based approach to sampling from the posterior, Eq.~\ref{eq:RML} is replaced by the following update step \citep{chen:13}
\begin{equation}
\begin{split}
\delta x_{\ell+1} & = - \frac{1}{(1+\lambda_\ell)} \Delta x_{\ell} \,  \Delta x_{\ell}\trp C_x\inv (x_\ell-x')  \\
 & \qquad - \Delta x_{\ell} \, \Delta d_{\ell}\trp  \left( (1+ \lambda_\ell) C_{d} + \Delta d_{\ell} \Delta d_\ell\trp \right)\inv
 \\
 &  \qquad \times \Biggl( g(m_\ell) + \epsilon'_i - d^o - \frac{1}{(1+\lambda)}  \Delta d_{\ell} \Delta x_{\ell}\trp 
 C_x\inv (x_\ell - x'_i)    \Biggr)
\end{split}
\label{eq:IES}
\end{equation}
where $\Delta x_{\ell} = \frac{(X_\ell - \bar{X}_\ell)}{\sqrt{(N-1)}}$ and similar for $\Delta d_{\ell}$. If this approach were applied directly, it can be seen that any update is restricted to the space spanned by the initial ensemble and the number of degrees of freedom available for calibration is $N_e-1$ where $N_e$ is the number of realizations in the initial ensemble. To avoid this limitation, localization is nearly always applied in large problems, but for clarity localization has been omitted in Eq.~\ref{eq:IES}. The main disadvantage of an ensemble-based methodology is the failure to handle strong nonlinearity well -- primarily because the same average sensitivity is used for all samples.

\subsubsection{Data assimilation -- hybrid IES} \label{sec:hybrid-IES}

The hybrid IES is a method that takes advantage of the ability of the RML to use different gain matrices for each sample and the ability of the IES to avoid the need for adjoint systems. As in the RML method, the (transpose of) sensitivity of data to model parameters in a non-centered parameterization is
\begin{equation}G\trp = G(x)\trp = \begin{bmatrix}
\nabla_{x} g_{1} & \nabla_{x} g_{2} & \ldots &
\nabla_{x} g_{N_d} \end{bmatrix}
= \nabla_{x} \big(g\trp \big)\end{equation}
and similarly define the sensitivity of data to the observable model parameters $m$, 
\begin{equation}
G_m\trp = G(m)\trp = \begin{bmatrix}
\nabla_{m} g_{1} & \nabla_{m} g_{2} & \ldots &
\nabla_{m} g_{N_d} \end{bmatrix}
= \nabla_{m} \big(g\trp \big).
\end{equation}
These two sensitivity matrices are related
\begin{equation}
G\trp = \nabla_{x} \big(g\trp \big) = \nabla_{x} \big(m\trp \big) \cdot \nabla_{m} \big(g\trp \big) = \nabla_{x} \big(m\trp \big) \cdot G_{m}\trp
\end{equation}
or 
\begin{equation}
G  = G_{m} \cdot   \left( \nabla_{x}  \big(m\trp \big)\right)\trp.
\end{equation}
For notational simplicity, the Jacobian of the observable model parameters with respect to the non-centered  hierarchical parameters is written as 
\begin{equation}
M_x=  \left( \nabla_{x}  \big(m\trp \big)\right)\trp
\end{equation}
where the dimension of $M_x$ is $N_m \times N_x$.
Substituting $G = G_m  M_x$ into the RML update expression (Eq.~\ref{eq:RML}) with ensemble representation of $G_m$ results in a  hybrid IES data assimilation approach
\begin{equation}
\begin{split}
\delta x 
& = - \frac{1}{(1+\lambda)}   (x-x')  - C_x M_x\trp  G_m\trp \left( (1+ \lambda) C_{d} + G_m M_xC_x M_x\trp  G_m\trp \right)\inv
\\
& \qquad \qquad \times \left(g(m)+\epsilon' - d^o - \frac{1}{(1+\lambda)} G_m M_x  (x - x')   \right) \\
& = - \frac{1}{(1+\lambda)}   (x-x')  \\
& \qquad - 
C_x M_x\trp  (\Delta m)\itp \, \Delta d\trp \left( (1+ \lambda) C_{d} + \Delta d \, (\Delta m)\inv M_x C_x M_x\trp  (\Delta m)\itp \, \Delta d\trp\right)\inv
\\
& \qquad \qquad \times \left(g(m)+\epsilon' - d^o - \frac{1}{(1+\lambda)} \Delta d \, (\Delta m)\inv M_x   (x - x')   \right) 
\end{split}
\label{eq:hybrid-IES}
\end{equation}
Note that in the hybrid IES method each ensemble member has its own Kalman gain matrix since the sensitivity $M_x$ of the physical model parameters $m$ to the non-centered parameters $x$ is specific to an ensemble member. Assuming that $m$ is determined from the known mean $m_{pr}$, the uncertain covariance $C_m$ and from the stochastic variable $z$, i.e.,
$m = m_{pr} + L(\sigma, a) z$ where $L L^T = C_m$.

\subsection{Factorization of covariance functions} \label{sec:cov_factorization}

The hybrid approach will only be useful if computation of the terms in Eq.~\ref{eq:hybrid-IES} is practical. First, note that $C_x$ is diagonal, so that it can be eliminated by scaling of the variables. Computing $M_x$ this way  relies on factorability of the covariance matrix. 
For functions in one-dimensional, the equivalent to defining $C_m = L L\trp$ is to define
\begin{equation}
C(r) = f \ast f\trp = \int_{-\infty}^{+\infty} f(s) f(r+s) \, ds
\end{equation}

For illustration,  consider the 
one-dimensional exponential covariance function
$C(r) = \sigma^2 e^{ - \mid r\mid  /a}$
where $r$ is the distance between two variables, $a$ is a measure of the range of the correlation and $\sigma^2$ is the variance. In one-dimensional, a symmetric factorization is 
$L_s(r) = \frac{ \sigma \sqrt{2} }{ \sqrt{a}   \pi}\ K_0(\mid r\mid  / a)$
where $K_0$ is the modified Bessel function of the second kind of order 0.
The factorizations are not unique, however, and it may be beneficial  to use a factorization for which the ``square root'' is not singular. A one-sided factorization that is always finite is
$L_a(r) = \sigma \sqrt{2/a} \ H(r) \exp(-r/a)$ where $H(r)$ is the Heaviside function.
Similarly, the symmetric factorization of the squared exponential or Gaussian covariance
$C(r) = \sigma^2 e^{ - r^2 /a^2}$ can be shown to be 
$L_s(x) = \sigma \left( \frac{ 4 }{ a^2  \pi} \right)^{1/4}   \exp(- 2 r^2 /a^2)$.

Similar factorizations can easily be derived for other covariance functions and for other dimensions \citep{oliver:95}. In this study, the factorization of the covariance function has been used to obtain an approximation of the factorization of the covariance matrix. A scale space implementation might be justified if additional accuracy was required \citep{lindeberg:90}. 

\section{Numerical examples}

Two numerical examples are presented to illustrate data assimilation for Gaussian models with uncertainty in the covariance. In the first example, the model is defined on a one-dimensional grid and the observations are of $m$. In the second example,  the uncertain permeability field in a two-dimensional porous medium is estimated by assimilation of a time series of water cut observations at 6 producing wells.

\subsection{One-dimensional linear Gaussian}

\begin{figure}[htbp]
\begin{subfigure}{0.45\textwidth}
\includegraphics[width=0.95\textwidth]{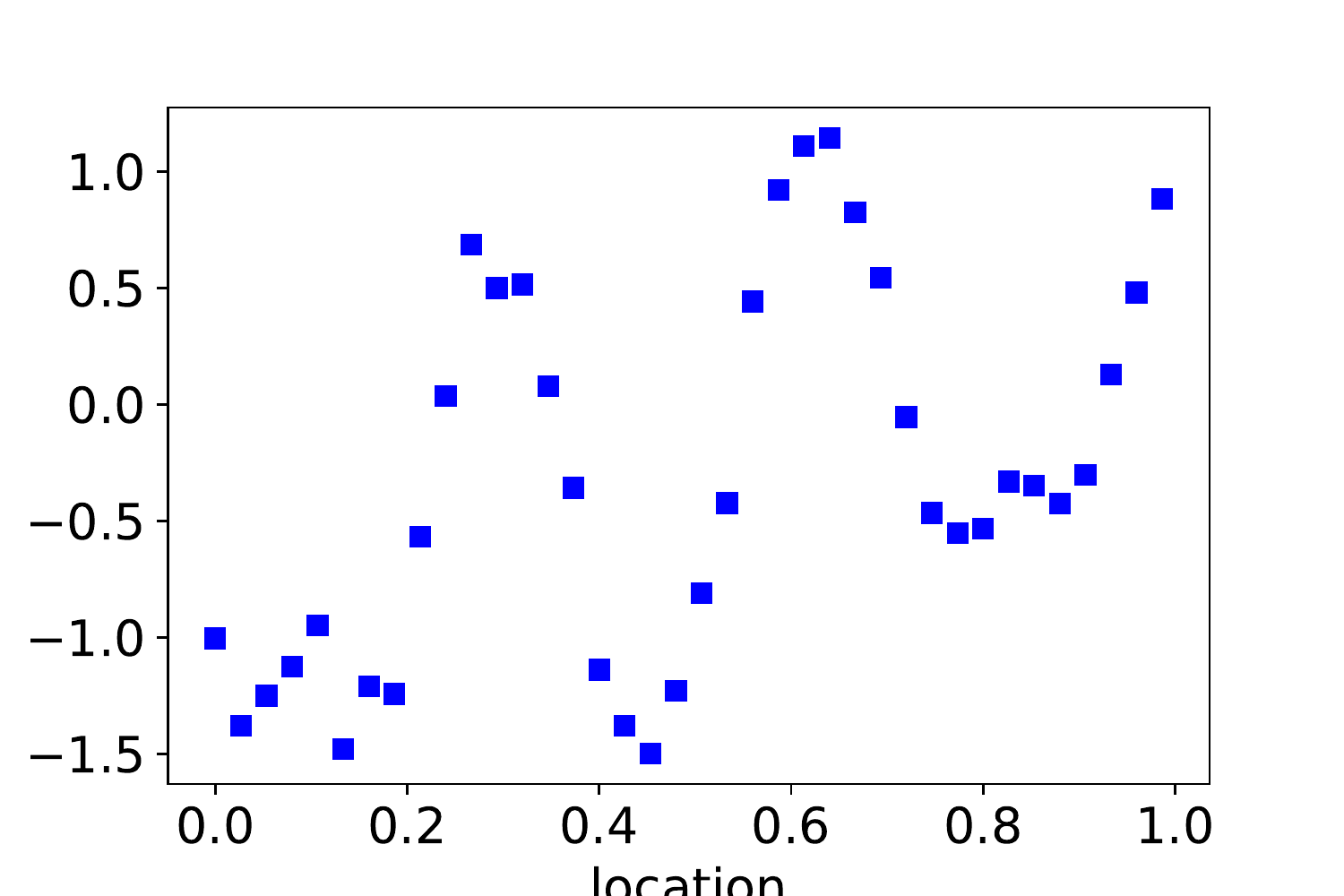} 
\caption{Noisy observations from the true data-generating model}
\label{fig:obs_1D}
\end{subfigure}
\hfill
\begin{subfigure}{0.45\textwidth}
\includegraphics[width=0.95\textwidth]{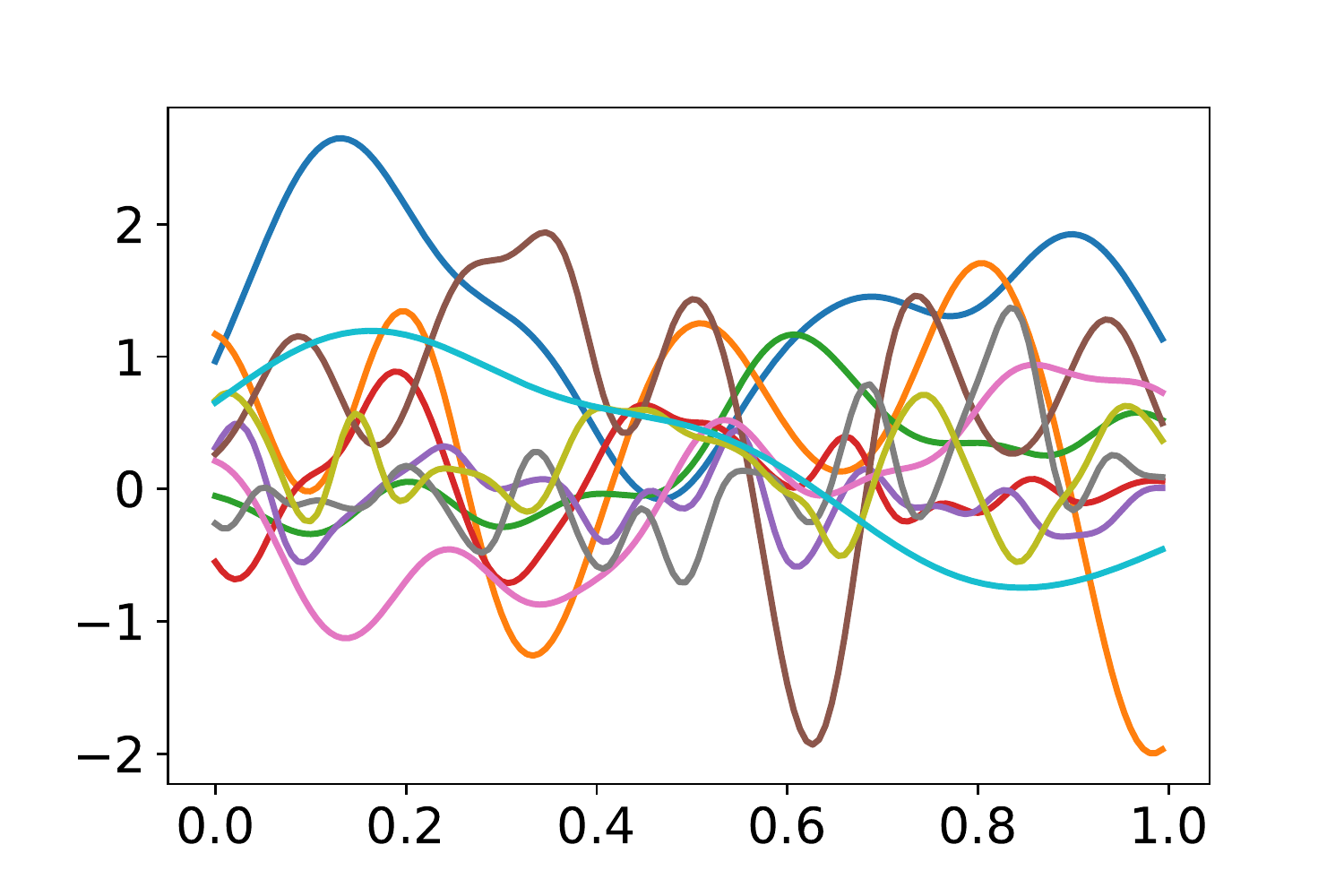} 
\caption{10 samples of $m$ from a hierarchical prior}
\label{fig:prior_1D}
\end{subfigure} 
\caption{The 1D linear test problem with hierarchical prior. }
\label{fig:obs_and_prior_1D}
\end{figure}

Consider a simple one-dimensional Gaussian random field, discretized on the interval $[0,1]$, into $n_m=150$ lattice points. 
The random variable $m$ is observed at every 4th lattice point with independent measurement noise characterized by $\sigma_d = 0.01$. 
The data-generating model has correlation range $a^{tr} =  0.100$ and standard deviation $\sigma_m^{tr} =  1.08$. The noisy observations are shown in Fig.~\ref{fig:obs_1D}.

The form of the covariance of the data-generating model $C_m$  (squared Gaussian) is assumed to be known -- only the correlation range $a$ and model variance $\sigma^2$ are uncertain. Because both parameters are required to be positive, log-normal distributions are assumed for each: $\theta_1 \equiv \log \sigma_m  \sim N(-0.22, 0.5^2)$ and $\theta_2 \equiv \log a \sim N(-2.3, 0.6^2)$.
Ten unconditional samples of $m$ from the prior are shown in Fig.~\ref{fig:prior_1D}. 

The posterior pdf can therefore be written as
\begin{multline}
p(z,a,\sigma\mid d) \propto \exp \biggl( -\frac{1}{2} (d-g(m(u,z)))^T C_d^{-1} (d-g(m(u,z))) \\ 
-\frac{1}{2} (\theta-\theta_{pr})\trp C_\theta\inv (\theta-\theta_{pr})
 -\frac{1}{2} z^Tz   \biggr)
\end{multline}
and the corresponding objective function for a minimization-based sampling approach is given by Eq.~\ref{eq:Jix}. Results from the three different data assimilation approaches described in Sec.~\ref{sec:data_assim_methods} (RML, IES and hybrid IES) are compared  with the exact distribution of hyperparameters obtained using marginalization for the centered parameterization \citep{rue:05,rue:07,fox:16} and with the true values of the hyperparameters.

\begin{figure}[htbp]
\begin{tabular}{ccc}
\includegraphics[width=0.32\textwidth]{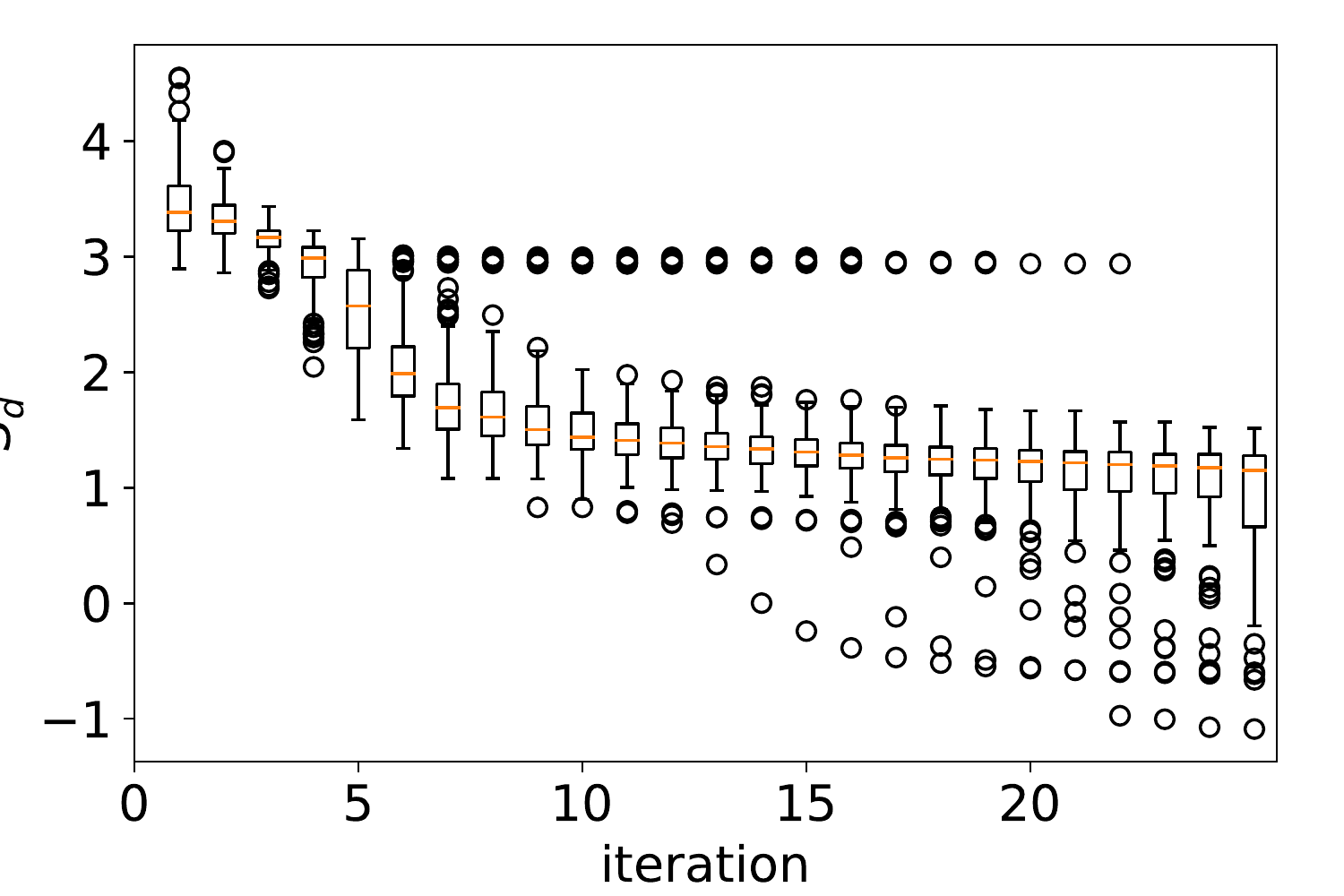}
&
\includegraphics[width=0.32\textwidth]{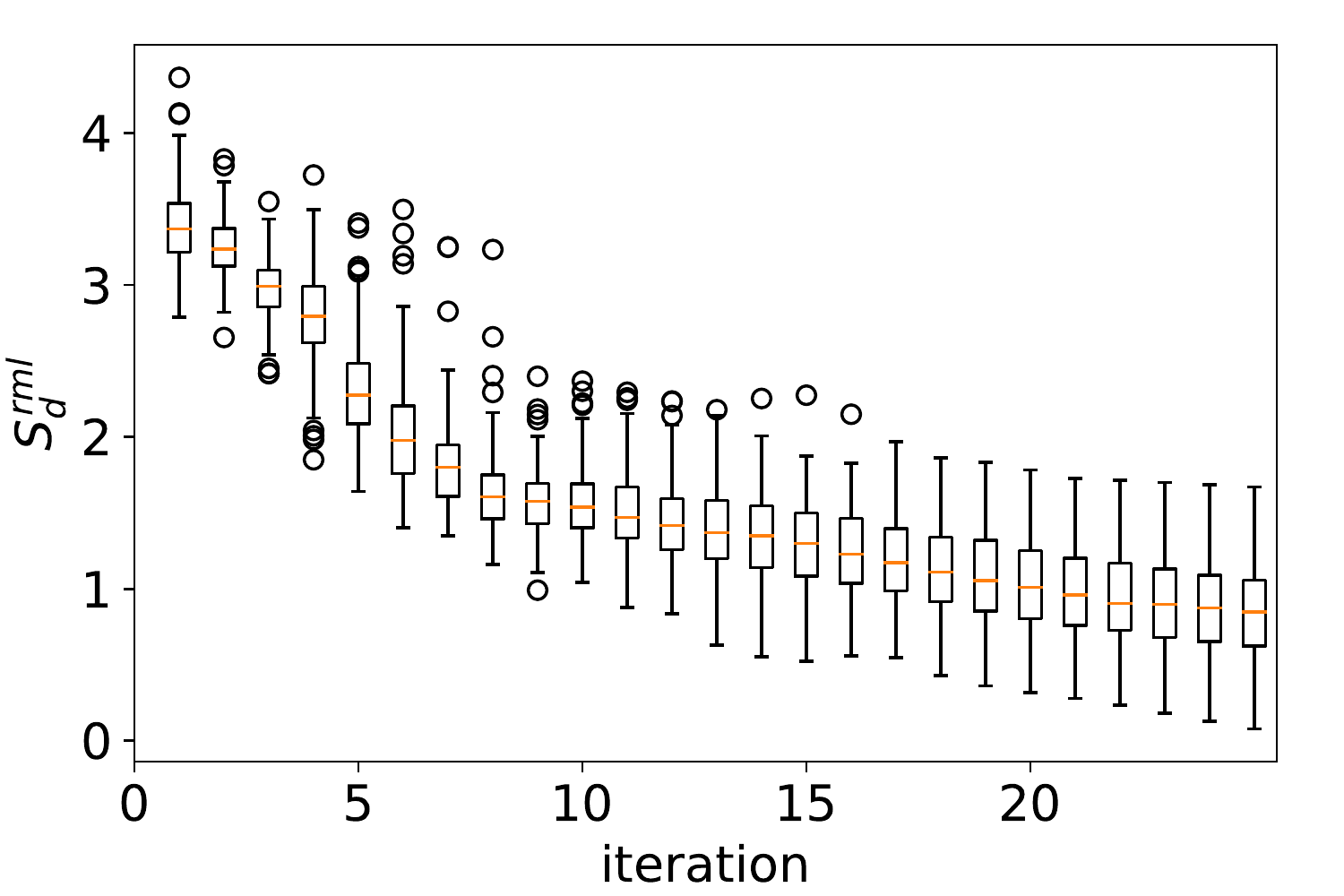}
&
\includegraphics[width=0.32\textwidth]{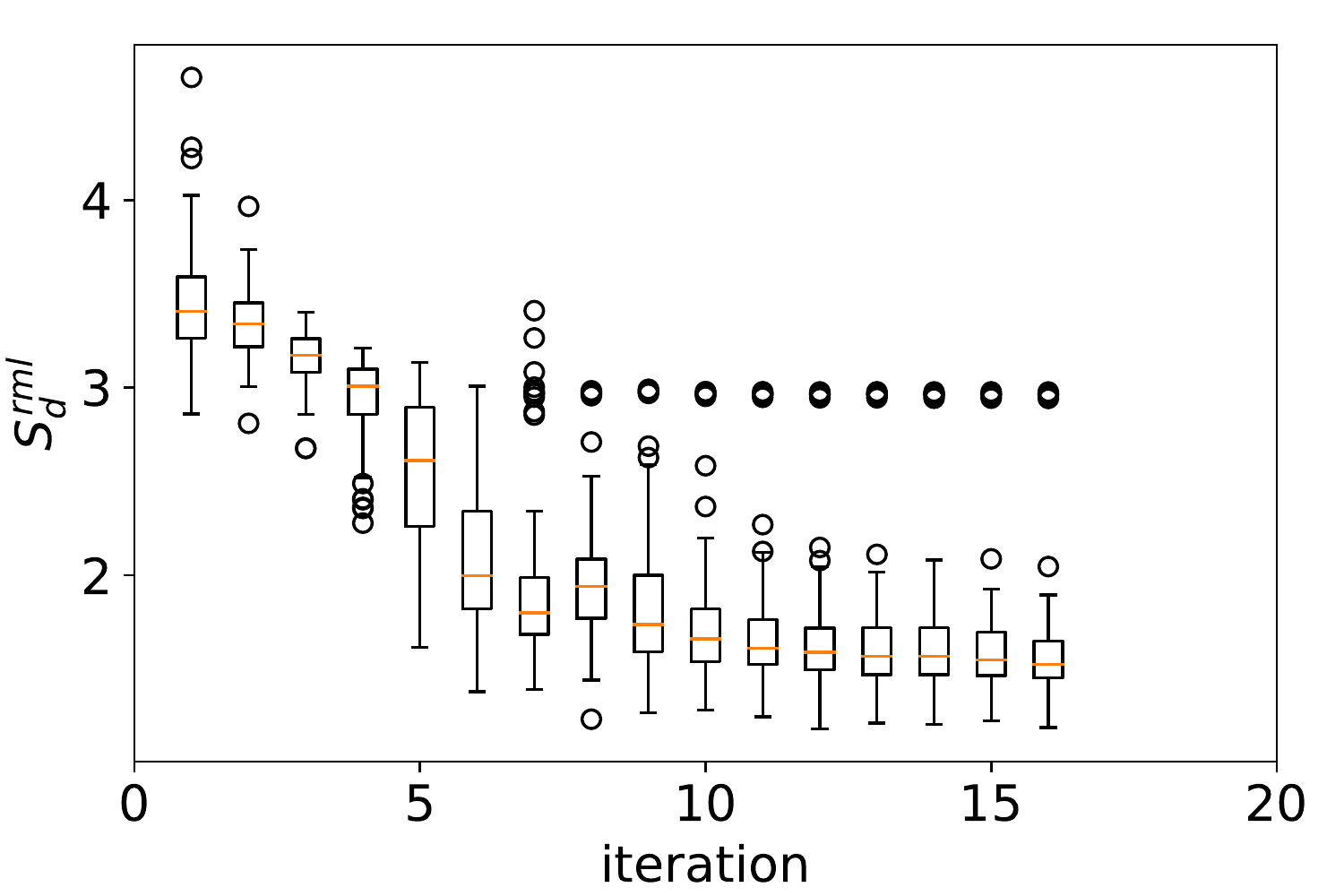}
\\
RML &  Iterative ES  & Hybrid IES
\end{tabular}
\caption{Reduction of squared data mismatch for three data assimilation algorithms.}
\label{fig:iteration_1D}
\end{figure}

Note that although RML does not use information from the ensemble for data assimilation -- each realization is generated independently -- 100 RML samples were generated for the comparison. It is necessary to set a few algorithmic parameters for the minimizations. For RML with Levenberg-Marquardt minimization, an initial value of $\lambda = 5000$ was used. It was decreased by a factor of 4 if the objective decreased. Otherwise, it was increased by a factor of 4. For each minimization, the initial model was chosen to be the sample from the prior.
For IES and hybrid IES, the  initial value of $\lambda$ was based on the initial mean value of the data mismatch objective function \citep{chen:13}. An ensemble size of 200 was used for IES and an ensemble size of 100 was used for hybrid IES.

\begin{figure}[htbp]
\begin{tabular}{cccc}
\raisebox{2ex}{\rotatebox{90}{range, $a$}} 
&
\includegraphics[width=0.3\textwidth]{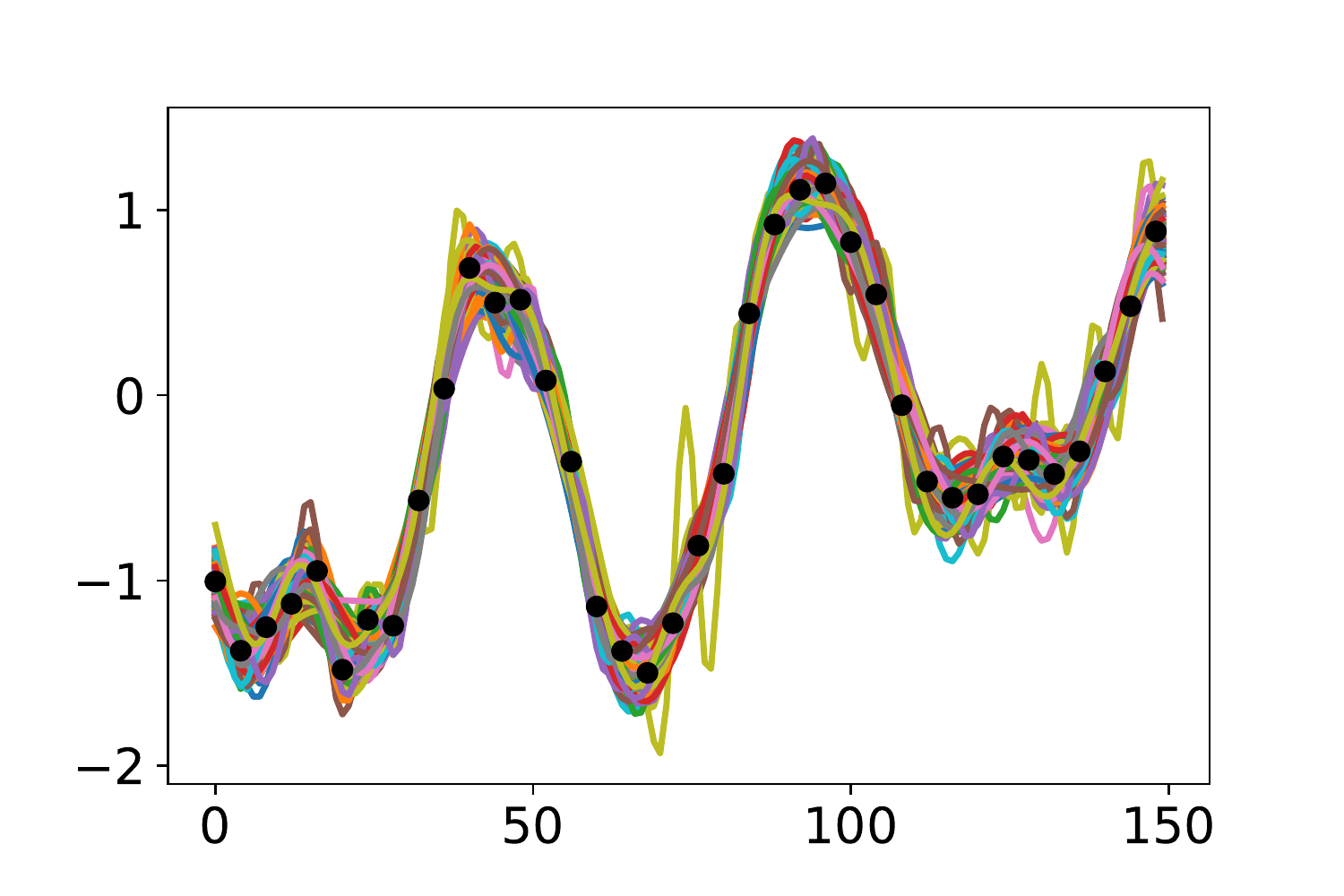}
&
\includegraphics[width=0.3\textwidth]{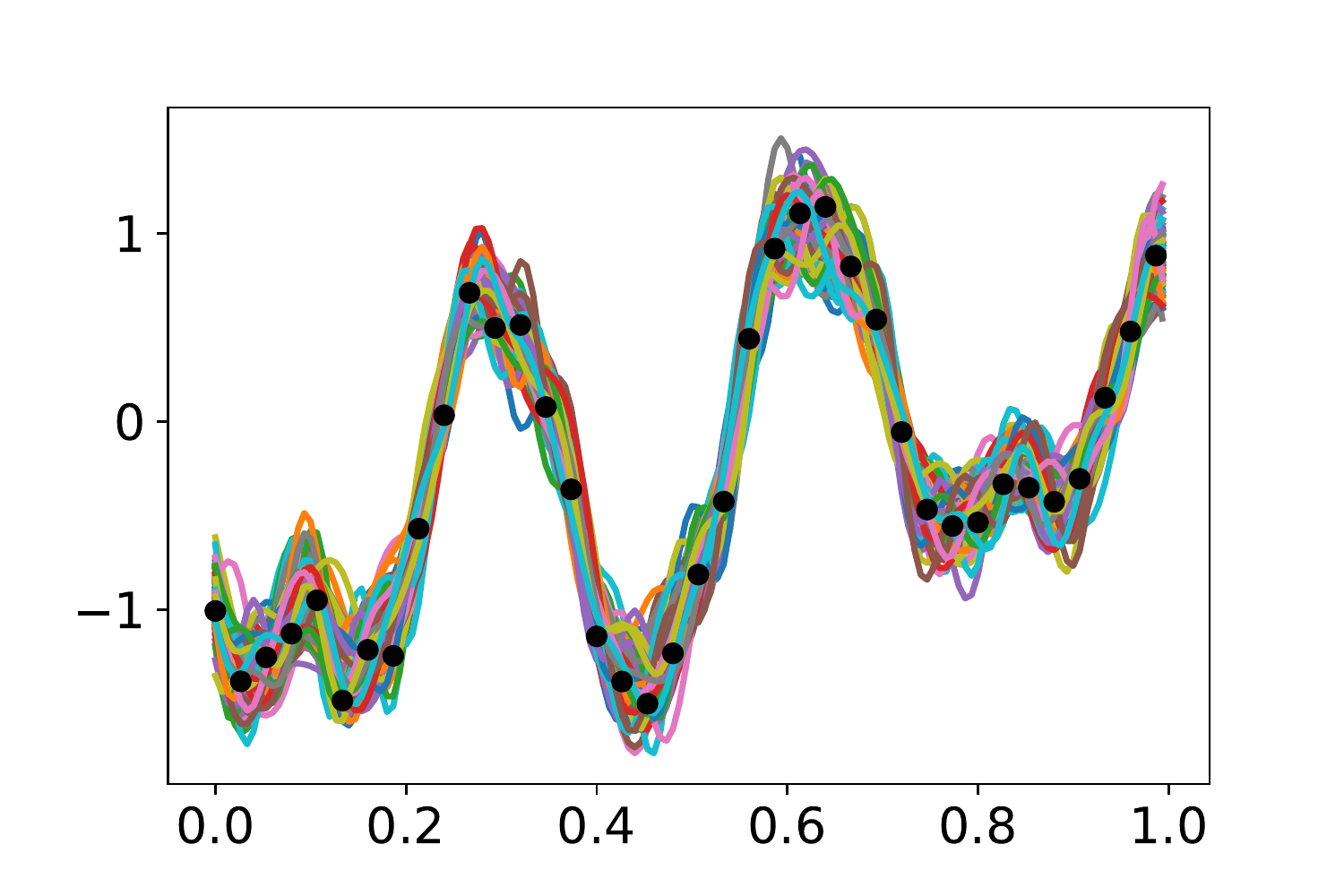}
&
\includegraphics[width=0.3\textwidth]{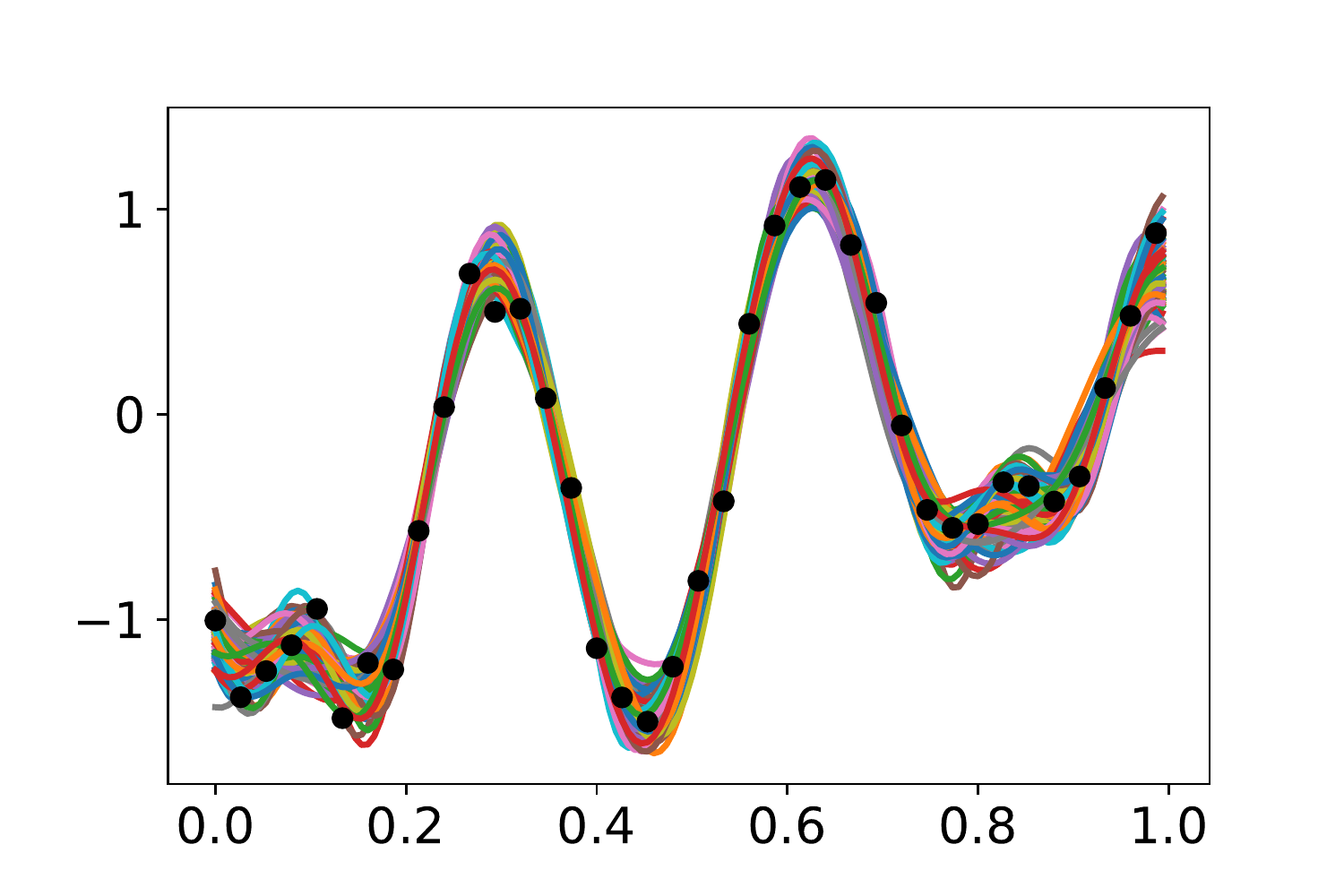}
\\
\raisebox{2ex}{\rotatebox{90}{range, $a$}} 
&
\includegraphics[width=0.3\textwidth]{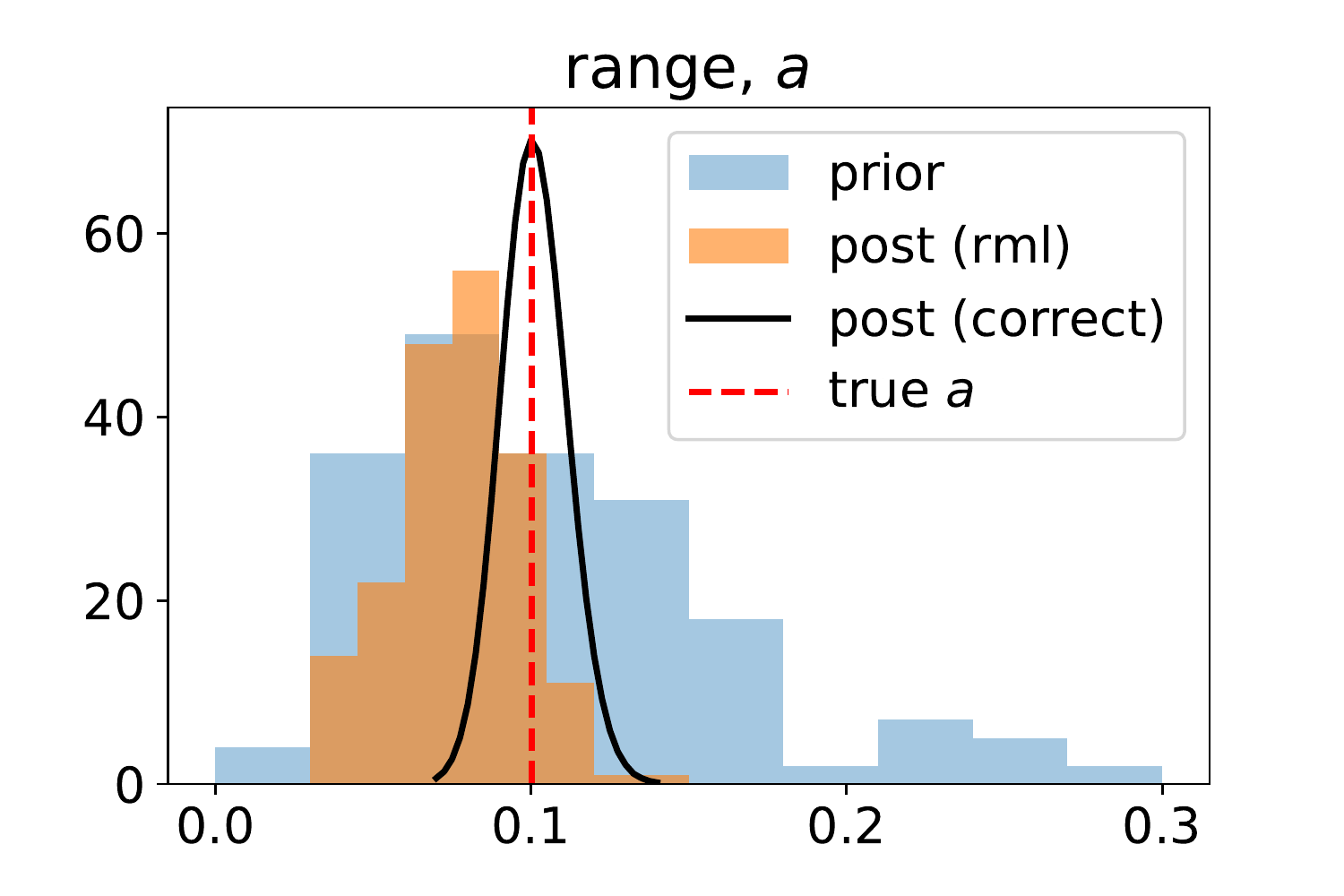}
&
\includegraphics[width=0.3\textwidth]{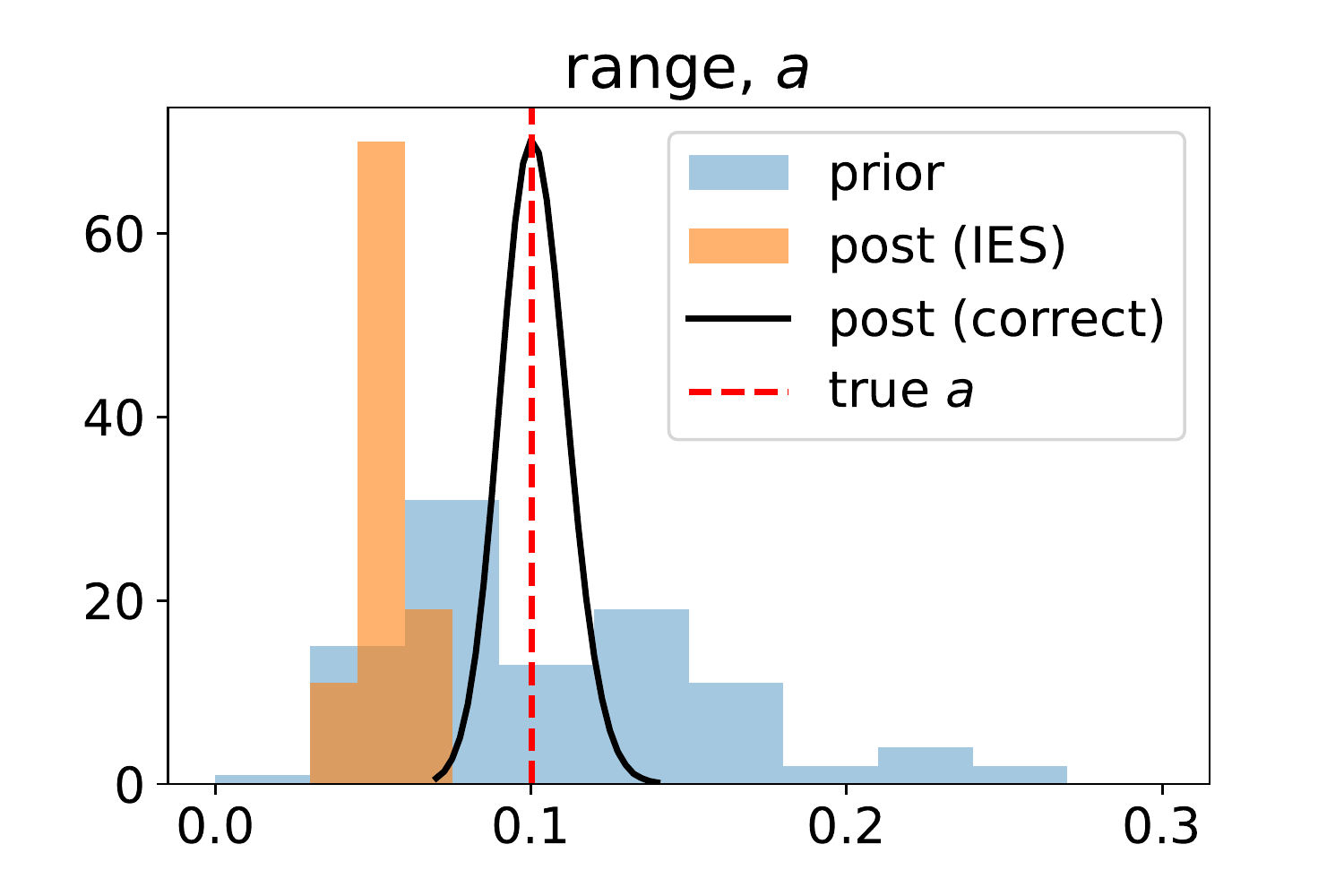}
&
\includegraphics[width=0.3\textwidth]{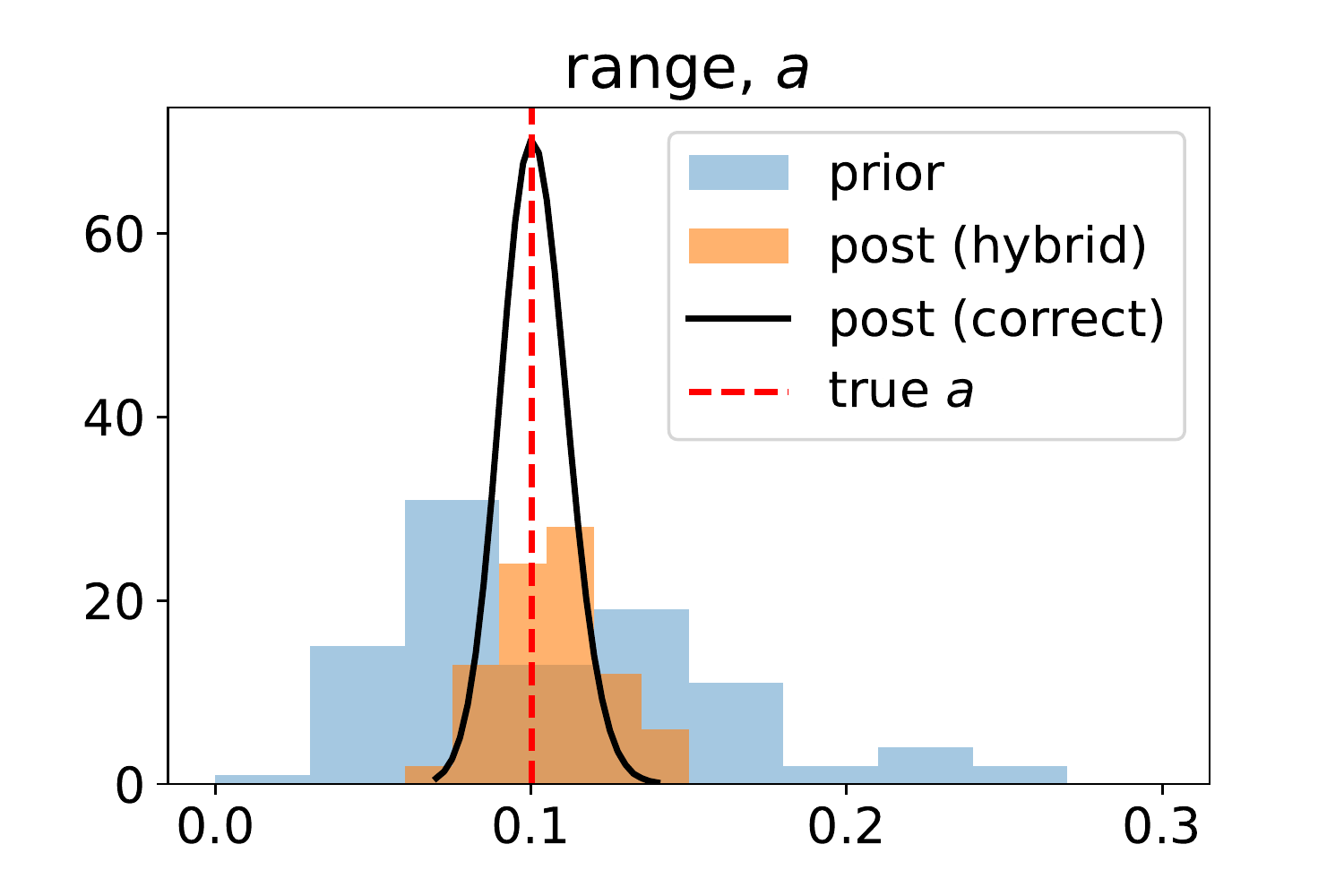}
\\
& RML &  Iterative ES  & Hybrid IES
\end{tabular}
\caption{Samples of $m$ from the posterior for three data assimilation methods (top row) and comparison of prior and posterior distributions of hyperparameters (bottom row).}
\label{fig:posterior_m_1D}
\end{figure}

The reduction in the  mismatch of simulated data to perturbed observations was rapid initially for all three methods, although a few realizations failed to converge for both RML and for the hybrid IES (Fig.~\ref{fig:iteration_1D}). The smallest mismatch values with perturbed observations were achieved by RML, with values of the realizations of the data objective function achieving values 
\[S_d^{rml}(m) = \frac{1}{2} \sum_{i=1}^{N_d} \frac{(g(m) - (d^o + \epsilon))^2}{\sigma_d^2} < 1\] 
in several cases. While this objective is useful for monitoring convergence, it is more useful for model checking to evaluate the squared data mismatch between predicted data from the calibrated models and the actual data;  the expected value of that metric is 
\[ E[S_d^{obs}(m)]  = \frac{1}{2} \sum_{i=1}^{N_d} \frac{(g(m) - d^o )^2}{\sigma_d^2} = \frac{1}{2} N_d =19. \]
The average values from all three methods are close to the expected value.

The posteriori realizations of $m$ from all three methods also look generally plausible, although it appears that realizations from RML and IES have correlation lengths that are generally shorter that the correlation length in the data-generating model (Fig.~\ref{fig:posterior_m_1D}).

\subsection{Fluid flow example} \label{sec:2D_flow}

In the second example, the data-generating system is a two-dimensional ($30\times 15$) aniso\-tropic porous medium with two-phase (oil and water), immiscible, incompressible flow driven by two injectors and 6 producing wells.  The permeability field that generates the data is a draw from a prior model in which the angle of the principal axis of anisotropy is 0.93 radians, the range parameter for the longest correlation length is 1.0 and the ratio of correlation ranges in the two principal directions is 6.0. The covariance type for log-permeability  is set to be ``Gaussian'', i.e. squared exponential (Eq.~\ref{eq:2D_gaussian}) with standard deviation 2.0. The porosity is uniform.

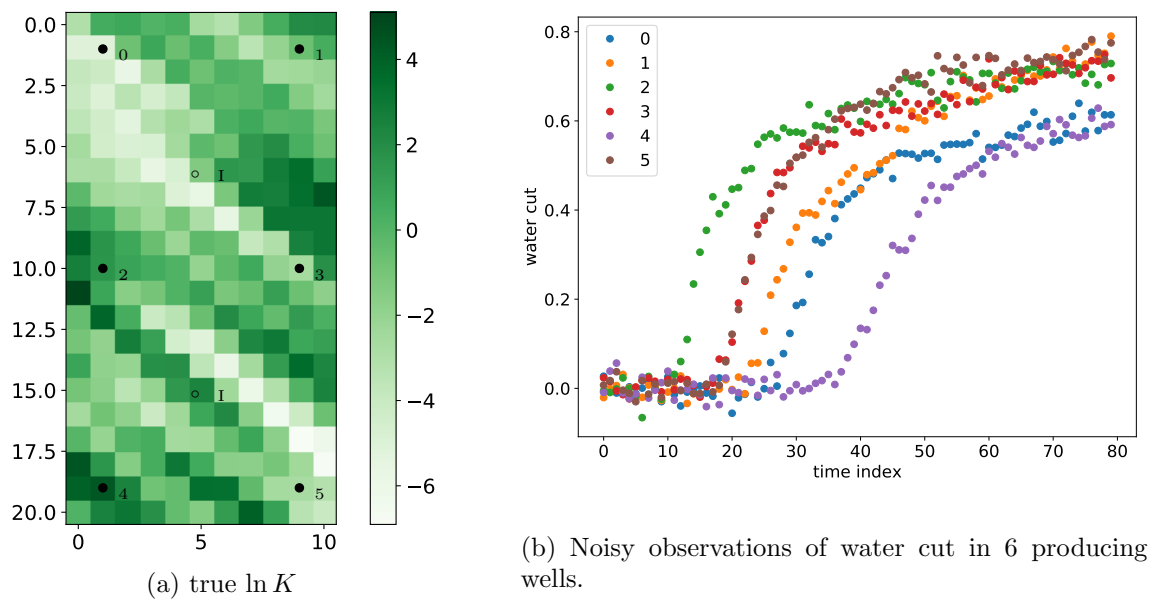
\begin{figure}[htbp]
\begin{subfigure}{0.375\textwidth}
\begin{overpic}[width=1.0\textwidth]{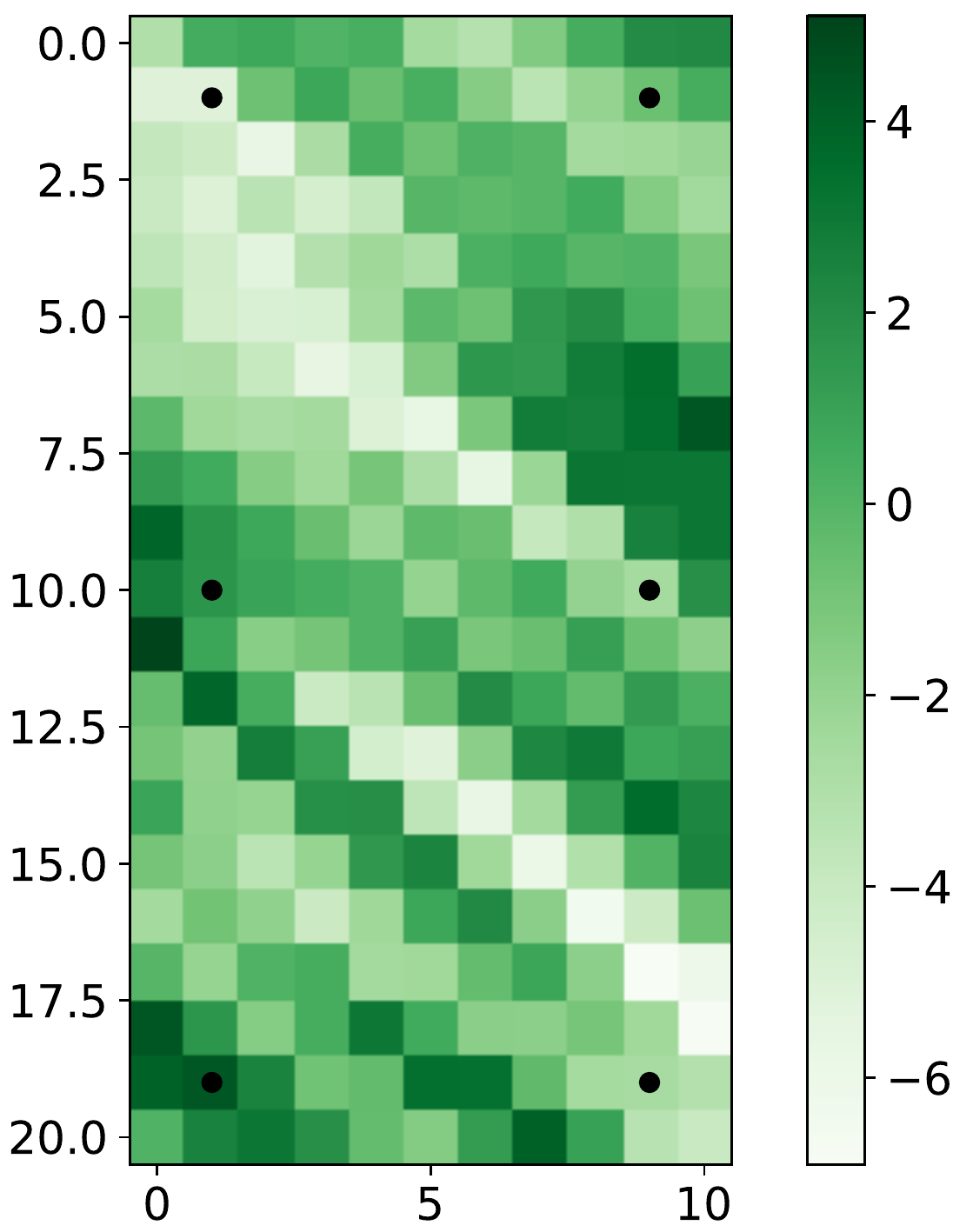}
\put(20,90){\tiny 0}
\put(20,50){\tiny 2}
\put(20,10){\tiny 4}
\put(55.5,90){\tiny 1}
\put(55.5,50){\tiny 3}
\put(55.5,10){\tiny 5}
\put(38,68){\tiny I}
\put(33,68.5){\tiny $\circ$}
\put(38,28){\tiny I}
\put(33,28.5){\tiny $\circ$}
\end{overpic}
\caption{true $\ln K$}
\label{fig:true_lnK}
\end{subfigure}
\hfill
\begin{subfigure}{0.55\textwidth}
\includegraphics[width=1.0\textwidth]{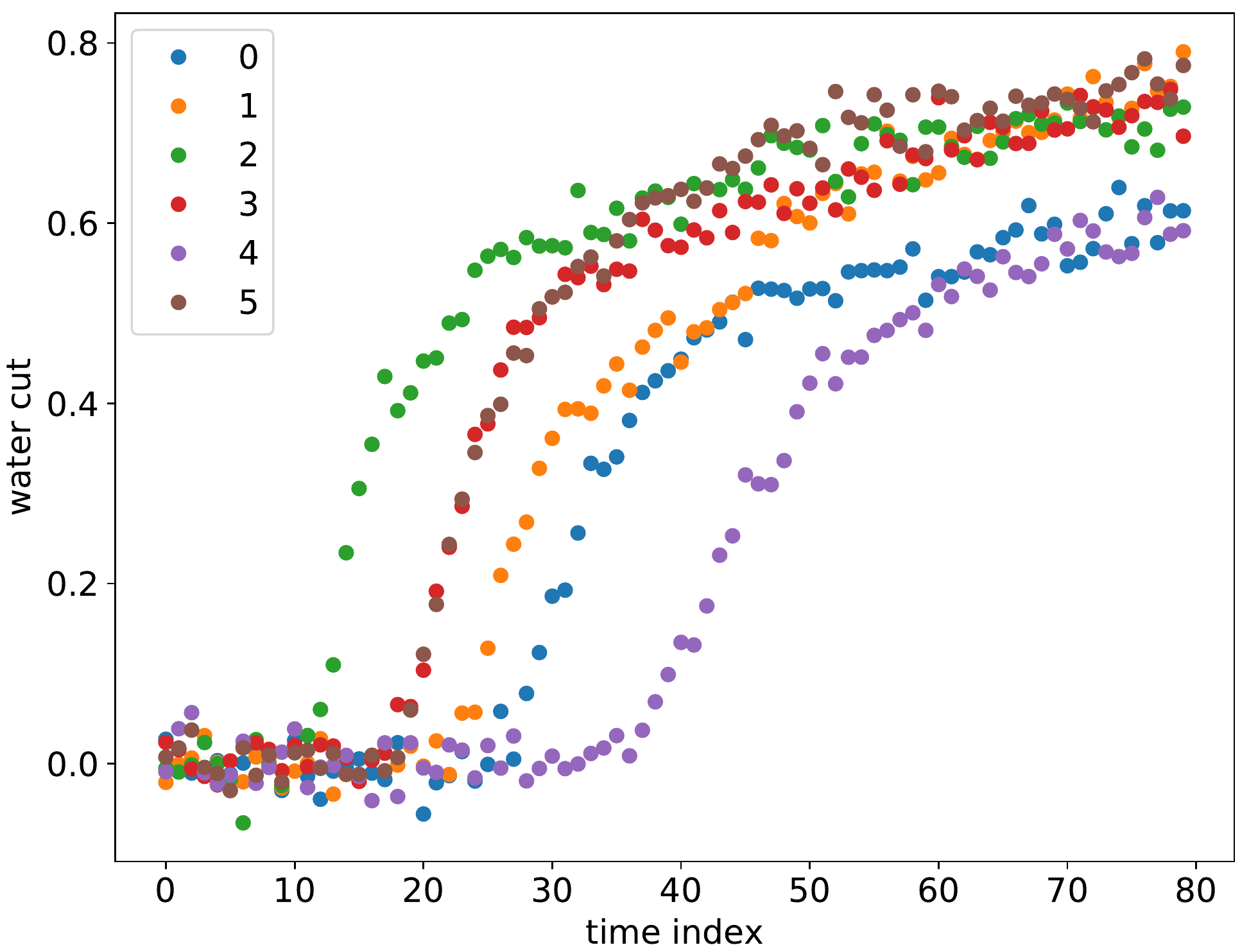} \\
\caption{Noisy observations of water cut in 6 producing wells.}
\label{fig:2D_obs}
\end{subfigure}
\caption{2D reservoir flow problem with uncertain prior covariance.}
\label{fig:2D_flow}
\end{figure}

\begin{figure}[htbp]
\centering
\begin{tabular}{ccc} 
\includegraphics[width=0.25\textwidth]{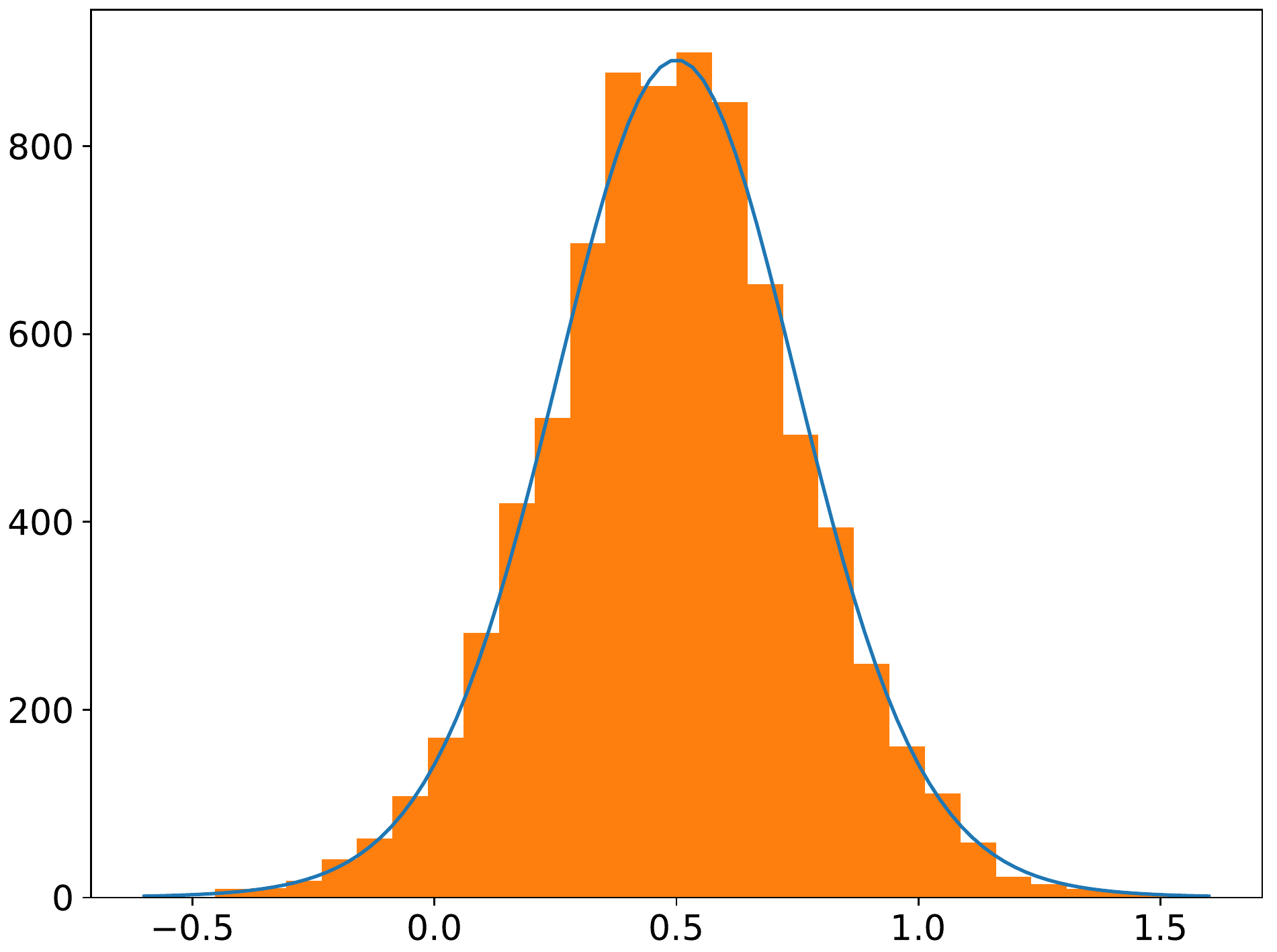} &
\includegraphics[width=0.25\textwidth]{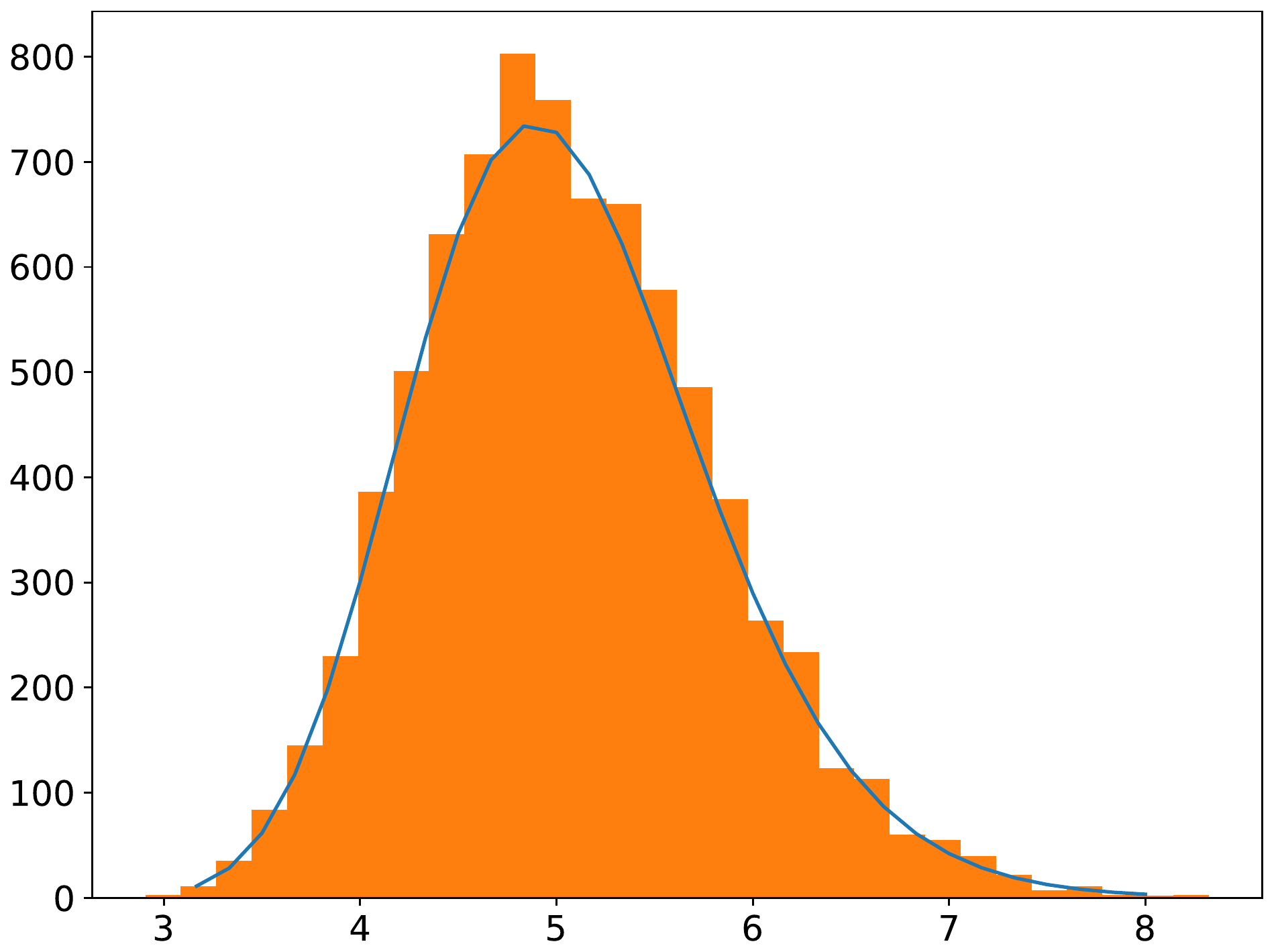} &
\includegraphics[width=0.25\textwidth]{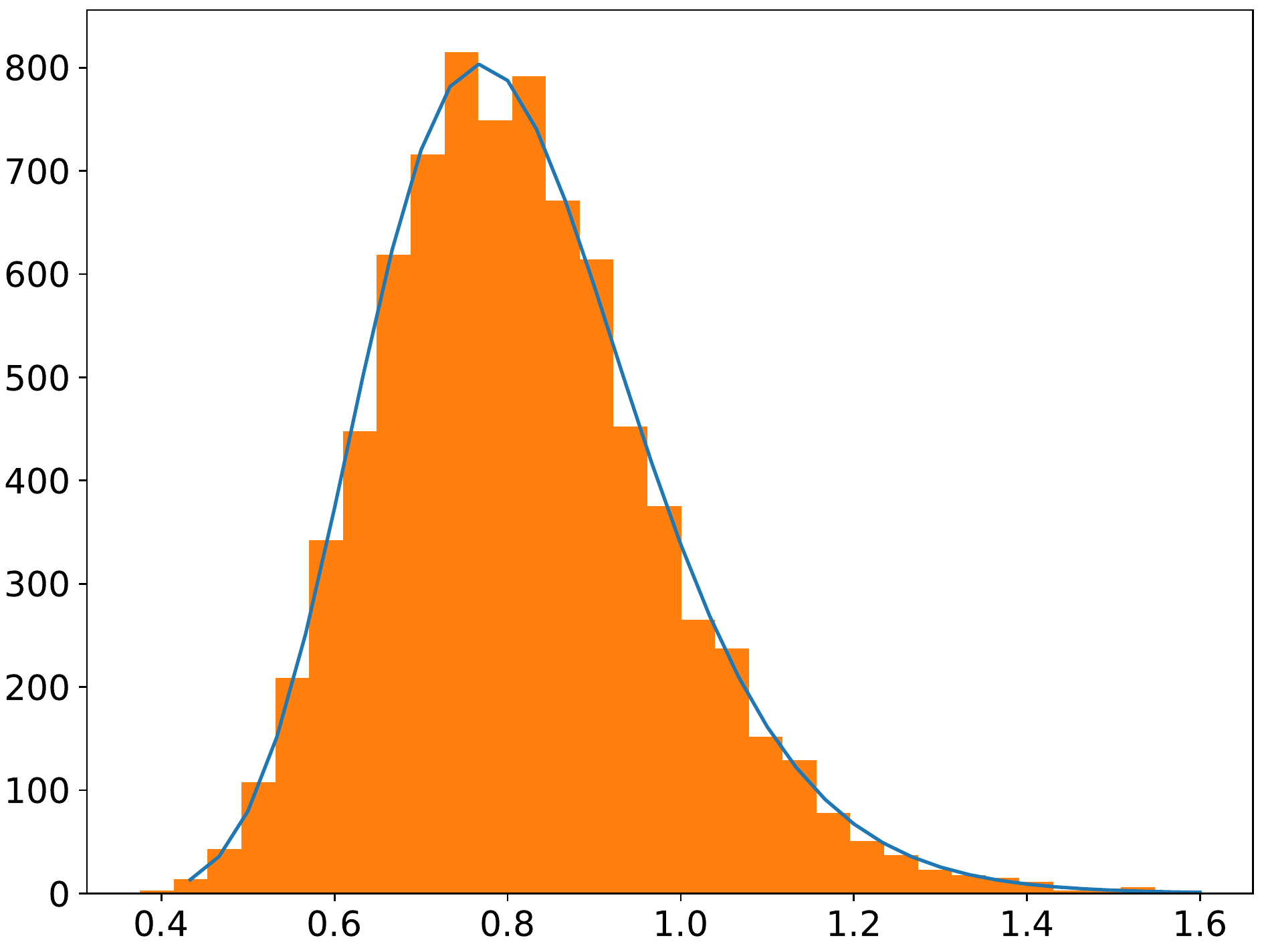} \\
orientation &  range ratio  & range 
\end{tabular}
\caption{Prior uncertainty in each of the hyperparameters of the hierarchical model for the two-dimensional flow problem.}
\label{fig:2D_flow_hyperparameter_priors}
\end{figure}

The data are measurements of water cut (fraction of produced fluid that is water) as a function of time in the producing wells.  Although the permeability is highly variable, the wells are controlled by total flow rate which is identical for all producers. Figure~\ref{fig:true_lnK} shows the  permeability field that was used to generate the observations and Figure~\ref{fig:2D_obs} the noisy observations of water cut  at each of the wells. The errors in the observations are independent Gaussian with standard error 0.02. 
For data assimilation, the prior covariance for log-permeability has uncertainty in the orientation of the principal axes of anisotropy and in the range of the correlation in each of the two principal directions. The prior uncertainties for each of the hyperparameters of the covariance are shown as histograms of sampled values in Fig.~\ref{fig:2D_flow_hyperparameter_priors}. The priors for correlation length and ratio of correlation ranges are both assumed to be log-normal. The prior for orientation of the principal axes of the covariance for permeability is Gauss-von Mises (Eq.~\ref{eq:Gauss-vonMises}), which is close to Gaussian when the variance is relatively small.

\subsection{Data assimilation results} \label{sec:data_assimilation_2Dflow}

Data assimilation with the hierarchical parameterization was performed using two methods: a standard Levenberg-Marquardt iterative ensemble smoother and a hybrid IES, also with Levenberg-Marquardt regularization. In both approaches, the parameters being updated are $z$ (dimension 450), and the three hyperparameters of the covariance function. The updates of $z$ were localized in the IES approach using the Gaspari-Cohn correlation function with a taper range that was the same as the true correlation range. Localization was not used in the hybrid IES approach.
An ensemble size of 200 was used for IES, while a smaller ensemble ($N_e = 100$) was used for the hybrid IES approach. In both approaches, the starting value of the Levenberg-Marquardt parameter  $\lambda$, was selected based on the magnitude of the initial data mismatch  \citep{chen:13}. If the average squared data mismatch decreased in an iteration, the value of $\lambda$ was decreased by a factor of 4. If the average square data mismatch increased, then $\lambda$ was multiplied by a factor of 4. Iterations were stopped if $\lambda$ 
increased in two successive iterations, or if the number of iterations exceeded 25, or if the magnitude of the reduction in data mismatch was too small.

\begingroup
\setlength{\tabcolsep}{2pt} 
\renewcommand{\arraystretch}{1.} 
\begin{figure}
\begin{tabular}{cccccc} 
\raisebox{2ex}{\rotatebox{90}{\scriptsize{prior}}} &
\includegraphics[width=0.18\textwidth]{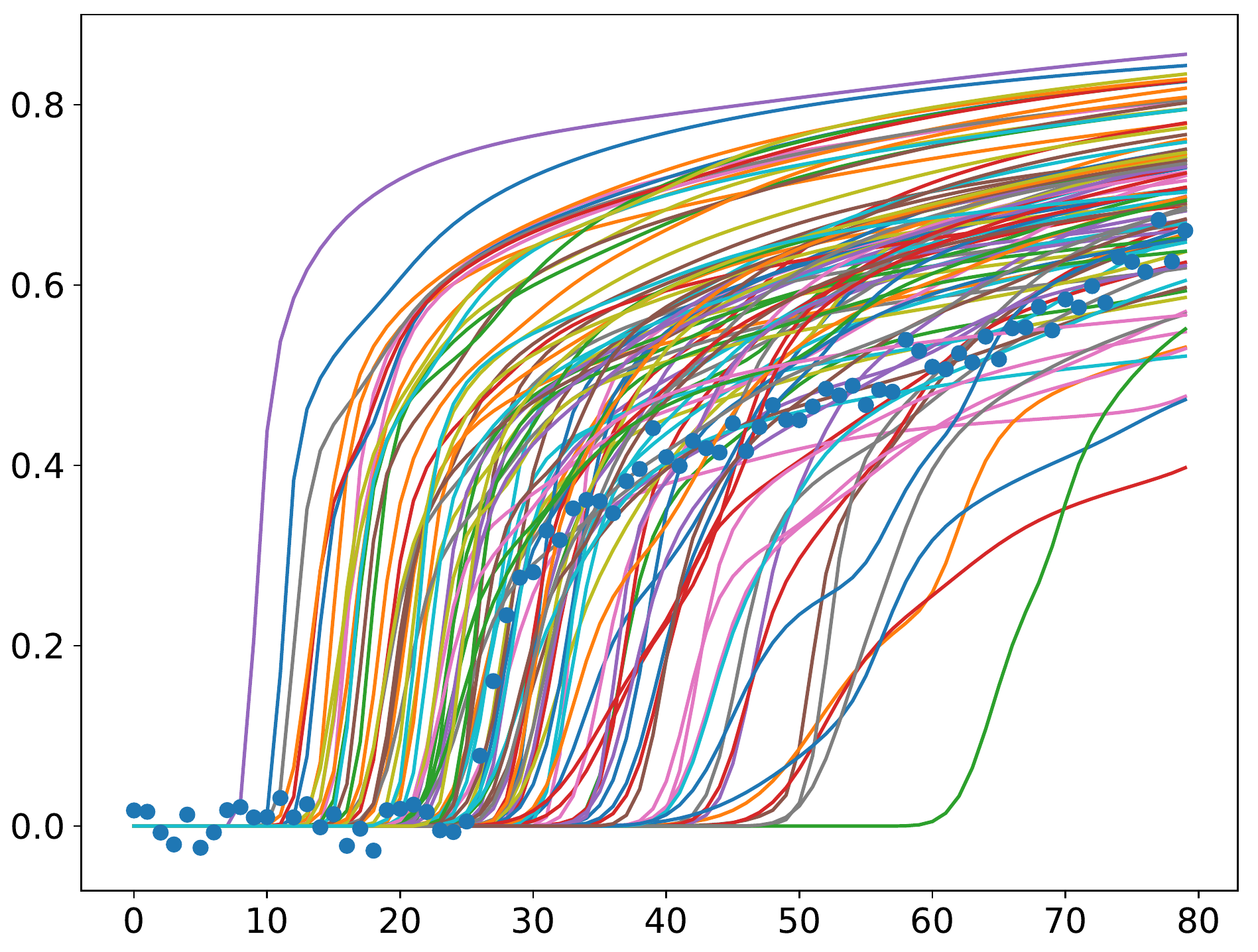} 
&
\includegraphics[width=0.18\textwidth]{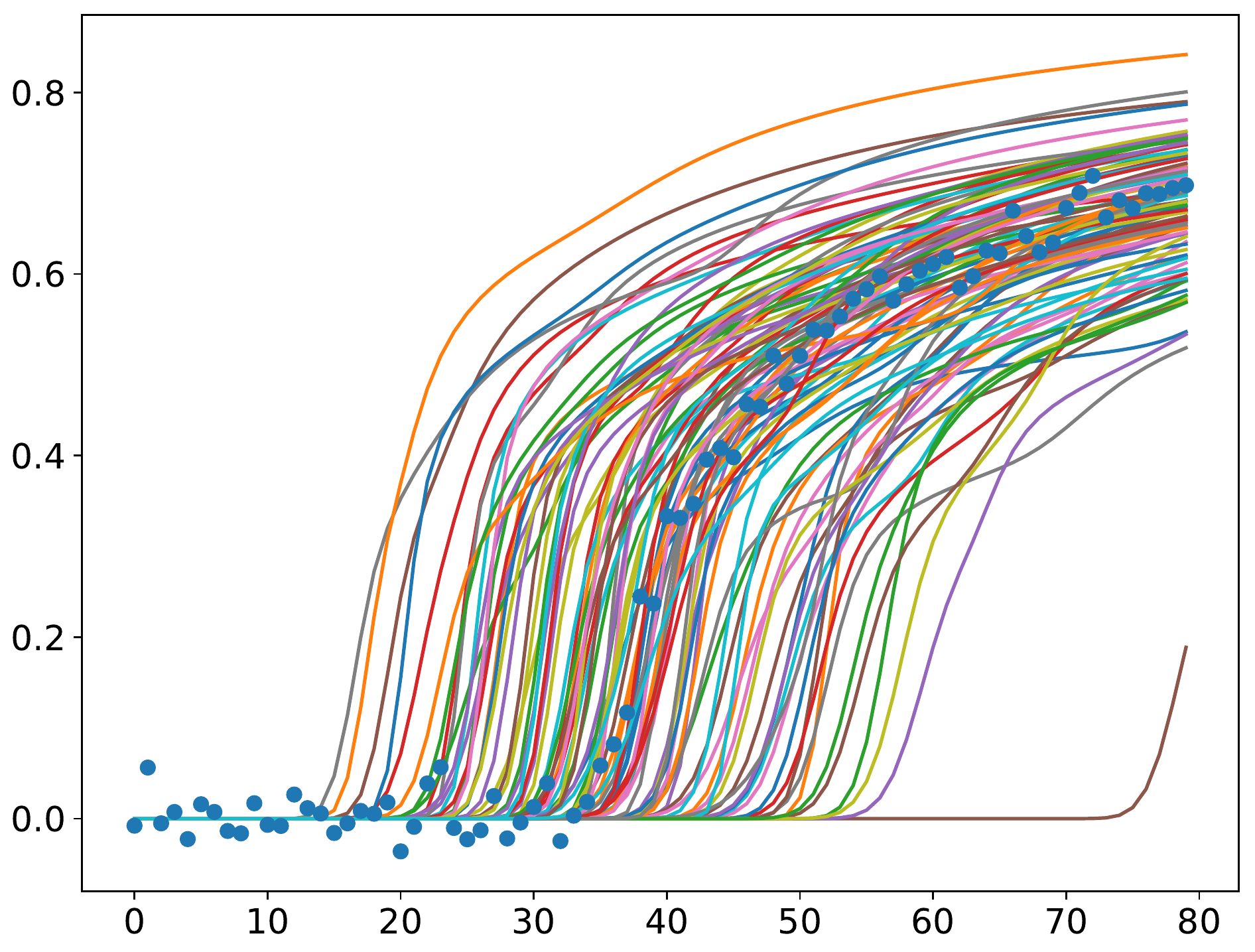}
&
\includegraphics[width=0.18\textwidth]{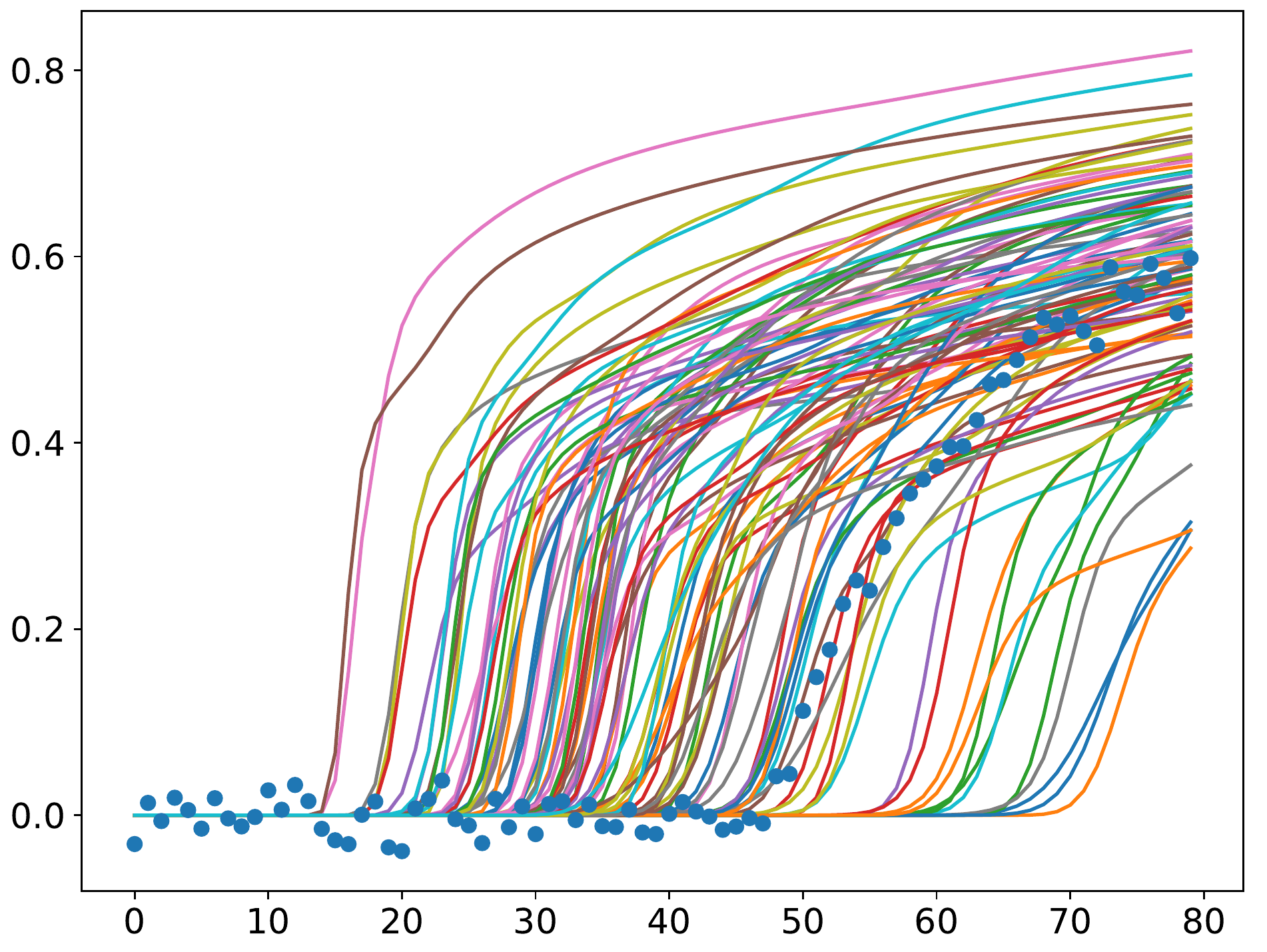}
&
\includegraphics[width=0.18\textwidth]{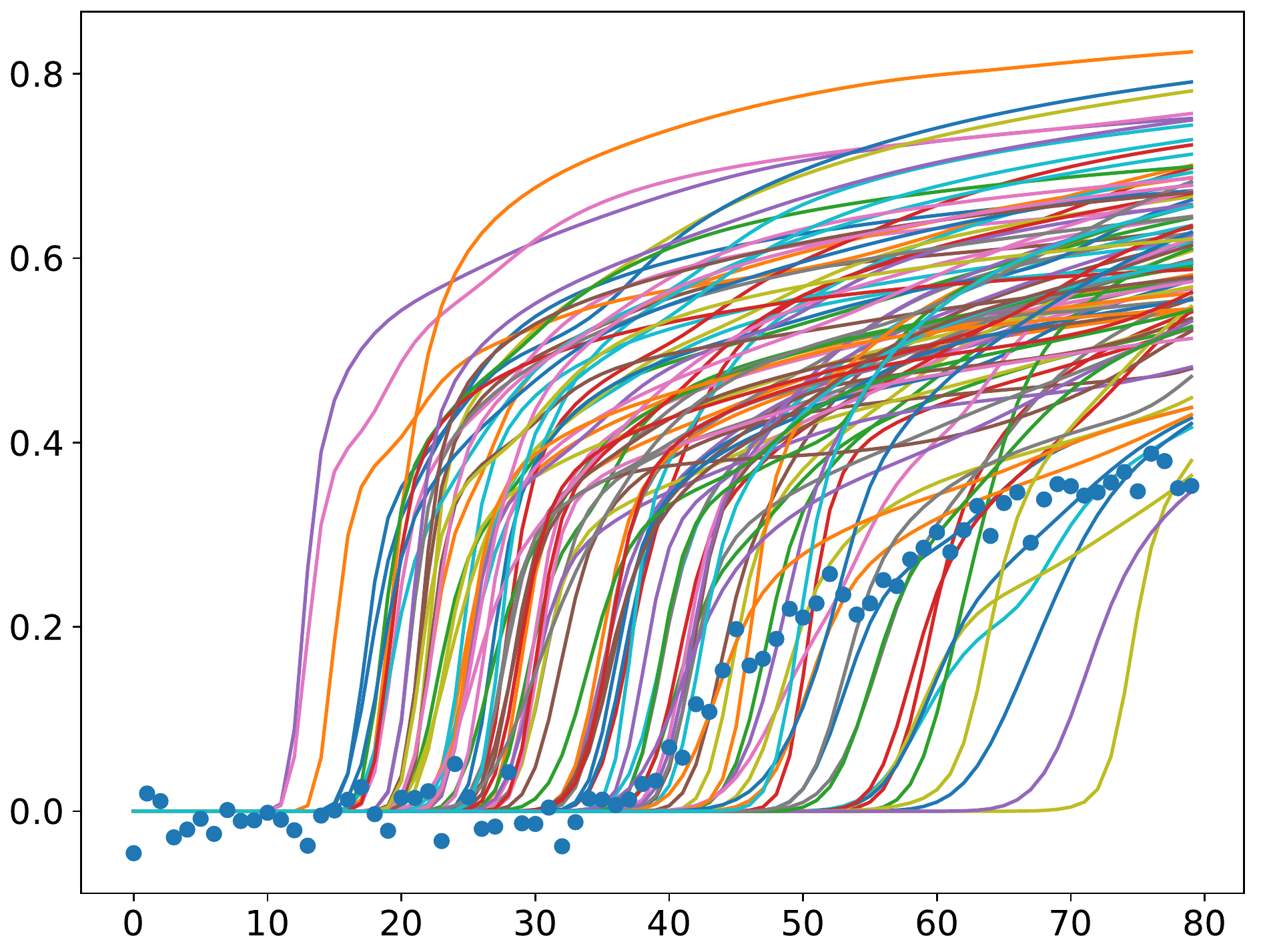} 
&
\includegraphics[width=0.18\textwidth]{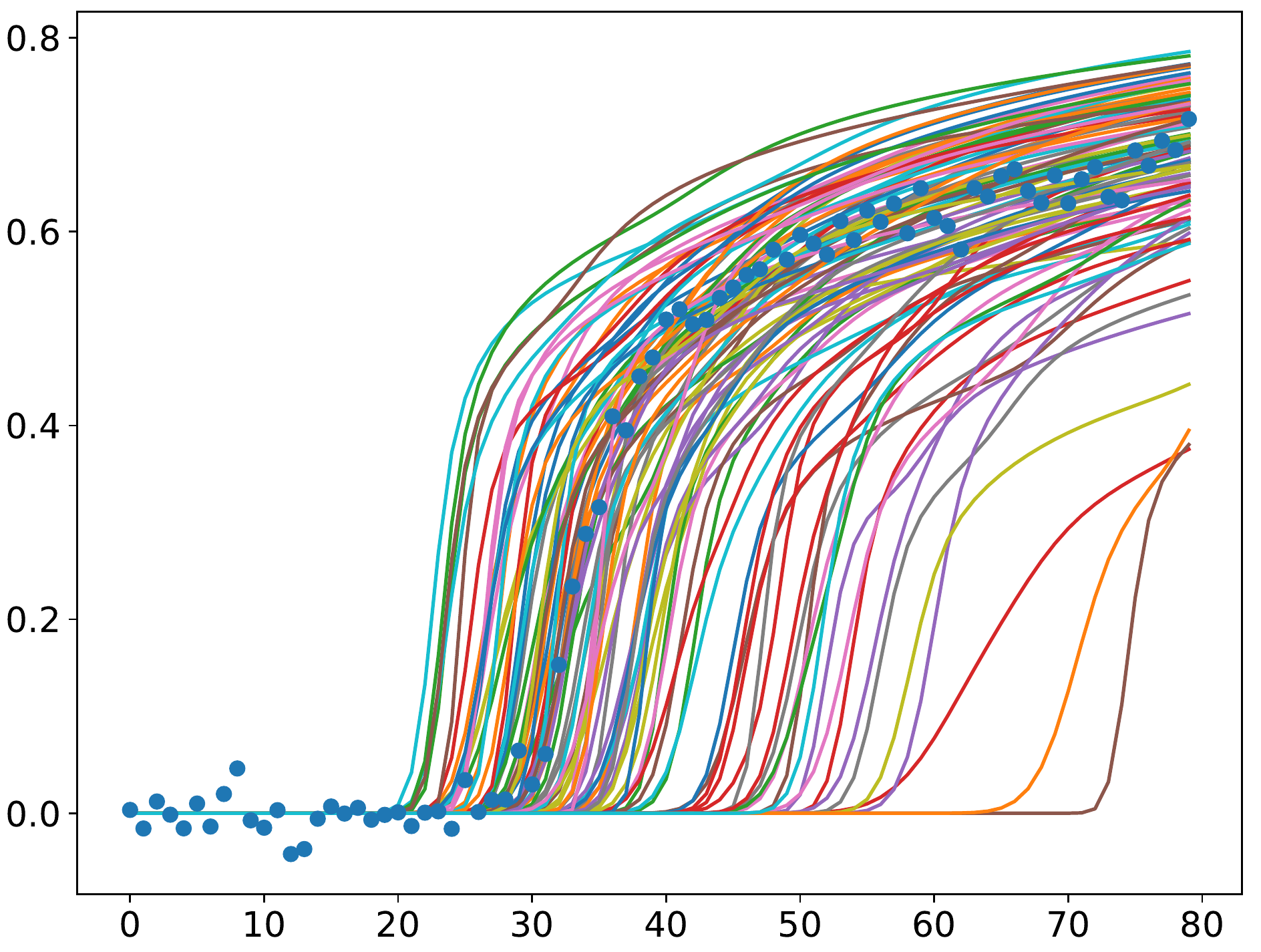}
\\
\raisebox{3.5ex}{\rotatebox{90}{\scriptsize{IES}}} 
&
\includegraphics[width=0.18\textwidth]{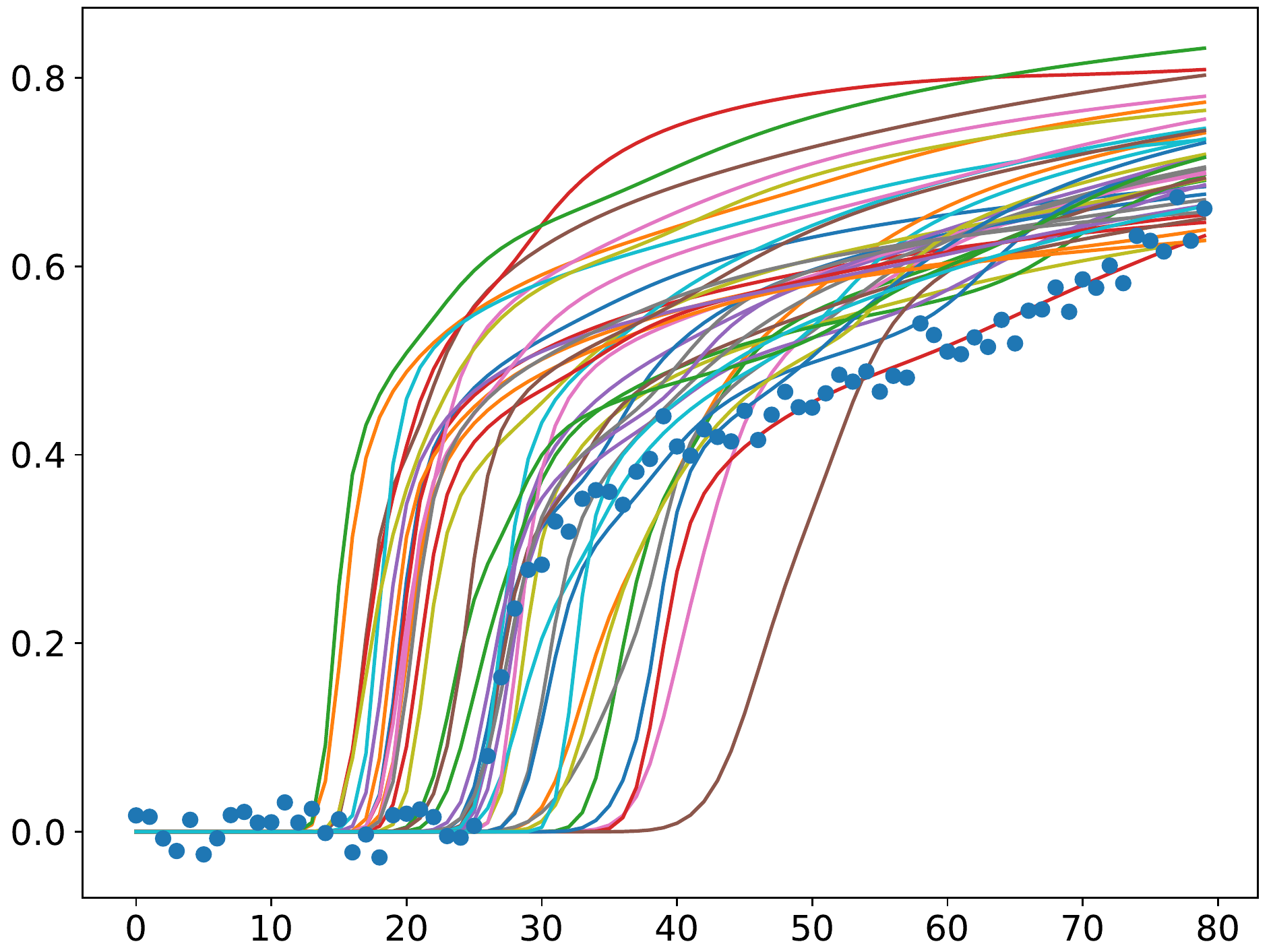} 
&
\includegraphics[width=0.18\textwidth]{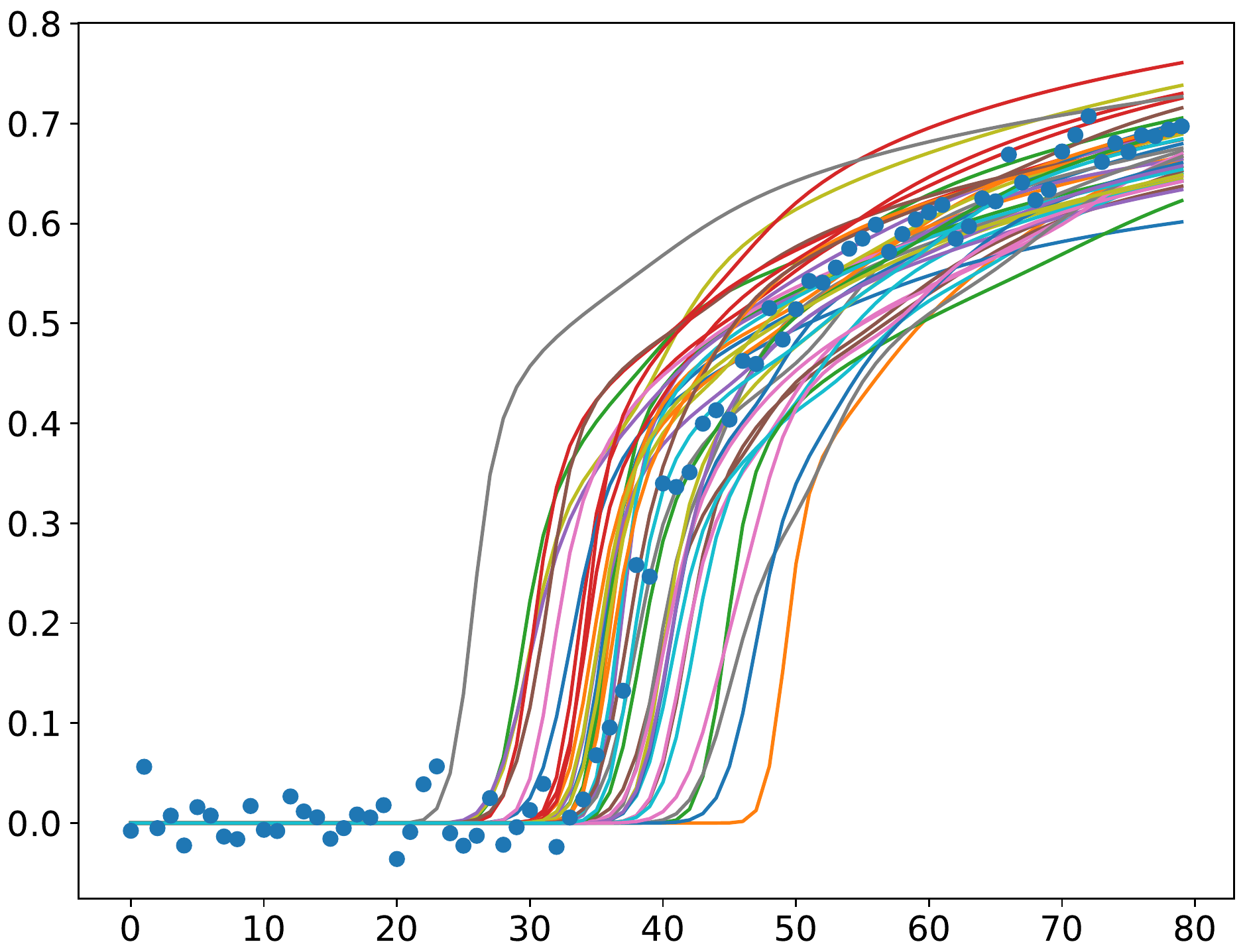} 
&
\includegraphics[width=0.18\textwidth]{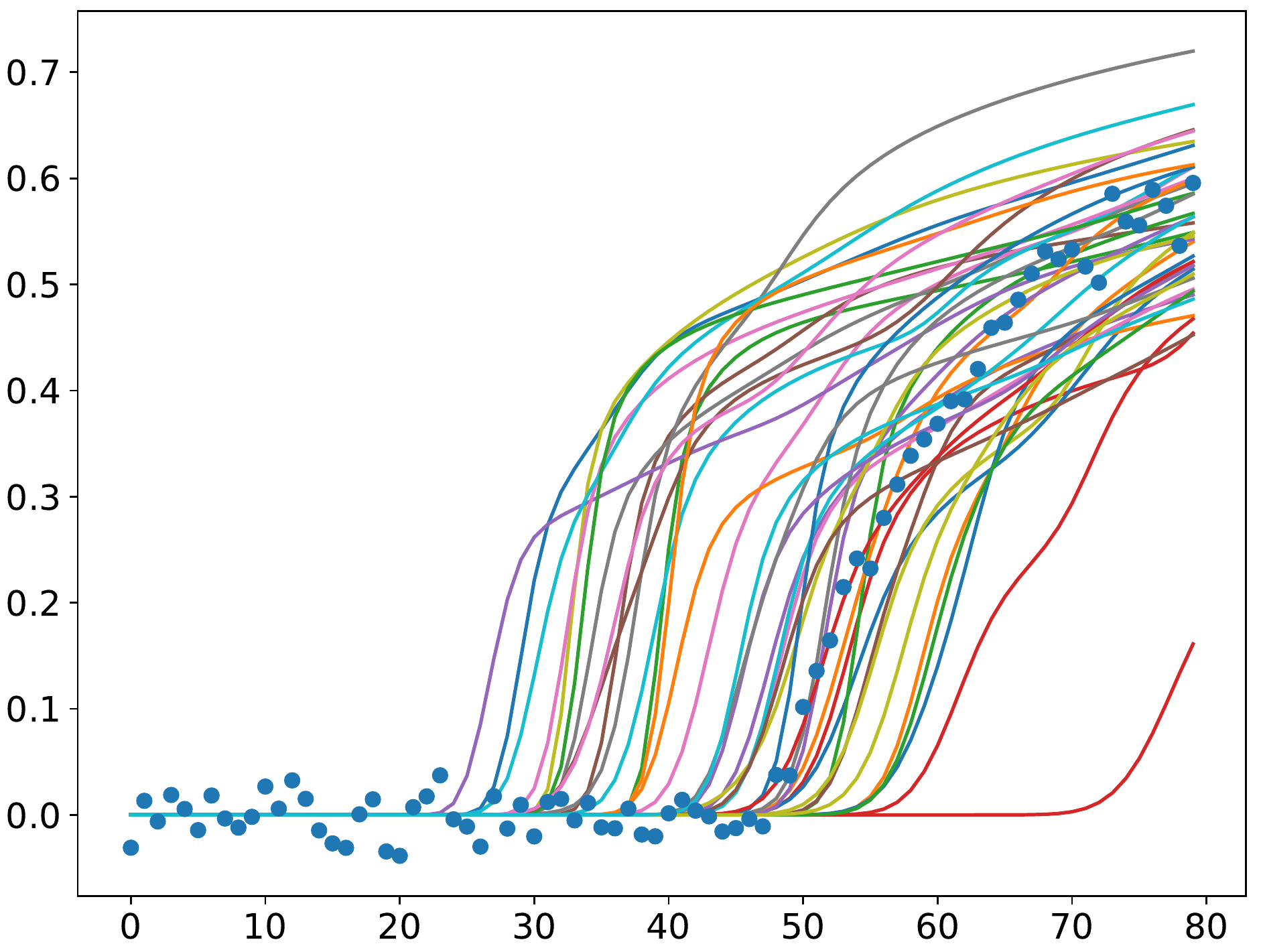} 
&
\includegraphics[width=0.18\textwidth]{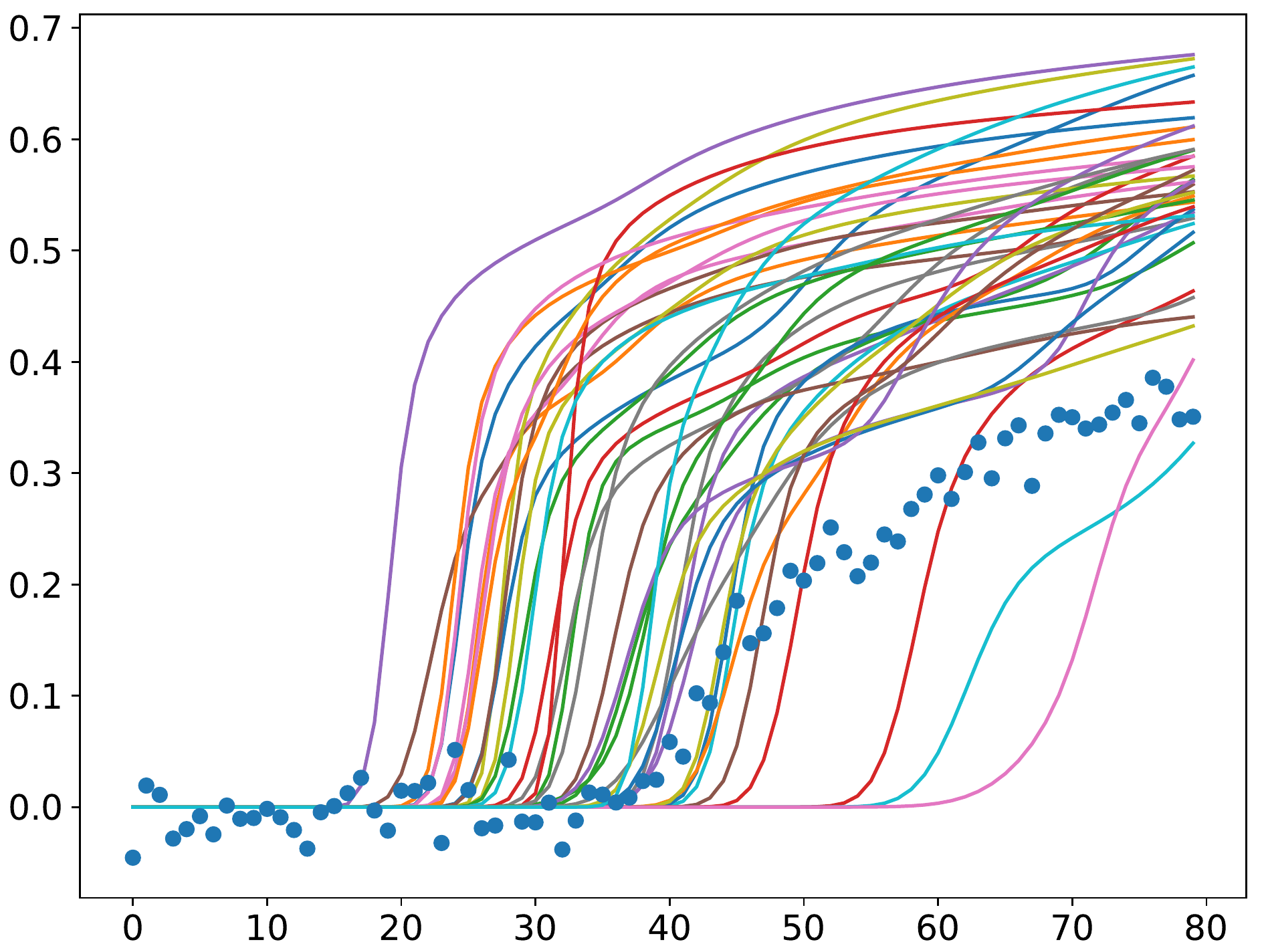} 
&
\includegraphics[width=0.18\textwidth]{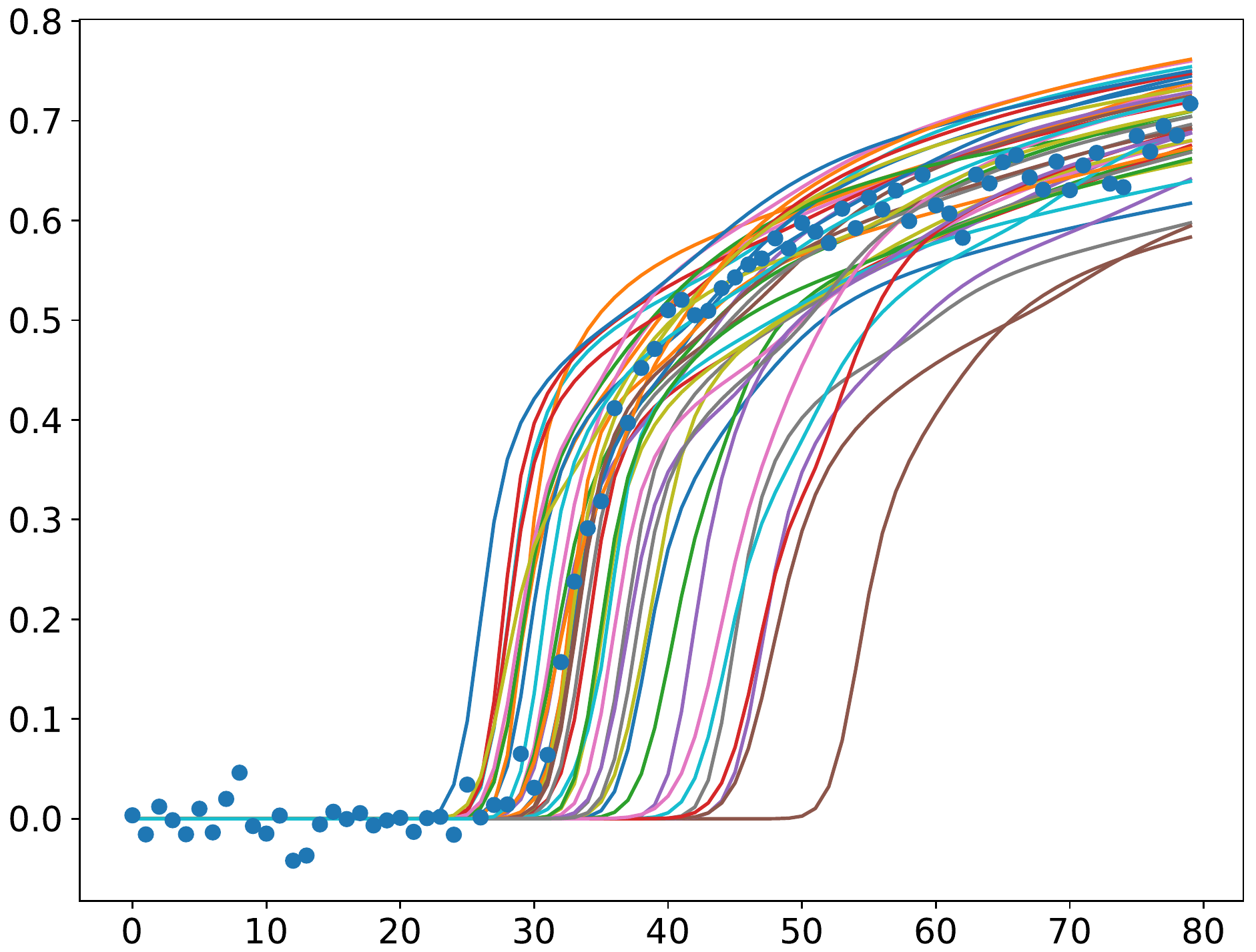} 
\\
\raisebox{0.5ex}{\rotatebox{90}{\scriptsize{hybrid IES}}} 
&
\includegraphics[width=0.18\textwidth]{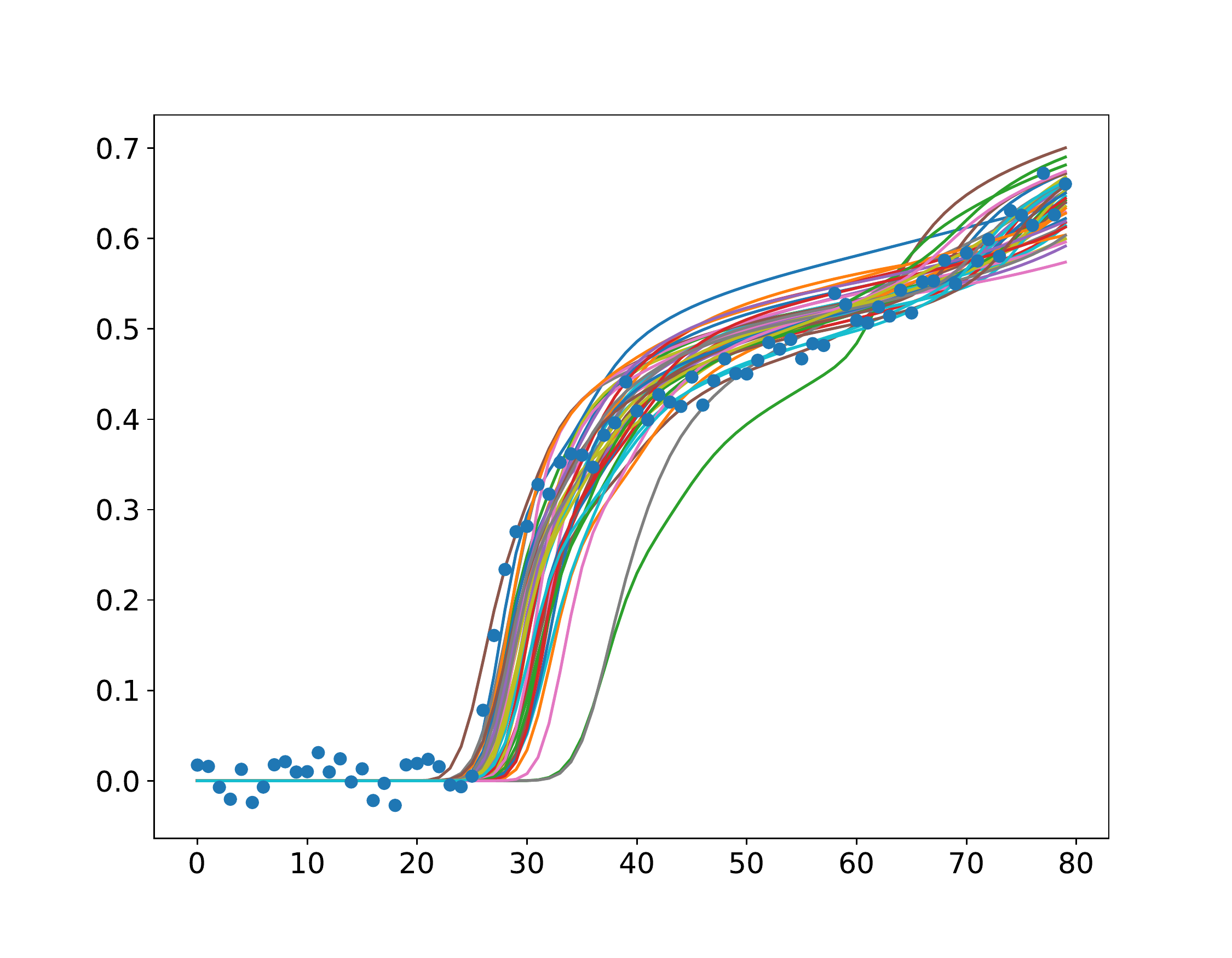}
&
\includegraphics[width=0.18\textwidth]{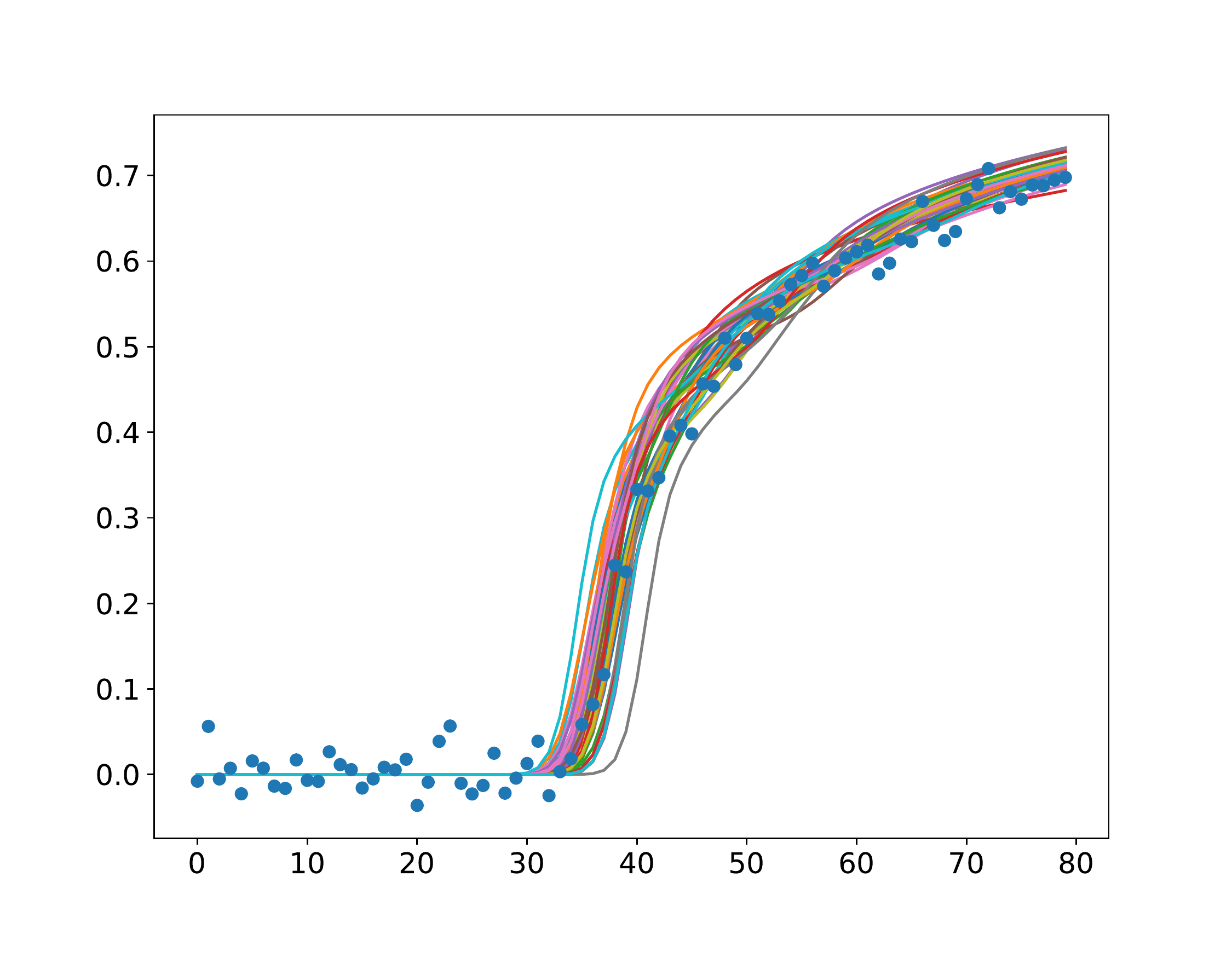}
&
\includegraphics[width=0.18\textwidth]{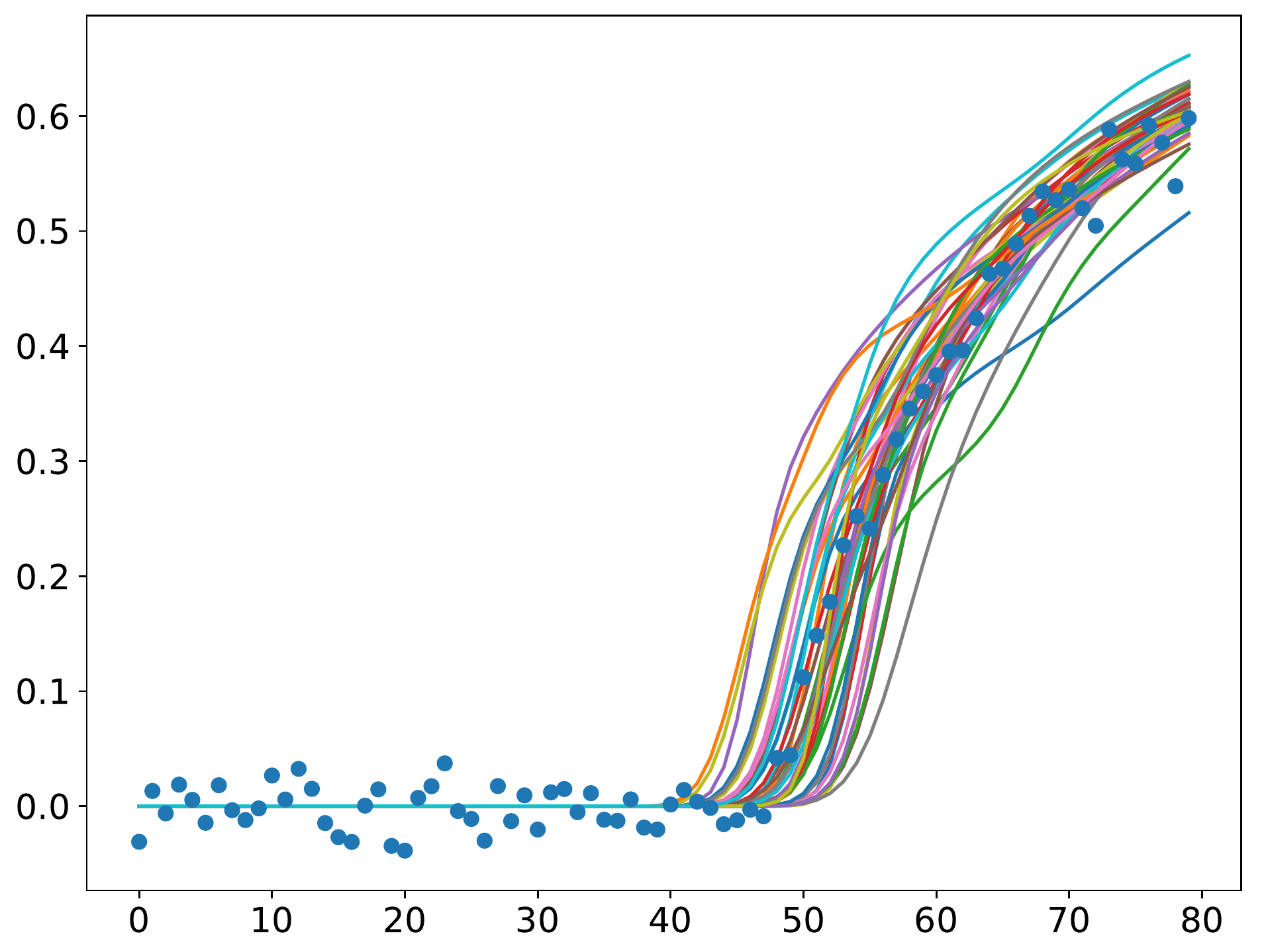}
&
\includegraphics[width=0.18\textwidth]{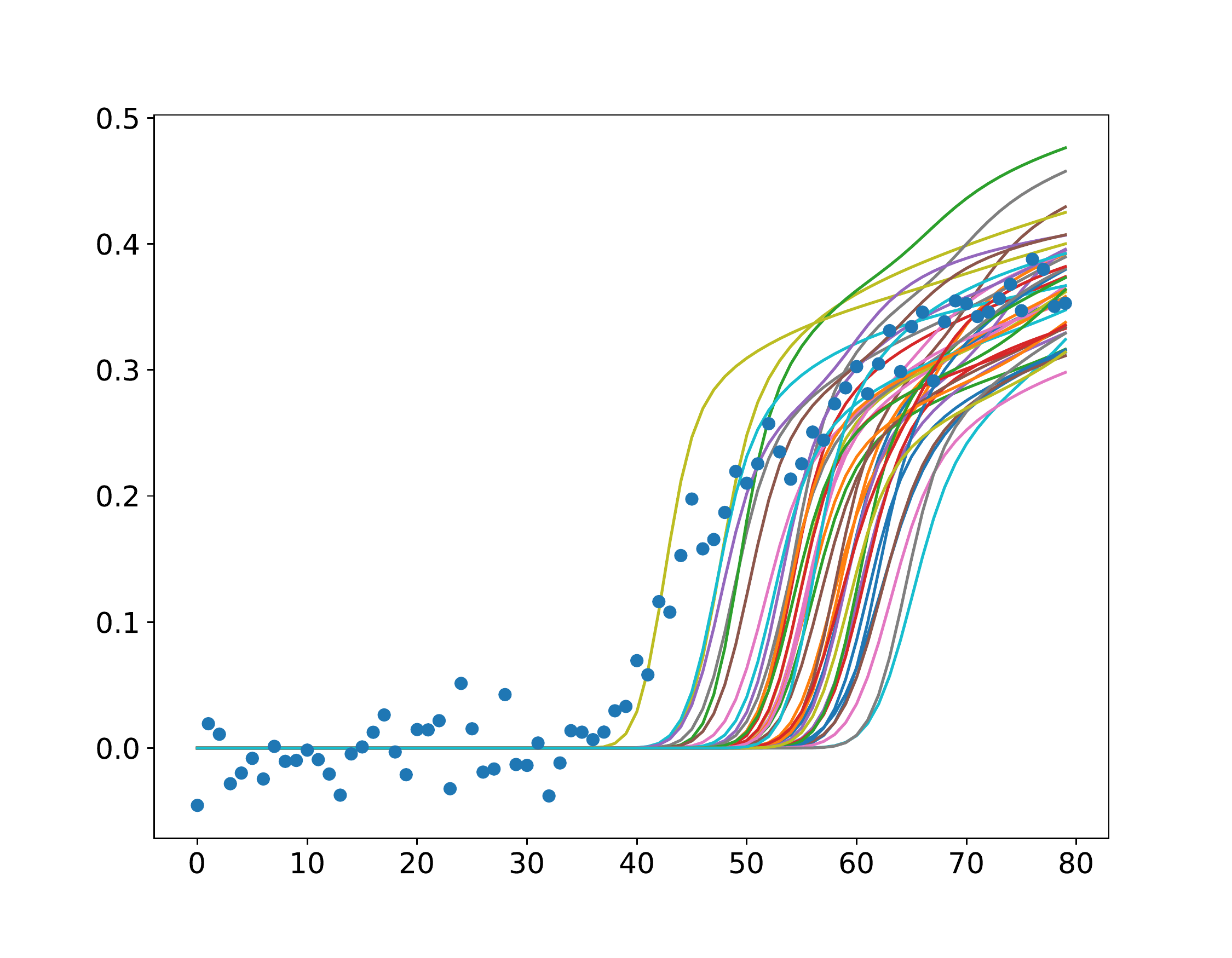}
&
\includegraphics[width=0.18\textwidth]{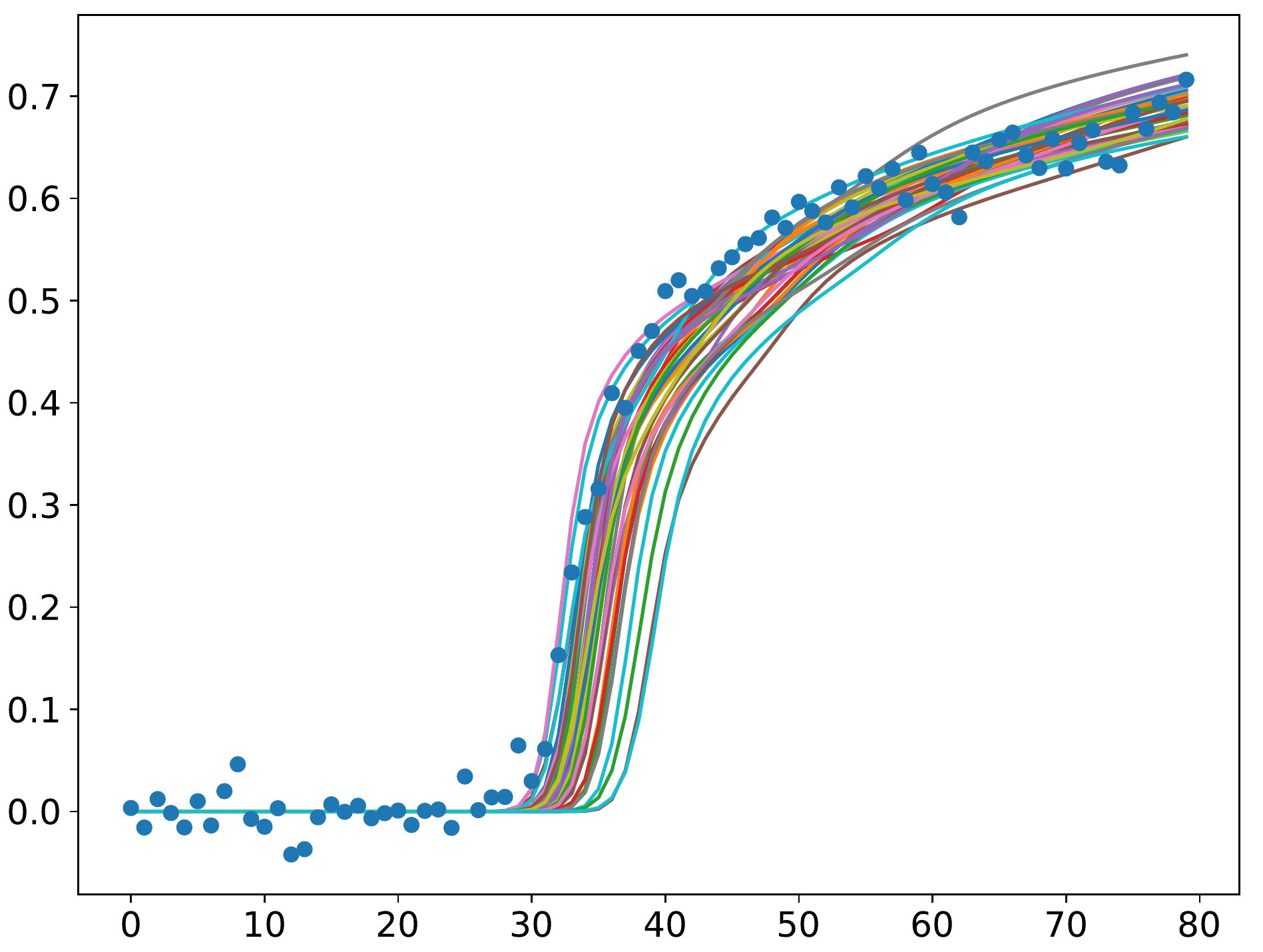}
\\
 &
Prod 0 & Prod 1 & Prod 2 & Prod 3 & Prod 4
\end{tabular}
\caption{Data assimilation with the hybrid IES. Watercut vs time for 40 realizations from the prior and from the posterior for five producers. Blue dots are noisy observations of water cut.}
\label{fig:prod_history}
\end{figure}
\endgroup

Prior and posterior predictive distributions of water cut are compared with observed water cut in Fig.~\ref{fig:prod_history} for 5 of the 6 producers. The prior predictive distribution (top row) ``covers'' the observations although the  prior distribution for producer 3 appears to be inconsistent with trends in the data. The posterior predictive distribution computed using IES (middle row) has reduced spread compared to the prior distribution, but the mismatch with actual observations (blue dots) is much larger than would be expected for conditional realizations. The hybrid IES (bottom row) generates posteriori predictions that are generally consistent with both the observations and with the measurement error, with the exception that the match with data in producer 3 is poor.

\begin{figure}[htbp]
\centering
\begin{overpic}[width=0.45\textwidth]{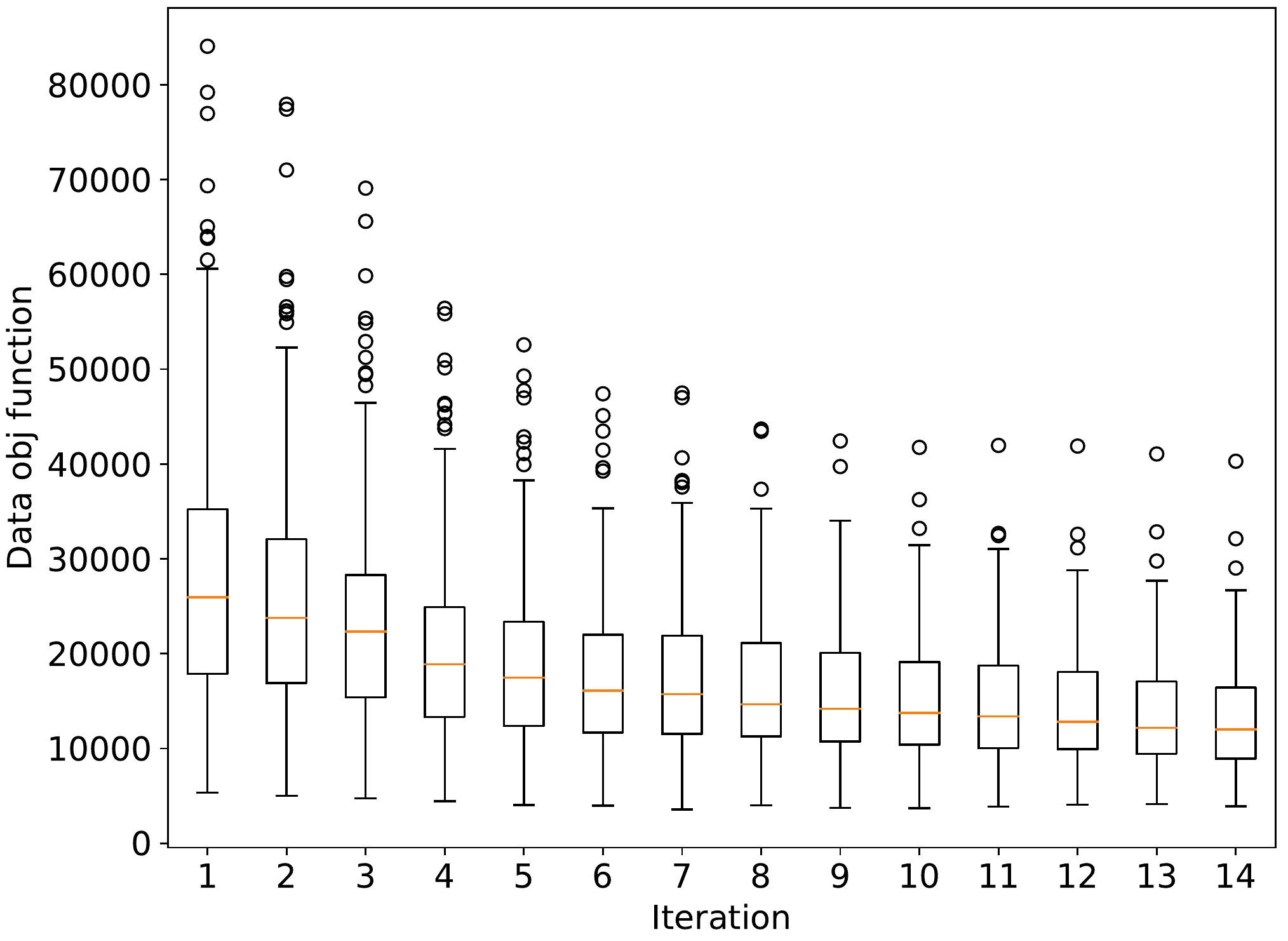}
\put(60,66){\footnotesize IES}
\end{overpic}
\begin{overpic}[width=0.45\textwidth]{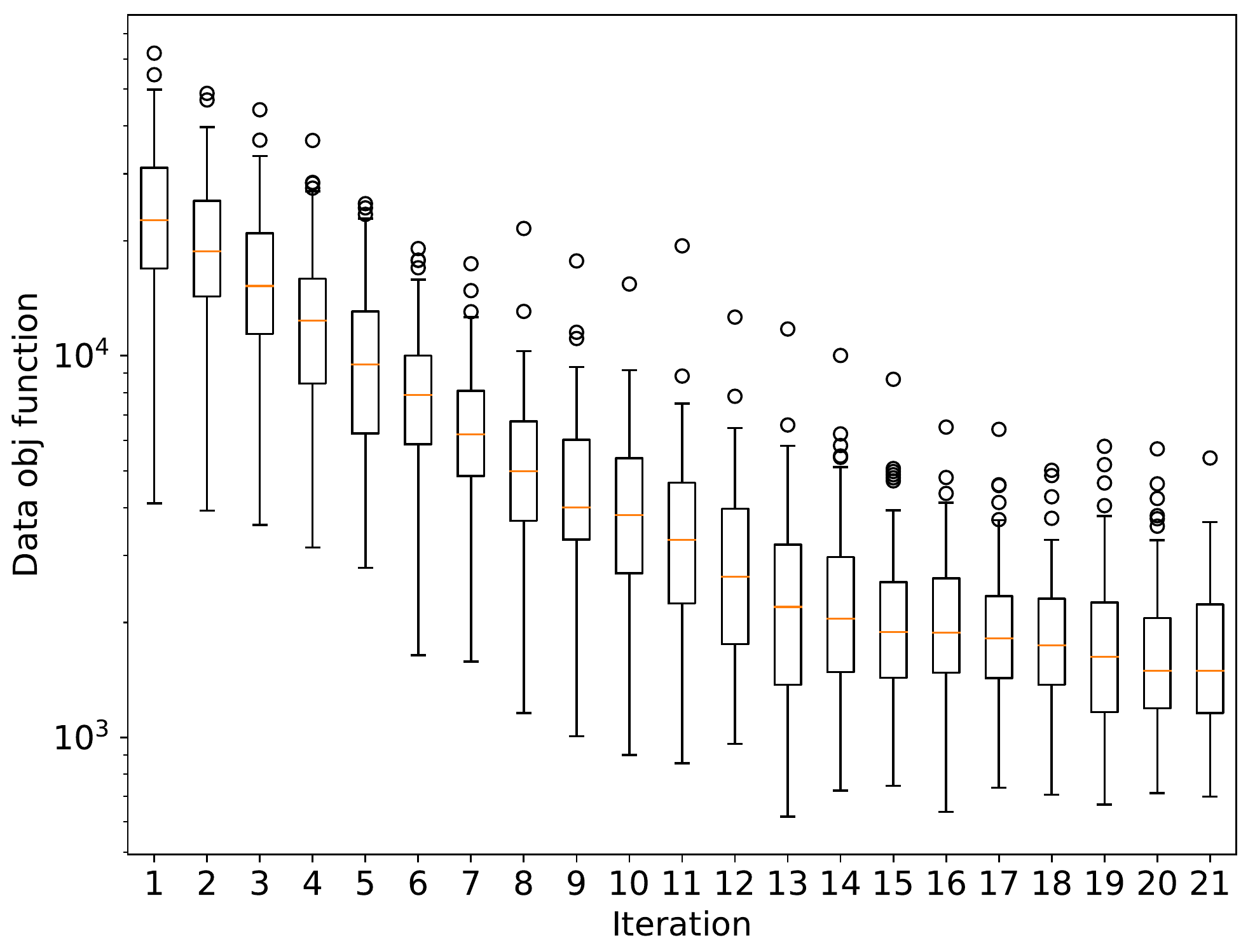}
\put(60,66){\footnotesize hybrid IES}
\end{overpic}
\caption{Reduction in squared mismatch with observed (not perturbed) data. Expected value at convergence is 240. Note that different scales are used for the two plots.} 
\label{fig:data_mismatch_flow}
\end{figure}

The improved match to the data obtained using the hybrid IES approach is confirmed quantitatively in Fig.~\ref{fig:data_mismatch_flow}. In both subfigures, the spread of squared data mismatch values is shown as a box which represents the range containing the central 50\% of the values. Note that the reduction in the squared data mismatch is slow for the IES and that the final mean value (13000) is much greater than the expected value for a properly calibrated ensemble (240). The hybrid IES converges much faster and ends at a much smaller mean value of data mismatch (1151). Although the final value is larger than expected, it is largely a result of the poor match to water cut observations in Producer 3.

\begin{figure}[htbp]
\centering
\begin{tabular}{c|c} 
\includegraphics[width=0.45\textwidth]{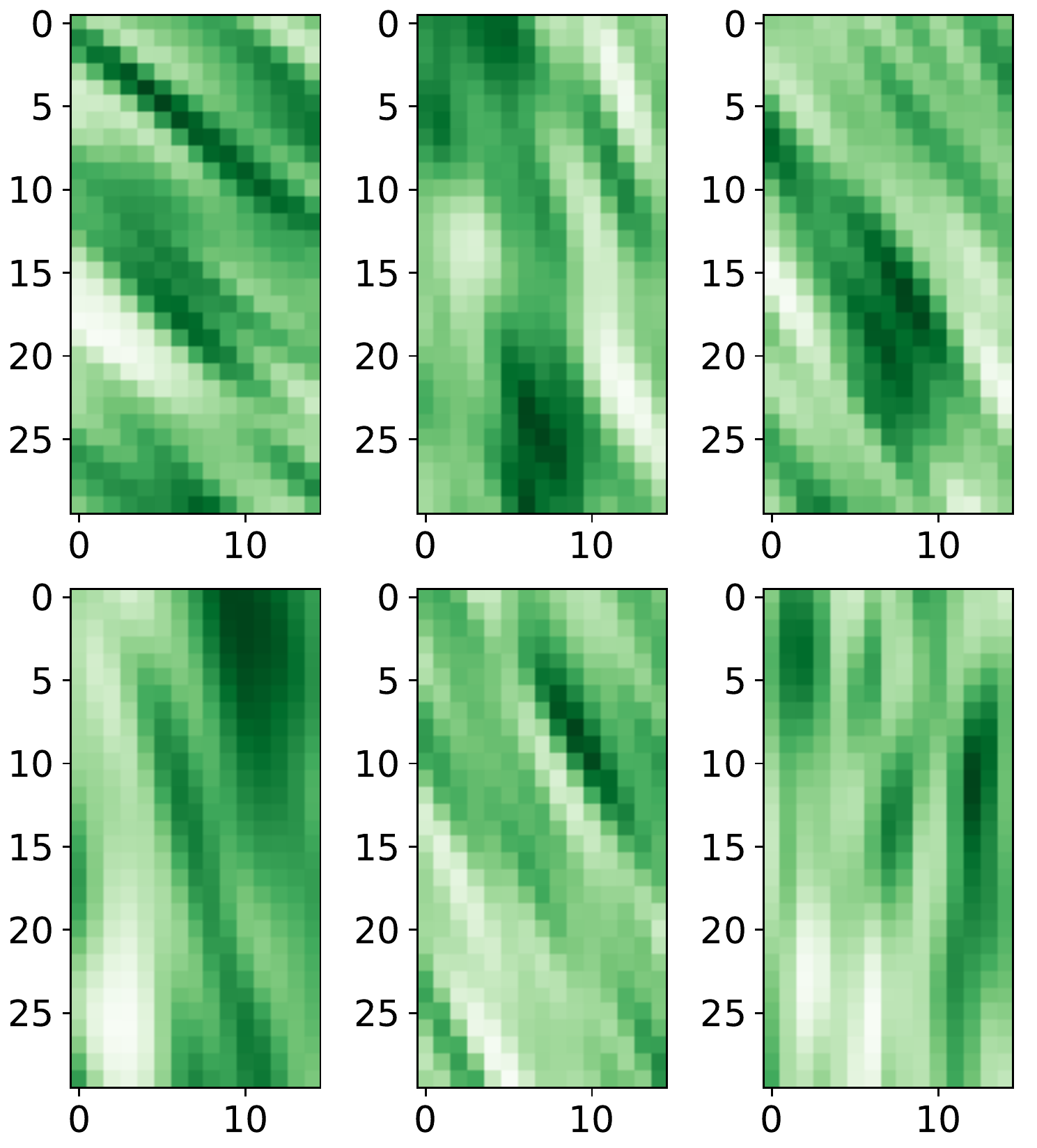} &
\includegraphics[width=0.45\textwidth]{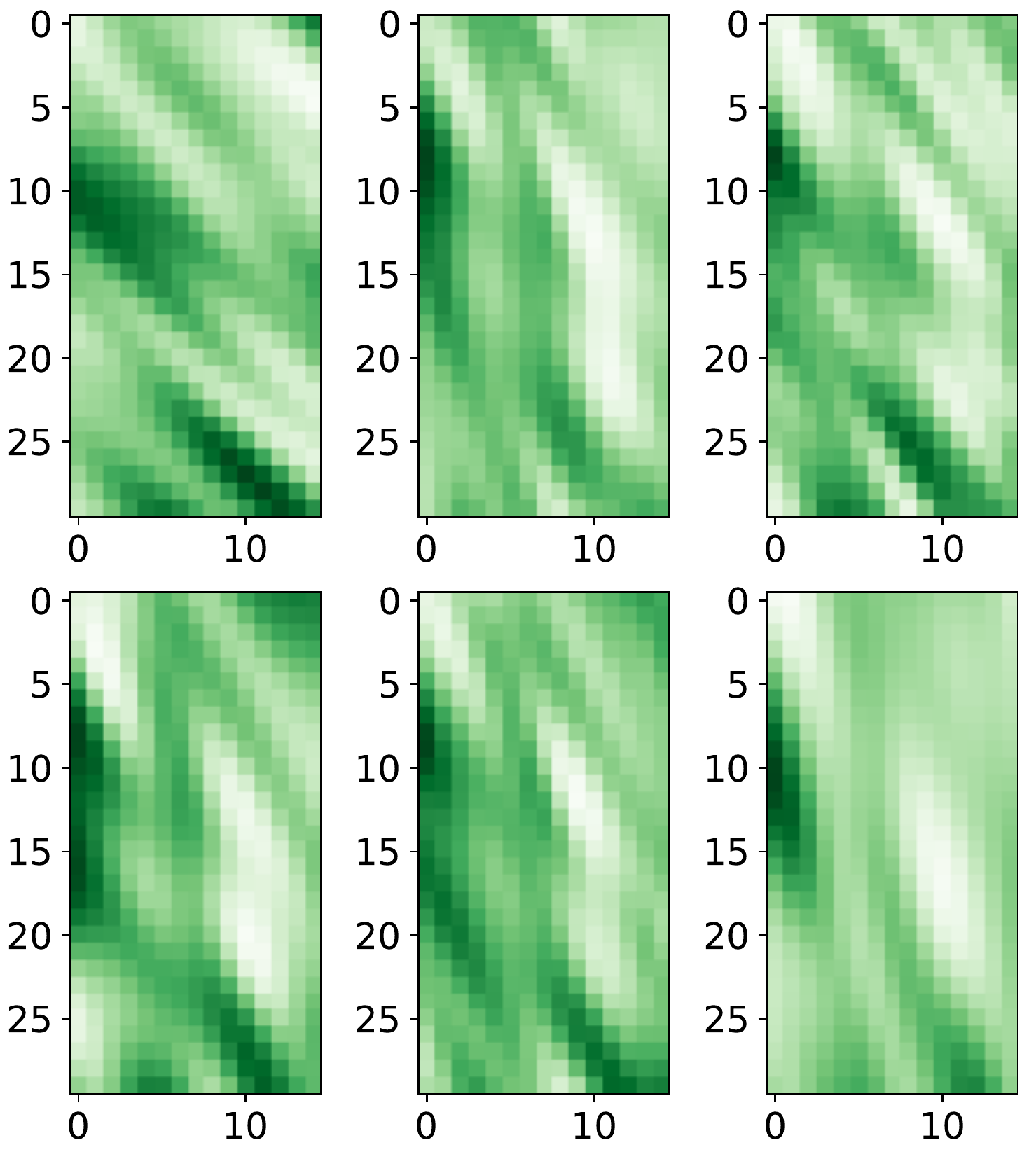} \\
samples from prior & samples from posterior
\end{tabular}
\caption{The first six realizations of the log-permeability field from the prior (left) and the corresponding realizations from the posterior (right). Darker greens correspond to larger values of log-permeability. }
\label{fig:samples_perm_2D}
\end{figure}

\begin{figure}[htbp]
\centering
\begin{tabular}{cccc} 
\includegraphics[width=0.31\textwidth]{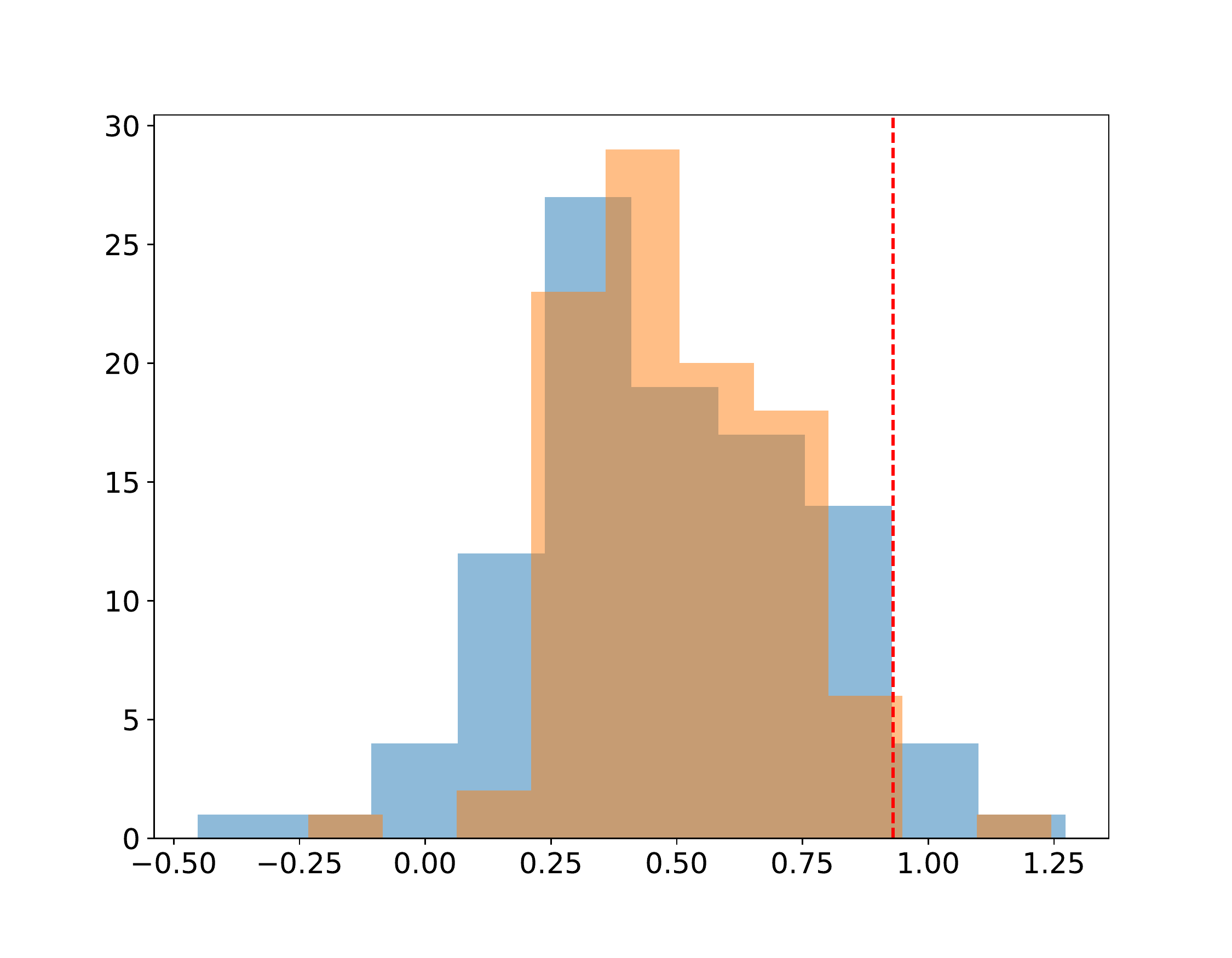} &
\includegraphics[width=0.31\textwidth]{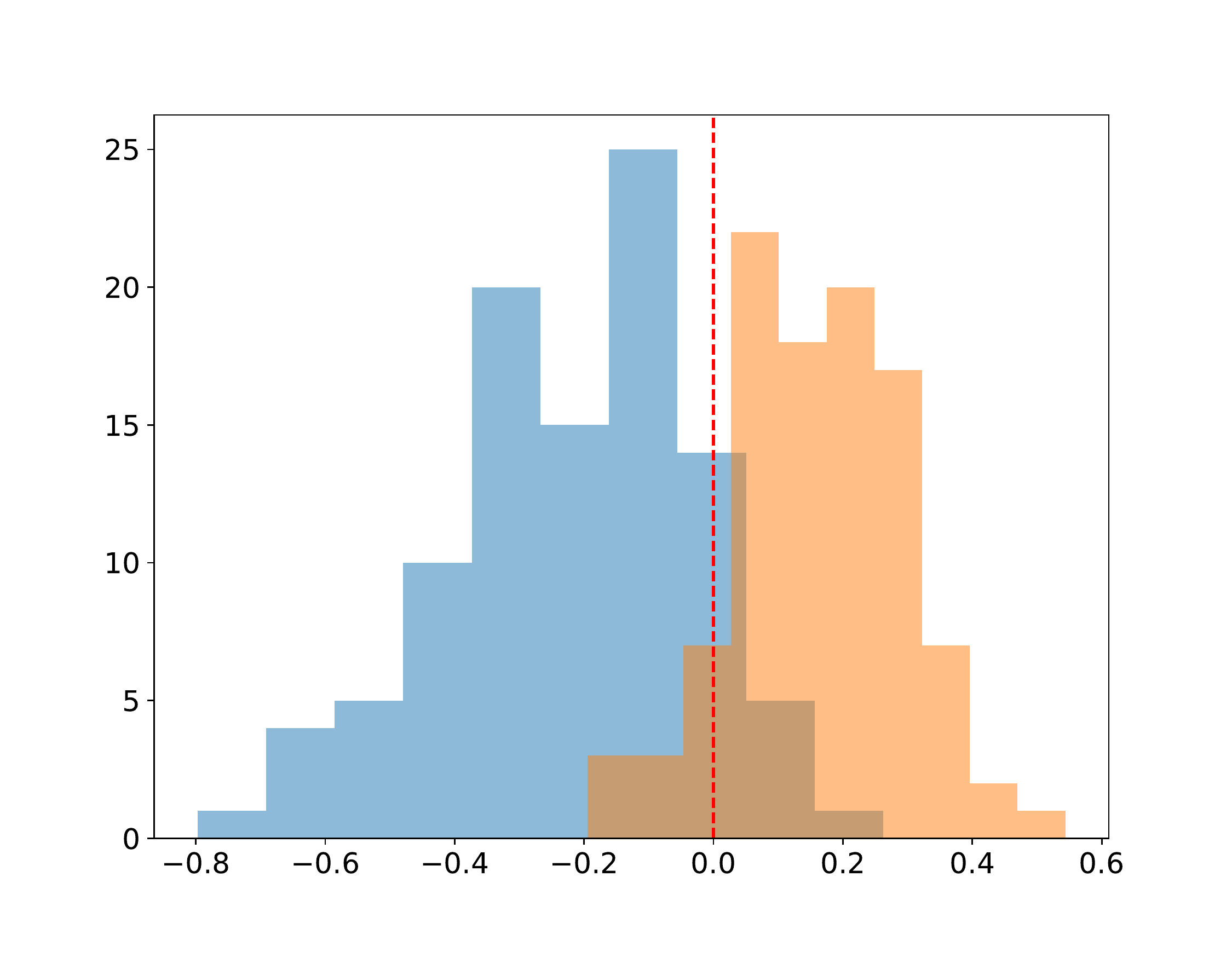} &
\includegraphics[width=0.31\textwidth]{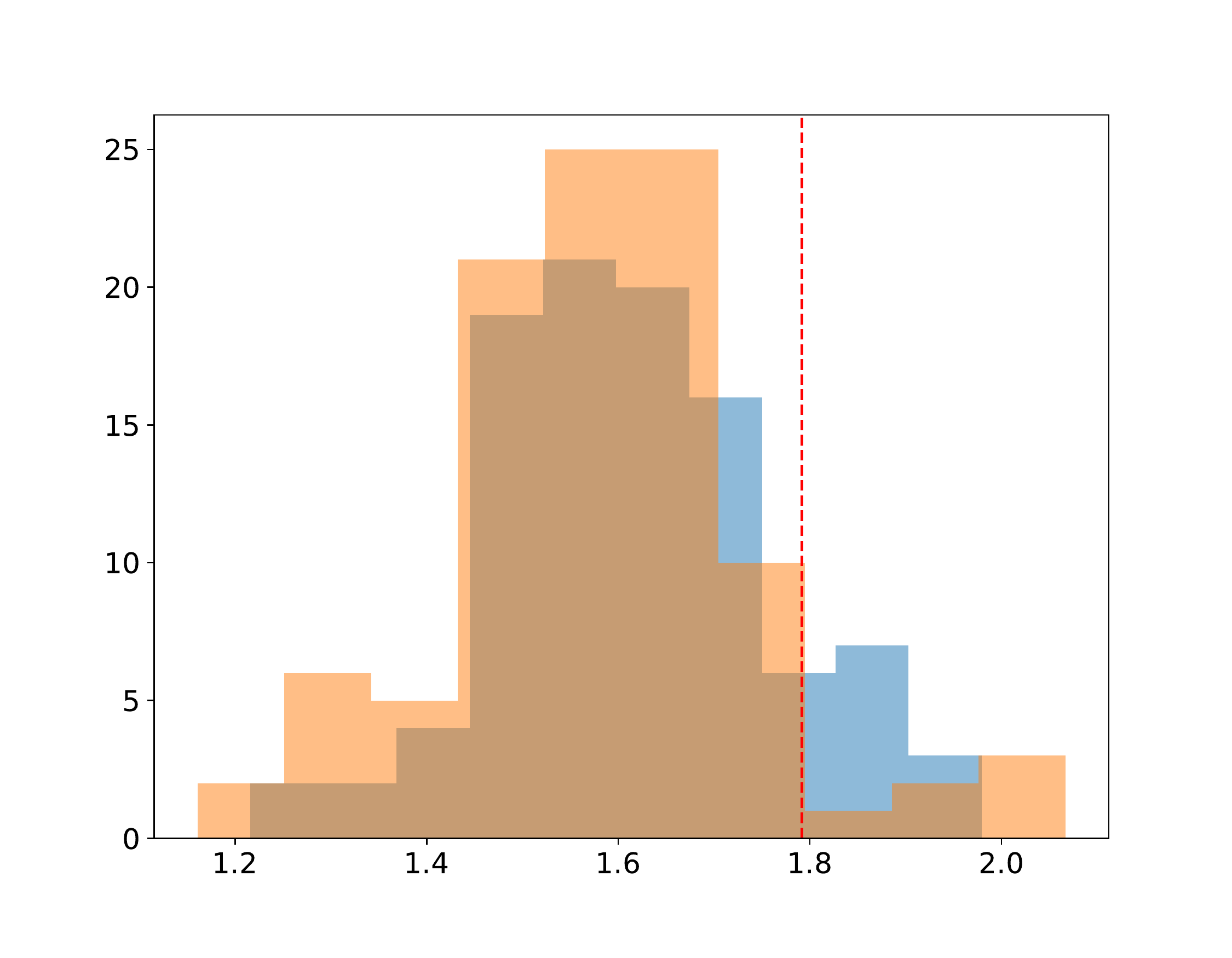} \\[-0.5em]
$\phi$ & $\ln \rho$ & $\ln \alpha$
\end{tabular}
\caption{Modification of distributions of hyperparameters after assimilation of flow data. Blue is prior distribution. Orange is posterior distribution. Red dashed lines show the values used in the data-generating model.} 
\label{fig:updates_hyperparameters}
\end{figure}

Figure~\ref{fig:samples_perm_2D} shows the first six ensemble members from the prior and the corresponding ensemble members from the posterior after data assimilation in the hierarchical model using the hybrid IES method. Six realizations are not sufficient to illustrate the results of the data assimilation, but by comparing corresponding realizations, it appears that the spread in orientation of the anisotropy is reduced and that the principal range is largely maintained. The effect of data assimilation on the hyperparameters is illustrated more quantittively in Fig.~\ref{fig:updates_hyperparameters} which compares the prior distribution for the three hyperparameters (blue), with the posterior marginal distributions (orange), and the values used in the data-generating model (dashed red line).  The posterior distribution of orientations ($\phi$) is narrower than the prior distribution, but the mean is nearly unchanged from the prior (Fig.~\ref{fig:updates_hyperparameters} (left)).  The range parameter ($\rho$) shows the largest influence of data assimilation. Short correlation ranges have been eliminated from the posterior while the mean of the posterior distribution has been shifted to a value that is larger than both the prior mean and the value that was used in the data generating model  (Fig.~\ref{fig:updates_hyperparameters} (center)). If the relationship of data to model hyperparameters was linear, one would expect the posterior mean to lie between the prior and the data-generating value, which is not what is seen here. The mean value of the ratio of correlation ranges in the principal directions ($\alpha$) has shifted to a slightly smaller value after data assimilation  (Fig.~\ref{fig:updates_hyperparameters} (right)), meaning that the mean correlation range in the second principal direction has increased slightly, but less than the range in the first principal direction.

\subsection{Discussion: benefit of hybrid approach}

\begin{figure}[htbp]
\begin{subfigure}{0.47\textwidth}
\includegraphics[width=\textwidth]{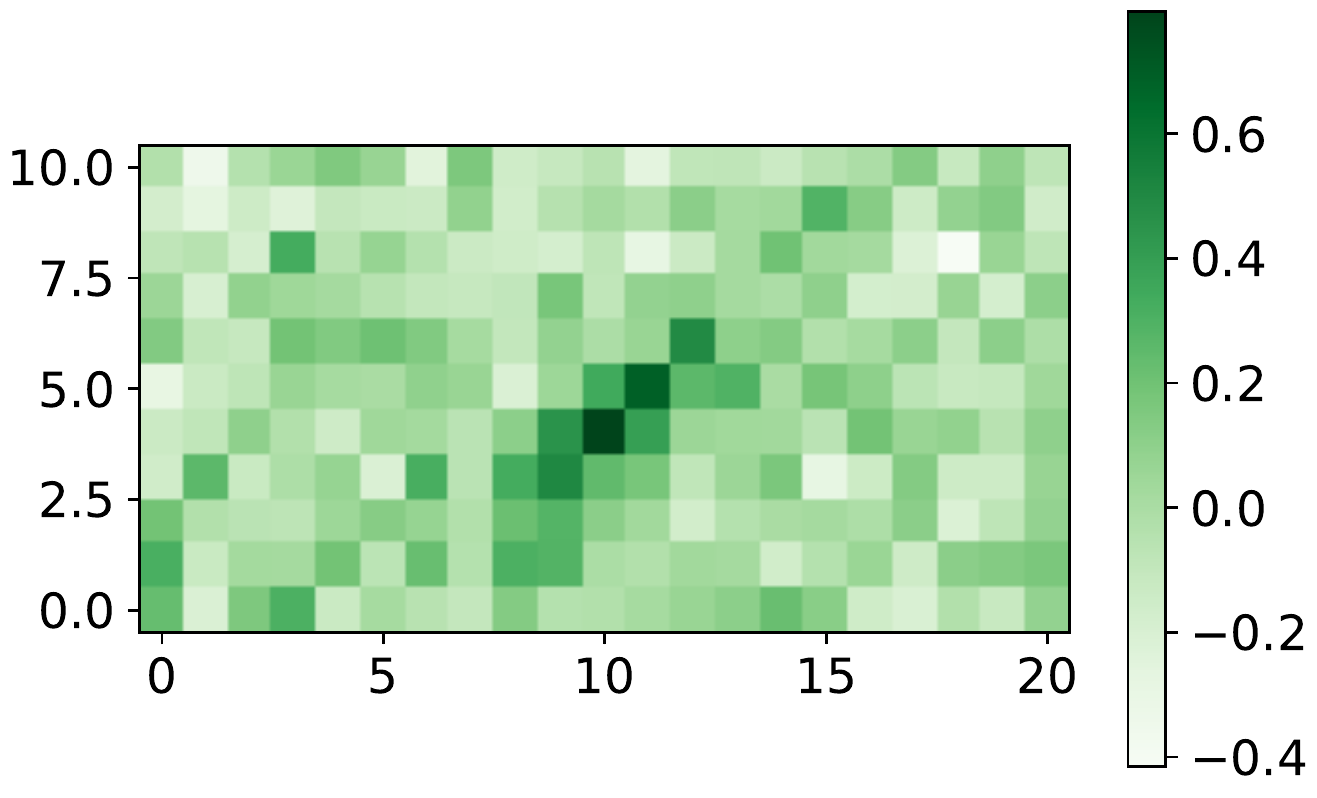} 
\caption{ensemble estimate of sensitivity using $N_e = 200$ (standard IES)} \label{fig:obs27_IES}
\end{subfigure}
\hfill
\begin{subfigure}{0.47\textwidth}
\includegraphics[width=\textwidth]{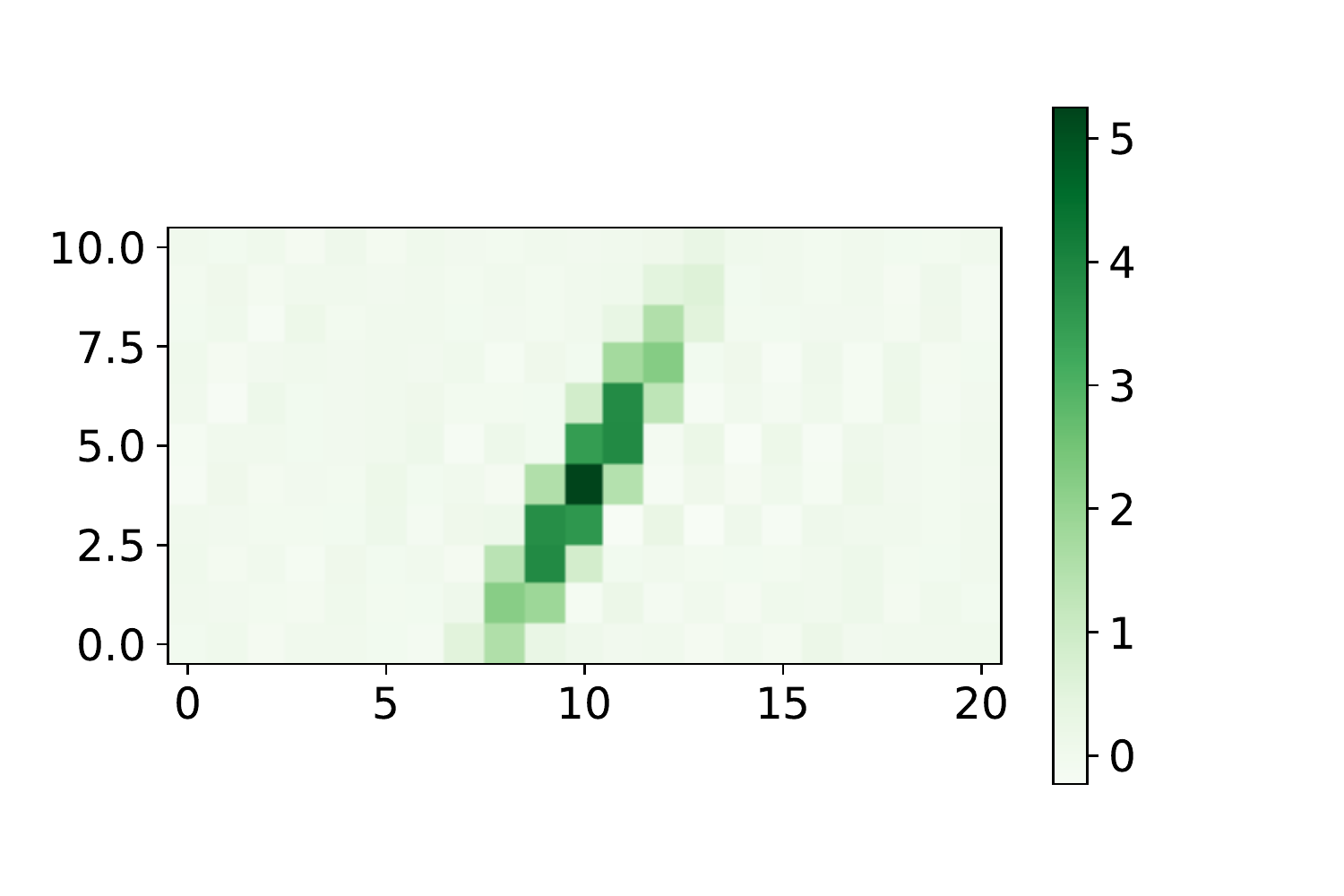} 
\caption{exact sensitivity (for realization 0)} \label{fig:obs27_exact}
\end{subfigure}
\caption{Sensitivity of observation of $m$ at (10,4) to values of $z$ at every cell. Note difference in color scales.}
\label{fig:obs27_sensitivity}
\end{figure}

In Section~\ref{sec:data_assimilation_2Dflow}, the IES method failed to assimilate data properly in a hierarchical model for which the anisotropy in the covariance was uncertain. The problem of estimating the permeability field from flow observations in this case is clearly nonlinear, but IES often works well for highly nonlinear problems so that does not appear to be the primary explanation. In this type of hierarchical problem,  it is relatively straightforward to see why a standard IES may not converge and why the hybrid IES works very well. To illustrate the relative performance of the two algorithms, consider a simpler two-dimensional hierarchical problem in which  an ensemble of realizations of $m$ are updated  from partial noisy observations of $m$. The prior covariance in this example has uncertain orientation and ranges, just as in the two-dimensional flow problem but with simpler observation and sensitivity. It is instructive in this case to look at the ensemble estimate of the  sensitivity, $\nabla_z m$, between observation of $m$ at a location near the center of the domain and $z$  for an ensemble  with 200 realizations (Fig.~\ref{fig:obs27_IES}) and compare it to the exact sensitivity for one particular realization (Fig.~\ref{fig:obs27_exact}).

The ensemble estimate of sensitivity  (Fig.~\ref{fig:obs27_IES}) suffers from several problems. The first is that although the ensemble size (200) is larger than typically used for data assimilation, the magnitudes of the spurious correlations are comparable  to the estimated magnitude at the observation location. Also, the magnitude of the ensemble estimate of the sensitivity at the observation location is almost an order of magnitude smaller than the correct value. Finally the region of significant sensitivity in the ensemble  estimate is different from the region in the exact calculation.
Increasing ensemble size would reduce the spurious correlations and improve the magnitude of sensitivity  at the peak, but the region of sensitivity  would not improve because the average orientation is different from the orientation in realization 0.

\begingroup
\setlength{\tabcolsep}{2pt} 
\renewcommand{\arraystretch}{1.} 
\begin{figure}[htbp]
\begin{tabular}{cccc}
\raisebox{2ex}{\rotatebox{90}{ensemble}} &
\includegraphics[width=0.31\textwidth]{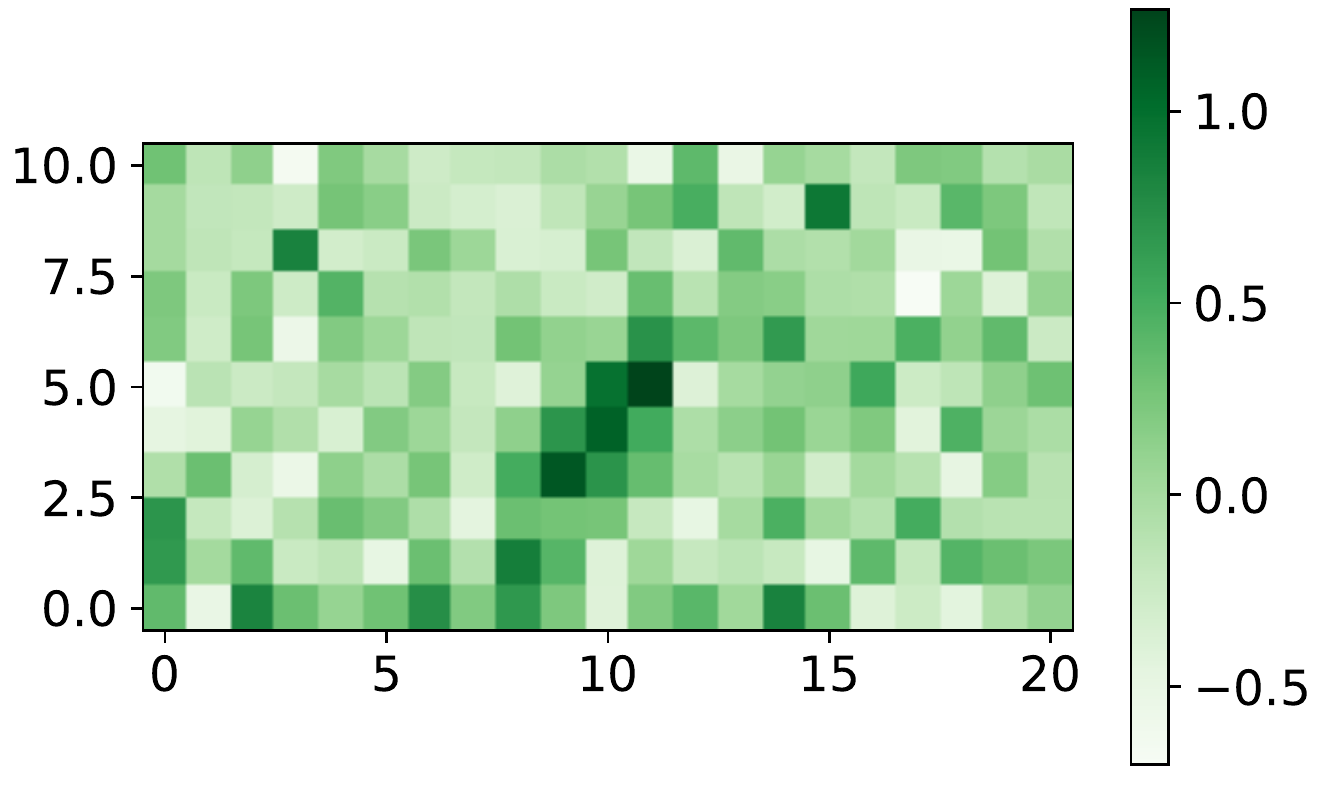} &
\includegraphics[width=0.31\textwidth]{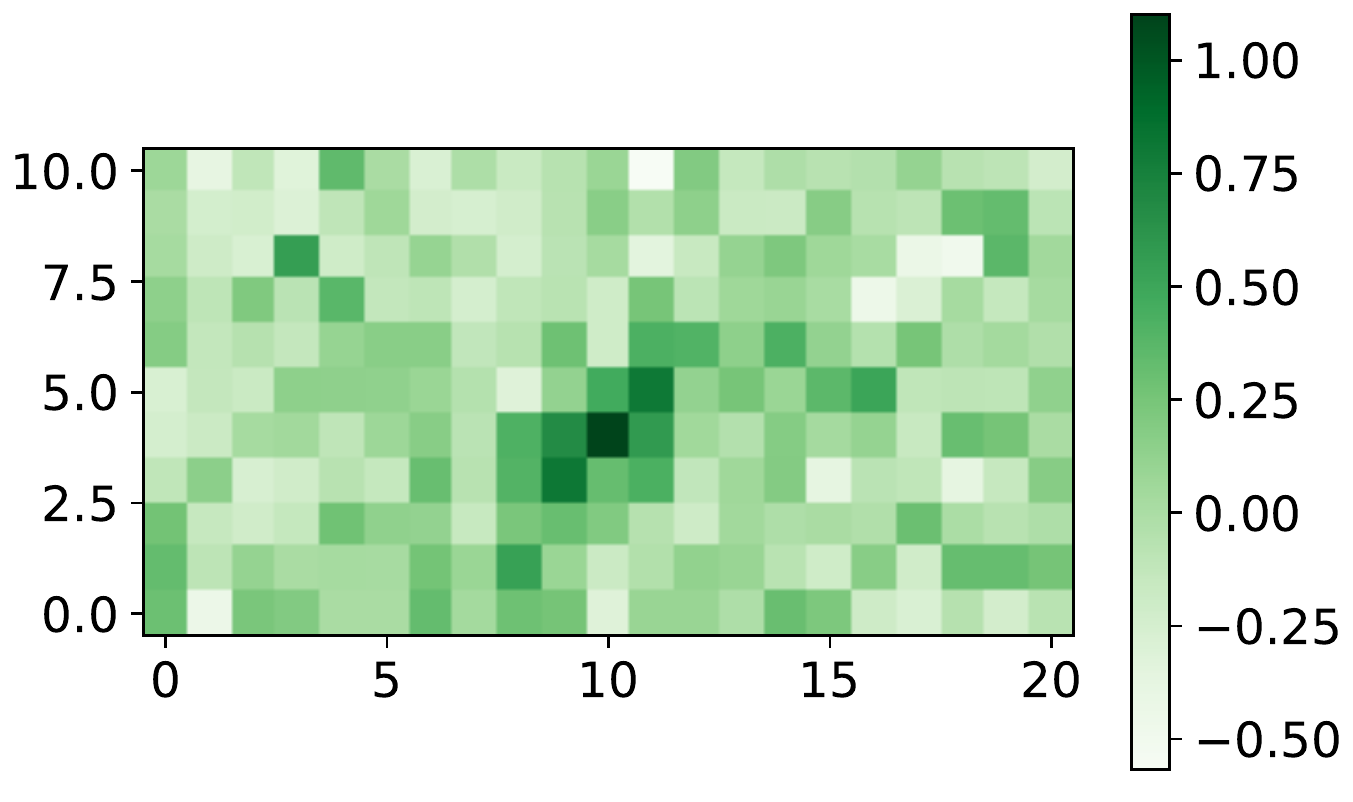} &
\includegraphics[width=0.31\textwidth]{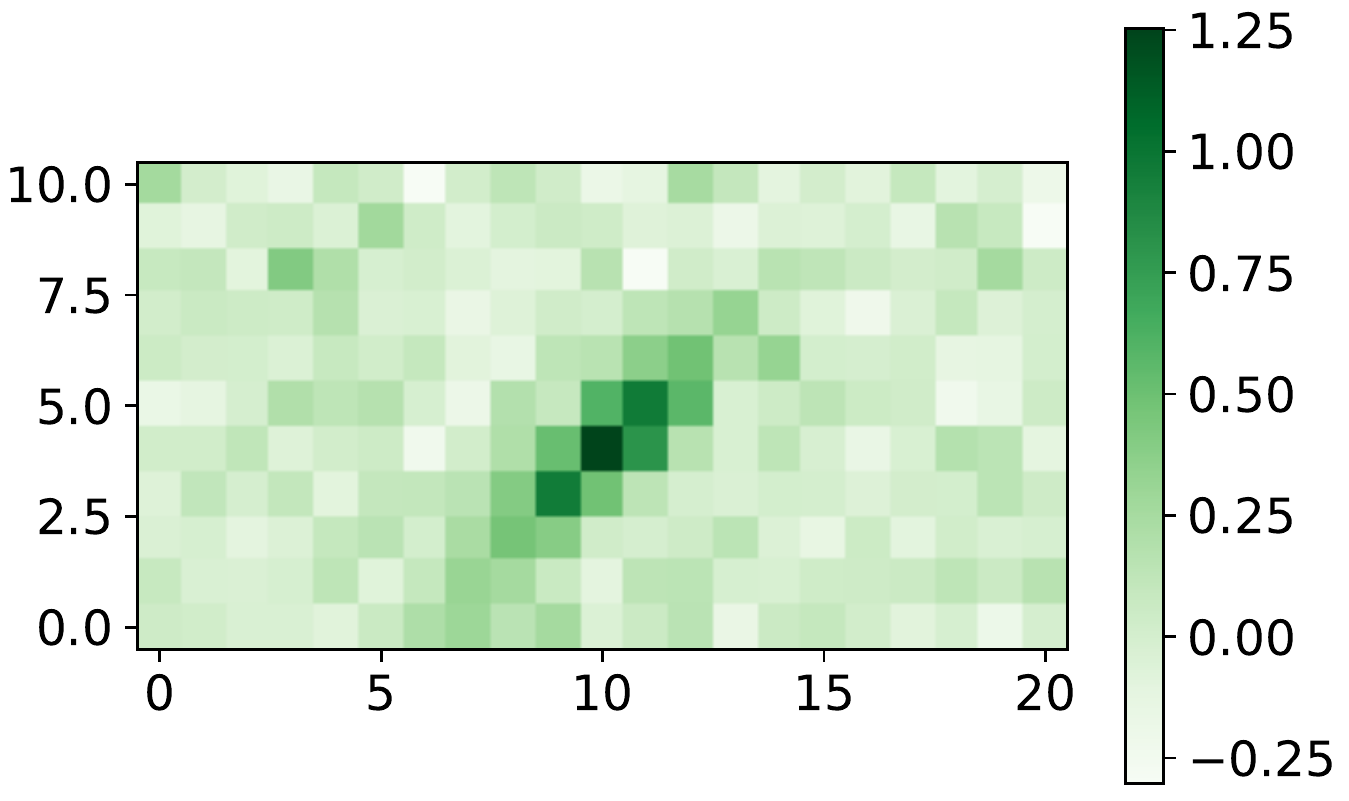} 
\\
\raisebox{4ex}{\rotatebox{90}{hybrid}} &
\includegraphics[width=0.31\textwidth]{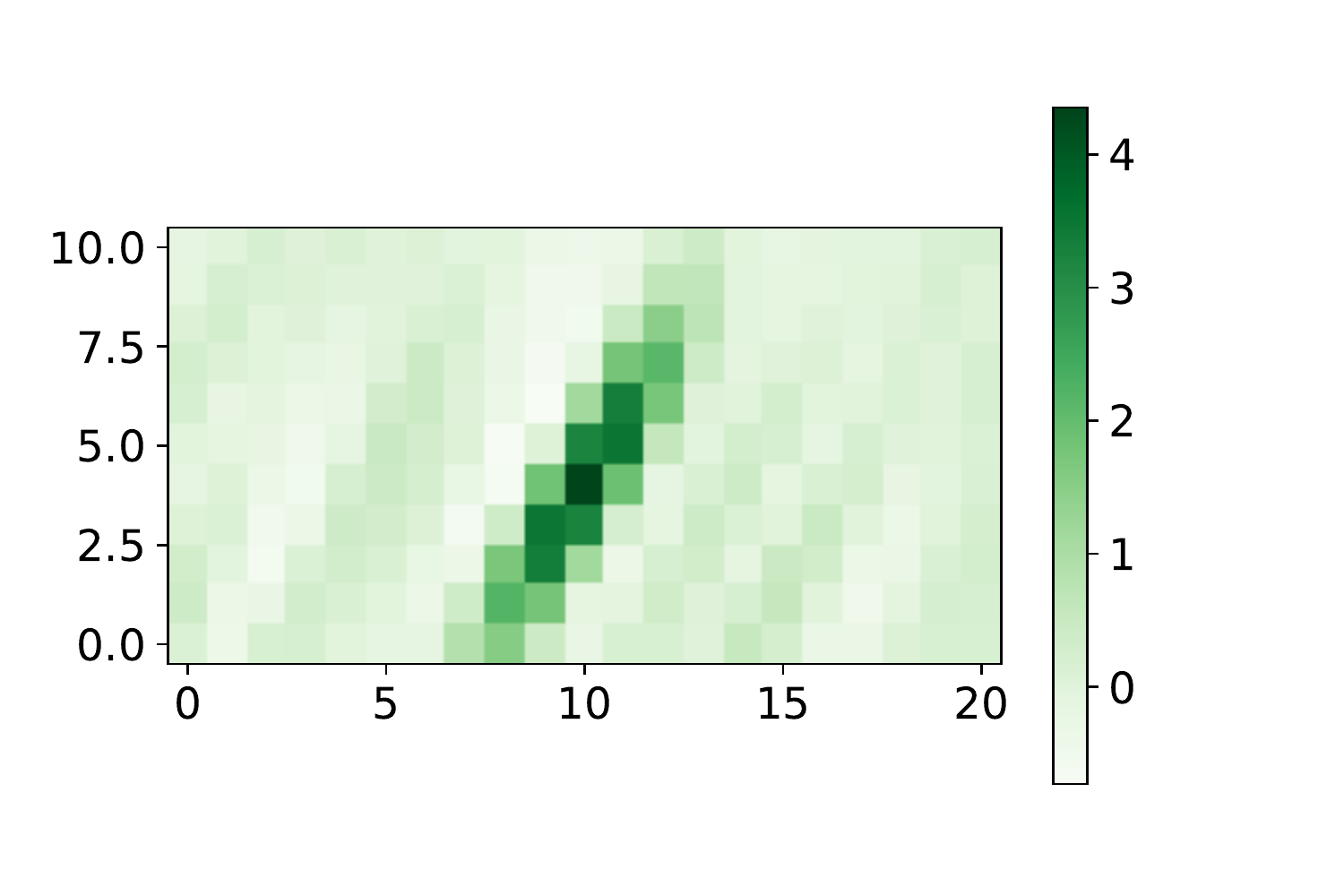} &
\includegraphics[width=0.31\textwidth]{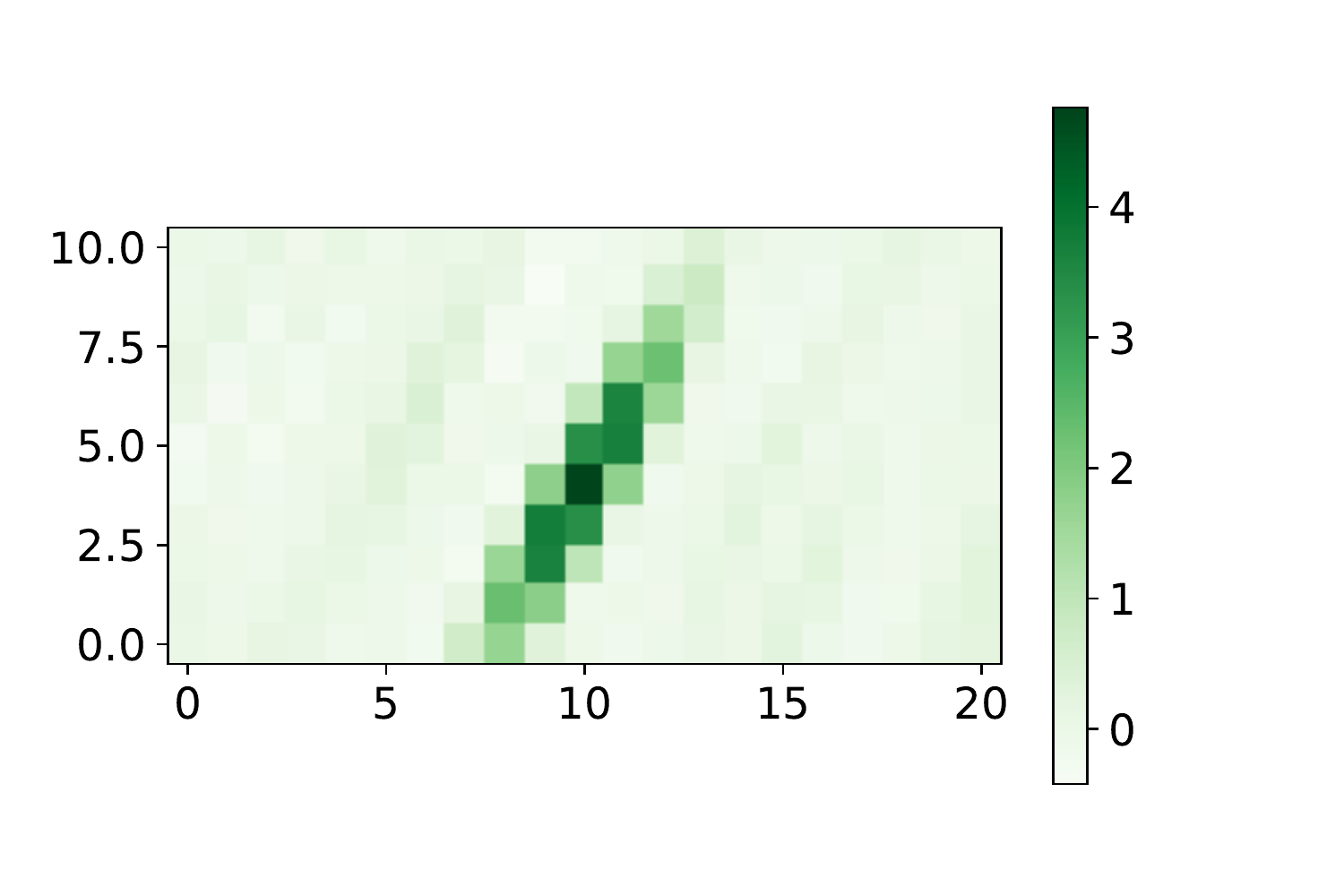} &
\includegraphics[width=0.31\textwidth]{hybrid_sens_d27_to_z_kappa4_Ne800.pdf}
\\
 & $N_e = 100$ &  $N_e = 200$ &  $N_e = 800$ 
\end{tabular}
\caption{Comparison of purely ensemble based estimates of the cross-covariance of observation at (10,4) to the values of $z$ (top row) to hybrid estimates of cross-covariance (bottom row).  }
\label{fig:obs27_cross-covariance}
\end{figure}
\endgroup

The effect of ensemble size on the estimation of the cross-covariance between the observation of $m$ at (10,4) and values of $z$ at every cell is shown in Fig.~\ref{fig:obs27_cross-covariance}. Because the covariance matrix for $z$ is the identity matrix, the cross-covariance should be the same as the sensitivity matrix if both are computed accurately. It can be seen, in fact, that the hybrid IES estimate of the cross-covariance for large ensemble size (Fig.~\ref{fig:obs27_cross-covariance}) is identical to the exact sensitivity shown in Fig.~\ref{fig:obs27_sensitivity}.  
Importantly, the estimates from the hybrid IES method with $N_e=100$ is much better than the standard IES estimate with ensemble size $N_e=800$. And again, the bias in the cross-covariance from the ensemble is large, indicating that the global estimate is not representative of the local value in this highly nonlinear problem. 

The reasons for the remarkably good results for the hybrid IES in this example are twofold. 
First,  $C_z = I$ for the non-centered hierarchical parameterization. The hybrid IES method uses the exact $C_z$, while the traditional IES uses a low-rank approximation of $C_z$, which suffers from spurious correlations.
Second, the hybrid method uses the chain rule to compute part of the sensitivity analytically. This eliminates the bias that appears when the mean hyperparameters do not match the hyperparameters of the individual realization.

\subsubsection{Summary}


In an attempt to increase the robustness of data assimilation against model misspecification, a hierarchical Gaussian model was introduced for history matching of flow data. The non-centered parameterization used here is relatively simple, but allows uncertainty in several important parameters of the prior covariance that are typically fixed during history matching. Because we assumed that the prior covariance was stationary, the total number of uncertain parameters was only slightly larger than the number of uncertain parameters in standard data assimilation but the \emph{effective} number of uncertain parameters is much larger, as the grid-based variables are independent and identically distributed (iid) in the non-centered hierarchical parameterization.

Unfortunately,  updating model parameters  using a standard ensemble Kalman-like method was not effect when the number of model cells was relatively large compared to the size of the ensemble, even though it is common in data assimilation for the number of model parameters to be much larger than the ensemble size. There appears to be two reasons for the increased difficulty with the use of data assimilation for the hierarchical model. The first is that the effect of spurious correlations is much larger with the non-centered parameterization and localization  based on the range of the prior covariance was not useful when the grid-based parameters were iid. Secondly, the addition of hyperparameters makes the problem more nonlinear so that the use of average sensitivities (or Jacobians) computed from the ensemble do not represent local sensitivities accurately. Although the problem of spurious correlations could conceptually be solved by increasing the ensemble size, that does not solve the problem of incorrect sensitivities. 

To solve these problems while retaining the advantages of the iterative ensemble smoothers (no need to derive the adjoint system and no need to store very large matrices), a hybrid IES was developed in which the ``difficult sensitivities'' such as the sensitivities of water production rate to permeability are still computed approximately using ensemble of model parameters and the ensemble of model predictions. The ``easy sensitivities'' such as the sensitivity of the permeability to the hyperparameters or to the latent iid variable $z$ are computed analytically. In a two-dimensional problem with uncertainty in the orientation of the principal axes of the covariance, hybrid estimates of $G$ and $CG\trp$ were much less noisy than estimates computed directly from the ensemble.
Additionally,  the hybrid estimates of sensitivities $G$ and cross-covariances $CG\trp$ are specific to a realization and will ill allow convergence to the correct value as ensemble size increases for linear observation operators. Finally, since each realization has its own Kalman gain matrix, the methodology is less limited by Gaussian assumptions
in the standard IES approaches.
The hybrid method provides greater robustness against nonlinearity if some of the nonlinearity is in the relationship between $z$ and $m$, as it is for the hierarchical model and also for the truncated plurigaussian model \citep{oliver:18a}.

In this paper, the hybrid IES method was tested on  a one-dimensional linear problem and on a two-dimensional-flow problem  with hierarchical parameterizations. 
In the one-dimensional problem, the
hierarchical parameterization provided uncertainty in the range and variance of the prior covariance.  The number of parameters was small compared to the ensemble size in this case, so all methods performed similarly, although the standard IES tended to produce correlation ranges that were too short.
In the two-dimensional flow problem, there was 
uncertainty in the orientation of the anisotropic covariance  and in the correlation range in the two principle directions. 
The data assimilation using  hybrid IES gave results that were much better than IES. In this problem, the number of model parameters was much greater than the number of ensemble members. The hierarchical model with hybrid IES data assimilation provided updated models with data mismatch magnitudes that were close to the expected values in all wells except one. 

The comparison of data assimilation with a hierarchical prior to data assimilation with a ``good'' and a ``bad'' prior, showed results with the hierarchical parameterization were slightly worse than results with the good prior and much better than results with the bad prior. But the good results for the hierarchical method were only obtained in conjunction with the use of the hybrid IES method of data assimilation. When the standard IES was used with the hierarchical parameterization, the mismatch to data was nearly as large as the prior mismatch.

\section{Acknowledgments}

Dean Oliver acknowledges financial support from the NORCE research cooperative research project ``Assimilating 4D Seismic Data: Big Data Into Big Models'' which is funded by industry partners, Equinor Energy AS, Lundin Energy Norway AS, Repsol Norge AS, Shell Global Solutions International B.V., TotalEnergies E\&P Norge AS and Wintershall Dea Norge AS, as well as the Research Council of Norway  through the Petromaks2 program (NFR project number: 295002).

\section{Gauss-Newton update with circular hyperparameter}
\label{sec:circular_hyperparameters}

In Sec.~\ref{sec:hybrid-IES},  the hybrid IES data assimilation algorithm is described in somewhat general terms, assuming that the prior densities for the hyperparameters of the covariance were Gaussian. Here additional background is provided for the case of geometric anisotropy in which the prior for the orientation is Gauss-von Mises.
The physical model parameters $m$ for the two-dimensional flow problem in Sec.~\ref{sec:2D_flow} are determined from the  mean $m_{pr}$ (assumed known), the uncertain covariance function $C_m$ (which is a function of correlation range $\rho$, ratio of ranges $\alpha$, and orientation of anisotropy, $\phi$) and from the stochastic variable $z$:
\begin{equation}
m = m\sbr{pr} + L(\rho, \alpha, \phi) z.
\end{equation}
Here $L L\trp = C_m$, $\rho \in (0, \infty)$ is the range parameter of the covariance, $\alpha \in (0, \infty)$ is the stretch parameter, and $\phi \in [-\pi/2, \pi/2)$ is the orientation of the principal axis of the covariance.

Because $\rho$ and $\alpha$ are required to be positive, I have chosen log-normal prior distributions for $u_1 \equiv \ln \rho$, 
\begin{equation}
p(u_1) \propto \exp\left(-\frac{1}{2} \frac{(u_1-\mu_{u_1})^2}{\sigma_{u_1}^2} \right),
\end{equation}
and for $u_2 \equiv \ln \alpha$,
\begin{equation}
p(u_2) \propto \exp\left(-\frac{1}{2} \frac{(u_2-\mu_{u_2})^2}{\sigma_{u_2}^2} \right).
\end{equation}
The prior distribution for the orientation $\phi$ cannot be Gaussian as it is defined on the interval $[-\pi/2, \pi/2)$. 
A Gauss von Mises (GVM) distribution  with parameters $\mu$ and $\kappa$ is a reasonable choice for the  probability density function of a random variable $\theta$  on the circle:
\begin{equation}
g(\phi; \mu, \kappa) = \frac{1}{ \pi I_0(\kappa)} \exp [\kappa \cos 2 (\phi - \mu_\phi )].
\label{eq:Gauss-vonMises}
\end{equation} 
where $I_0(\kappa)$ is a modified Bessel function of first kind of order 0 \citep{jammalamadaka:01}. Note that the prior for $\phi$ is the only prior that is not Gaussian (although for large values of $\kappa$ it is approximately Gaussian or a mixture of Gaussians).
Finally, the prior for $z$ is
\begin{equation} 
p(z) \propto \exp\left(-\frac{1}{2} z\trp z \right),
\end{equation}
hence, the posterior pdf for $\rho, \alpha, \phi, z$ can be written as
\begin{multline}
p(z,u,\phi \mid d) \propto \exp \left( -\frac{1}{2} (d-g(m(u,\phi , z)))\trp C_d\inv (d-g(m(u,\phi, z))) \right) \\
\times \exp \left(-\frac{1}{2} (u-\mu_{u})\trp C_u\inv (u-\mu_{u}) -\frac{1}{2} z\trp z \right) \exp [\kappa \cos 2 (\phi - \mu_\phi )]. 
\end{multline}
where I have defined $u=[u_1,u_2]$ and $C_u = \begin{bmatrix} \sigma_{u_1}^2 & 0 \\ 0 & \sigma_{u_1}^2 \end{bmatrix}$.
Computation of the maximum a posteriori point can be found from the maximizer of the  negative logarithm of the posterior pdf,
\begin{multline}
S(z,u,\phi) = \frac{1}{2} (d-g(m(u,\phi,z)))\trp C_d\inv (d-g(m(u,\phi,z))) \\ 
+  \frac{1}{2} u\trp C_u\inv u  +\frac{1}{2} z\trp z - \kappa \cos 2 (\phi - \mu_\phi ).
\end{multline}

For an RML-like approximation to sampling, minimizers of 
\begin{multline}
S^\ast(z,u,\phi) = \frac{1}{2} (d-(g(m)+\epsilon^\ast))\trp C_d\inv (d-(g(m)+\epsilon^\ast))  +  \frac{1}{2} (u - u^\ast)\trp C_u\inv(u - u^\ast)  \\ + \frac{1}{2} (z-z^\ast)\trp (z-z^\ast) - \kappa \cos 2 (\phi - \phi^\ast )
\end{multline}
are computed, where $\epsilon^\ast \sim N[0,C_d]$ is a sample of the observation error, $u^\ast \sim N[\mu_u,C_u]$, etc.

An efficient way to find a minimizer, is to solve for $\nabla S = 0$, where
\begin{equation}
\nabla S  = - G\trp C_d\inv (d-(g(m)+\epsilon^\ast)) + \begin{bmatrix}  (z-z^\ast)  \\
C_u\inv(u - u^\ast)  \\
2 \kappa \sin 2 (\phi - \phi^\ast ) \end{bmatrix}
\label{eq:grad_S}
\end{equation}
where $G = D_{z,u,\phi}g$ instead of the sensitivity with respect to $m$, as is more typical.
The Gauss-Newton approximation of the Hessian is
\begin{equation}
\nabla (\nabla S)\trp \approx G\trp C_d\inv G +
\begin{bmatrix} I_z & 0 & 0 \\
0 & C_u\inv & 0 \\
0 & 0 & 4 \kappa \cos 2 (\phi - \phi^\ast )\end{bmatrix}
\end{equation}
but  
\begin{equation}
\nabla (\nabla S)\trp \approx G\trp C_d\inv G +
\begin{bmatrix} I_z & 0 & 0 \\
0 & C_u\inv & 0 \\
0 & 0 & 4 \kappa \end{bmatrix}
\end{equation}
is used for data assimilation, as it is always positive definite.
Define the prior covariance matrix for  the model parameters as
\begin{equation}
\mathbf{C}_x =
\begin{bmatrix} I_z & 0 & 0 \\
0 & C_u & 0 \\
0 & 0 & 1/(4 \kappa) \end{bmatrix}.
\end{equation}
With this definition of the prior covaraince,  Eq.~\ref{eq:grad_S} can be rewritten
\begin{equation}
\nabla S  =  G\trp C_d\inv (g(m)+\epsilon^\ast - d) + \mathbf{C}_x\inv \begin{bmatrix}  (z-z^\ast)  \\
(u - u^\ast) \\
\frac{1}{2} \sin 2 (\phi - \phi^\ast ) \end{bmatrix}.
\end{equation}
After substitution, the $\ell$th Levenberg-Marquardt update for $x$  can then be written as
\begin{equation}\begin{split}
\delta x_{\ell+1} &  = - \frac{1}{(1+\lambda_\ell)}  \begin{bmatrix}  (z_\ell-z^\ast)  \\
(u_\ell - u^\ast) \\
\frac{1}{2} \sin 2 (\phi_\ell - \phi^\ast ) \end{bmatrix} \\
& \qquad - C_x G_\ell\trp \left( (1+ \lambda_\ell) C_{d} + G_\ell C_x G_\ell\trp \right)\inv
\\
& \qquad \qquad \times \left(g(m_\ell)+\epsilon^\ast - d - \frac{1}{(1+\lambda)} G_\ell \begin{bmatrix}  (z_\ell-z^\ast)  \\
(u_\ell - u^\ast) \\
\frac{1}{2} \sin 2 (\phi_\ell - \phi^\ast ) \end{bmatrix}   \right)
\end{split}\end{equation}
which corresponds to Eq.~\ref{eq:RML}, the only difference being the explicit treatment of uncertainty in orientation as a Gauss von Mises distribution.

\section{Sensitivity of $m$ to parameters -- geometric anisotropy}

In the hybrid method, derivatives of the square root  $L$  of the prior covariance matrix with respect to the hyperparameters are required. In order to obtain numerical results, attention is restricted to the case of geometric anisotropy \citep[][sec.~2.5.2]{chiles:99}, in which the anisotropy is defined by rotation and stretching of the coordinate system. Assuming that in the new coordinate system, the covariance can be written as
$\cov(x,x') = \cov(r)$
where $r$ is the distance between points $x$ and $x'$ after transformation.

\subsubsection{The coordinate transformation for geometric anisotropy}

The new coordinates $x^\ast$ are defined such that $x^\ast = A x$ where
\begin{equation}\begin{split}
A & = \begin{bmatrix}1 & 0 \\ 0 & \alpha \end{bmatrix} \begin{bmatrix}\cos \phi & \sin \phi \\ -\sin \phi & \cos \phi \end{bmatrix} \\
& =
\begin{bmatrix}\cos \phi & \sin \phi \\ - \alpha \sin \phi & \alpha \cos \phi \end{bmatrix}
\end{split}\end{equation}
The distance (for covariance computation) is thus of the form $r=\sqrt{(x^\ast)\trp x^\ast} = \sqrt{x\trp H x}$ where
\begin{equation}
\begin{split}
H & = \begin{bmatrix}\cos \phi & - \alpha \sin \phi \\  \sin \phi & \alpha \cos \phi \end{bmatrix}
\begin{bmatrix}\cos \phi & \sin \phi \\ - \alpha \sin \phi & \alpha \cos \phi \end{bmatrix} \\
& = \begin{bmatrix}\cos^2 \phi + \alpha^2 \sin^2 \phi & (1-\alpha^2) \sin \phi \cos \phi \\ (1-\alpha^2) \sin \phi \cos \phi & \sin^2 \phi + \alpha^2 \cos^2 \phi \end{bmatrix}.
\end{split} \label{eq:H}
\end{equation}
Recall that $m = m\spr{pr} + L z$, where $C_x = L L\trp$ and 
\begin{equation} 
M_x  = \begin{bmatrix} L &  (\frac{\partial }{\partial u_1} L) z & (\frac{\partial }{\partial u_2} L) z &  (\frac{\partial }{\partial \phi} L) z \end{bmatrix} .
\label{eq:M_x}
\end{equation}
For the flow problem in Sec.~\ref{sec:2D_flow}, it was assumed that the covariance took the form
\begin{equation}
C(r) = \sigma^2 e^{-3r^2/\rho^2}.
\label{eq:2D_gaussian}
\end{equation}
The convolution ``square root'' in 2D is  
\begin{equation}
f(r)  = \frac{2 \sigma \sqrt{3} }{\rho \sqrt{\pi}}   \exp\Bigl( -\frac{6 r^2}{\rho^2} \Bigr).
\end{equation}
The matrix square root, $L$, will simply be the discrete version of this functional form and effects of boundaries will be ignored. 

\subsection{Derivative of $f$ wrt $\phi$}

\begin{equation}\begin{split}
\frac{\partial f}{\partial \phi} 
&   = \frac{2 \sigma \sqrt{3}}{\rho \sqrt{\pi}} \Bigl( - \frac{6}{\rho^2} \Bigr) \exp \Bigl( -\frac{6 r^2}{\rho^2} \Bigr) \frac{\partial r^2}{\partial \phi} 
\\
& = - \frac{4 \sigma \sqrt{27}}{\rho^3 \sqrt{\pi}}
\exp \Bigl( -\frac{6 r^2}{\rho^2} \Bigr) \Bigr( x\trp \frac{\partial H}{\partial \phi} x \Bigl) 
\\
&  = - \frac{6 }{\rho^2 }
\Bigr( x\trp \frac{\partial H}{\partial \phi} x \Bigl) f(r).  
\end{split}\end{equation}
From Eq.~\ref{eq:H}, it is straightforward to show that
\begin{equation}
\frac{\partial H}{\partial \phi}  =
(1-\alpha^2) \begin{bmatrix} - \sin 2 \phi & \cos 2 \phi  \\  \cos 2 \phi & \sin 2 \phi \end{bmatrix}
\end{equation}
(Note that if $\alpha = 1$ the covariance is isotropic and there is no sensitivity to orientation.)

\subsection{Derivative of $f$ wrt $\alpha$}

\begin{equation}\begin{split}
\frac{\partial f}{\partial \alpha} &  = \frac{2 \sigma \sqrt{3}}{\rho \sqrt{\pi}} \Bigl( - \frac{6}{\rho^2} \Bigr) \exp \Bigl( -\frac{6 r^2}{\rho^2} \Bigr) \frac{\partial r^2}{\partial \alpha} 
\\
& = - \frac{4 \sigma \sqrt{27}}{\rho^3 \sqrt{\pi}}
\exp \Bigl( -\frac{6 r^2}{\rho^2} \Bigr) \Bigr( x\trp \frac{\partial H}{\partial \alpha} x \Bigl)  
\end{split}\end{equation}
where
\begin{equation}
\frac{\partial H}{\partial \alpha}  =
 \alpha \begin{bmatrix} 1 - \cos 2 \phi &  - \sin 2 \phi  \\ -  \sin 2 \phi  & 1 + \cos 2 \phi \end{bmatrix}
\end{equation}

\subsection{Derivative of $f$ wrt $\rho$}

\begin{equation} 
\frac{\partial f}{\partial \rho} 
 = \left( \frac{12 r^2}{ \rho^2} -1  \right) \left( \frac{ 2 \sigma \sqrt{3}}{ \rho^2 \sqrt{\pi}}    \right) \exp \left( -\frac{6 r^2}{\rho^2} \right)  
\end{equation}
%



\begin{thebibliography}{}

\bibitem[\protect\astroncite{Ba et~al.}{2022}]{ba:22}
Ba, Y., de~Wiljes, J., Oliver, D.~S., and Reich, S. (2022).
\newblock Randomized maximum likelihood based posterior sampling.
\newblock {\em Computat. Geosci.}, 26(1):217--239.

\bibitem[\protect\astroncite{Burgers et~al.}{1998}]{burgers:98}
Burgers, G., van Leeuwen, P.~J., and Evensen, G. (1998).
\newblock Analysis scheme in the ensemble {K}alman filter.
\newblock {\em Mon. Weather Rev.}, 126(6):1719--1724.

\bibitem[\protect\astroncite{Chada et~al.}{2018}]{chada:18}
Chada, N.~K., Iglesias, M.~A., Roininen, L., and Stuart, A.~M. (2018).
\newblock Parameterizations for ensemble {K}alman inversion.
\newblock {\em Inverse Problems}, 34(5):055009.

\bibitem[\protect\astroncite{Chen and Oliver}{2010}]{chen:10a}
Chen, Y. and Oliver, D.~S. (2010).
\newblock Parameterization techniques to improve mass conservation and data
  assimilation for ensemble {K}alman filter ({SPE} 133560).
\newblock In {\em SPE Western Regional Meeting, 27--29 May 2010, Anaheim,
  California, USA}.

\bibitem[\protect\astroncite{Chen and Oliver}{2012}]{chen:12a}
Chen, Y. and Oliver, D.~S. (2012).
\newblock Ensemble randomized maximum likelihood method as an iterative
  ensemble smoother.
\newblock {\em Mathematical Geosciences}, 44(1):1--26.

\bibitem[\protect\astroncite{Chen and Oliver}{2013}]{chen:13}
Chen, Y. and Oliver, D.~S. (2013).
\newblock Levenberg-{M}arquardt forms of the iterative ensemble smoother for
  efficient history matching and uncertainty quantification.
\newblock {\em Computational Geosciences}, 17(4):689--703.

\bibitem[\protect\astroncite{Chil{\`e}s and Delfiner}{1999}]{chiles:99}
Chil{\`e}s, J.-P. and Delfiner, P. (1999).
\newblock {\em Geostatistics: Modeling Spatial Uncertainty}.
\newblock John Wiley \& Sons, New York.

\bibitem[\protect\astroncite{Dunlop et~al.}{2020}]{dunlop:20}
Dunlop, M.~M., Helin, T., and Stuart, A.~M. (2020).
\newblock Hyperparameter estimation in {Bayesian} {MAP} estimation:
  Parameterizations and consistency.
\newblock {\em The SMAI Journal of Computational Mathematics}, 6:69--100.

\bibitem[\protect\astroncite{Emerick}{2016}]{emerick:16}
Emerick, A.~A. (2016).
\newblock Analysis of the performance of ensemble-based assimilation of
  production and seismic data.
\newblock {\em Journal of Petroleum Science and Engineering}, 139:219--239.

\bibitem[\protect\astroncite{Emerick and Reynolds}{2013}]{emerick:13a}
Emerick, A.~A. and Reynolds, A.~C. (2013).
\newblock Ensemble smoother with multiple data assimilation.
\newblock {\em Computers \& Geosciences}, 55:3--15.

\bibitem[\protect\astroncite{Evensen}{1994}]{evensen:94}
Evensen, G. (1994).
\newblock Sequential data assimilation with a nonlinear quasi-geostrophic model
  using {Monte} {Carlo} methods to forecast error statistics.
\newblock {\em J. Geophys. Res.}, 99(C5):10143--10162.

\bibitem[\protect\astroncite{Fox and Norton}{2016}]{fox:16}
Fox, C. and Norton, R.~A. (2016).
\newblock Fast sampling in a linear-{G}aussian inverse problem.
\newblock {\em SIAM/ASA Journal on Uncertainty Quantification},
  4(1):1191--1218.

\bibitem[\protect\astroncite{Gneiting et~al.}{2010}]{gneiting:10}
Gneiting, T., Kleiber, W., and Schlather, M. (2010).
\newblock Mat{\'e}rn cross-covariance functions for multivariate random fields.
\newblock {\em Journal of the American Statistical Association},
  105(491):1167--1177.

\bibitem[\protect\astroncite{Houtekamer and Mitchell}{1998}]{houtekamer:98}
Houtekamer, P.~L. and Mitchell, H.~L. (1998).
\newblock Data assimilation using an ensemble {K}alman filter technique.
\newblock {\em Mon. Weather Rev.}, 126(3):796--811.

\bibitem[\protect\astroncite{Jammalamadaka and
  Sengupta}{2001}]{jammalamadaka:01}
Jammalamadaka, S.~R. and Sengupta, A. (2001).
\newblock {\em Topics in Circular Statistics}, volume~5.
\newblock World Scientific Publishing.

\bibitem[\protect\astroncite{Kitanidis}{1995}]{kitanidis:95}
Kitanidis, P.~K. (1995).
\newblock Quasi-linear geostatistical theory for inversing.
\newblock {\em Water Resour. Res.}, 31(10):2411--2419.

\bibitem[\protect\astroncite{Li et~al.}{2010}]{li:10a}
Li, G., Han, M., Banerjee, R., and Reynolds, A.~C. (2010).
\newblock Integration of well-test pressure data into heterogeneous geological
  reservoir models.
\newblock {\em SPE Reservoir Evaluation \& Engineering}, 13(03):496--508.

\bibitem[\protect\astroncite{Lindeberg}{1990}]{lindeberg:90}
Lindeberg, T. (1990).
\newblock Scale-space for discrete signals.
\newblock {\em IEEE Transactions on Pattern Analysis and Machine Intelligence},
  12(3):234--254.

\bibitem[\protect\astroncite{Malinverno and Briggs}{2004}]{malinverno:04}
Malinverno, A. and Briggs, V.~A. (2004).
\newblock Expanded uncertainty quantification in inverse problems: Hierarchical
  {B}ayes and empirical {B}ayes.
\newblock {\em Geophysics}, 69(4):1005--1016.

\bibitem[\protect\astroncite{Moore and Doherty}{2005}]{moore:05}
Moore, C. and Doherty, J. (2005).
\newblock Role of the calibration process in reducing model predictive error.
\newblock {\em Water Resources Research}, 41(5).
\newblock W05020.

\bibitem[\protect\astroncite{Myrseth and Omre}{2010}]{myrseth:10}
Myrseth, I. and Omre, H. (2010).
\newblock Hierarchical ensemble {K}alman filter.
\newblock {\em SPE Journal}, 15(2):569--580.

\bibitem[\protect\astroncite{Oliver}{1995}]{oliver:95}
Oliver, D.~S. (1995).
\newblock Moving averages for {G}aussian simulation in two and three
  dimensions.
\newblock {\em Mathematical Geology}, 27(8):939--960.

\bibitem[\protect\astroncite{Oliver and Alfonzo}{2018}]{oliver:18}
Oliver, D.~S. and Alfonzo, M. (2018).
\newblock Calibration of imperfect models to biased observations.
\newblock {\em Comput. Geosci.}, 22(1):145--161.

\bibitem[\protect\astroncite{Oliver and Chen}{2018}]{oliver:18a}
Oliver, D.~S. and Chen, Y. (2018).
\newblock Data assimilation in truncated plurigaussian models: impact of the
  truncation map.
\newblock {\em Mathematical Geosciences}, 50(8):867--893.

\bibitem[\protect\astroncite{Oliver et~al.}{1996}]{oliver:96e}
Oliver, D.~S., He, N., and Reynolds, A.~C. (1996).
\newblock Conditioning permeability fields to pressure data.
\newblock In {\em Proceedings of the European Conference on the Mathematics of
  Oil Recovery, V}, pages 1--11.

\bibitem[\protect\astroncite{Oliver et~al.}{2008}]{oliver:08}
Oliver, D.~S., Reynolds, A.~C., and Liu, N. (2008).
\newblock {\em Inverse Theory for Petroleum Reservoir Characterization and
  History Matching}.
\newblock Cambridge University Press, Cambridge.

\bibitem[\protect\astroncite{Papaspiliopoulos
  et~al.}{2003}]{papaspiliopoulos:03}
Papaspiliopoulos, O., Roberts, G.~O., and Sk{\"o}ld, M. (2003).
\newblock Non-centered parameterisations for hierarchical models and data
  augmentation.
\newblock In Bernardo, J.~M., Bayarri, M.~J., Berger, J.~O., Dawid, A.~P.,
  Heckerman, D., Smith, A. F.~M., and West, M., editors, {\em Bayesian
  Statistics 7: Proceedings of the Seventh Valencia International Meeting},
  volume 307, pages 307--326. Oxford University Press, USA.

\bibitem[\protect\astroncite{Park et~al.}{2013}]{park:13}
Park, H., Scheidt, C., Fenwick, D., Boucher, A., and Caers, J. (2013).
\newblock History matching and uncertainty quantification of facies models with
  multiple geological interpretations.
\newblock {\em Comput. Geosci.}, 17(4):609--621.

\bibitem[\protect\astroncite{Reich}{2011}]{reich:11}
Reich, S. (2011).
\newblock A dynamical systems framework for intermittent data assimilation.
\newblock {\em {BIT} Numer Math}, 51(1):235--249.

\bibitem[\protect\astroncite{Roininen et~al.}{2019}]{roininen:19}
Roininen, L., Girolami, M., Lasanen, S., and Markkanen, M. (2019).
\newblock Hyperpriors for {M}at{\'e}rn fields with applications in {B}ayesian
  inversion.
\newblock {\em Inverse Problems and Imaging}, 13(1):1--29.

\bibitem[\protect\astroncite{Rue and Held}{2005}]{rue:05}
Rue, H. and Held, L. (2005).
\newblock {\em Gaussian Markov Random Fields: Theory and Applications}.
\newblock CRC Press.

\bibitem[\protect\astroncite{Rue and Martino}{2007}]{rue:07}
Rue, H. and Martino, S. (2007).
\newblock Approximate {B}ayesian inference for hierarchical {G}aussian {M}arkov
  random field models.
\newblock {\em Journal of Statistical Planning and Inference},
  137(10):3177--3192.
\newblock Special Issue: Bayesian Inference for Stochastic Processes.

\bibitem[\protect\astroncite{Scales and Tenorio}{2001}]{scales:01}
Scales, J.~A. and Tenorio, L. (2001).
\newblock Prior information and uncertainty in inverse problems.
\newblock {\em Geophysics}, 66(2):389--397.

\bibitem[\protect\astroncite{Stojkovic et~al.}{2017}]{stojkovic:17}
Stojkovic, I., Jelisavcic, V., Milutinovic, V., and Obradovic, Z. (2017).
\newblock Fast sparse {G}aussian {M}arkov random fields learning based on
  {C}holesky factorization.
\newblock In {\em Proceedings of the Twenty-Sixth International Joint
  Conference on Artificial Intelligence (IJCAI-17)}, pages 2758--2764.

\bibitem[\protect\astroncite{Tsyrulnikov and Rakitko}{2017}]{tsyrulnikov:17}
Tsyrulnikov, M. and Rakitko, A. (2017).
\newblock A hierarchical {B}ayes ensemble {K}alman filter.
\newblock {\em Physica D: Nonlinear Phenomena}, 338:1--16.

\bibitem[\protect\astroncite{Zhang and Oliver}{2011}]{zhang:11a}
Zhang, Y. and Oliver, D.~S. (2011).
\newblock History matching using a multiscale stochastic model with the
  ensemble {K}alman filter: a field case study.
\newblock {\em SPE Journal}, 16(2):307--317.

\bibitem[\protect\astroncite{Zhou et~al.}{2018}]{zhou:18}
Zhou, Q., Liu, W., Li, J., and Marzouk, Y.~M. (2018).
\newblock An approximate empirical {B}ayesian method for large-scale
  linear-{G}aussian inverse problems.
\newblock {\em Inverse Problems}, 34(9):095001.

\end{thebibliography}

\bibliographystyle{apa}

\end{document}